\documentclass[preprint]{emulateapj}

\usepackage{subfigure}
\usepackage{graphicx}

\received{}
\revised{}
\accepted{}

\shortauthors{Boyer et al.}
\shorttitle{``SAGE-SMC: Evolved Stars''}

\begin{document}

\title{Surveying the Agents of Galaxy Evolution in the Tidally-Stripped, Low Metallicity Small Magellanic Cloud (SAGE-SMC). II. Cool Evolved Stars}

\author{Martha~L.~Boyer\altaffilmark{1},
  Sundar~Srinivasan\altaffilmark{2},
  Jacco~Th.~van~Loon\altaffilmark{3},
  Iain~McDonald\altaffilmark{4},
  Margaret~Meixner\altaffilmark{1},
  Dennis~Zaritsky\altaffilmark{5}
  Karl~D.~Gordon\altaffilmark{1},
  F.~Kemper\altaffilmark{6,4},
  Brian~Babler\altaffilmark{7},
  Miwa~Block\altaffilmark{5},
  Steve~Bracker\altaffilmark{7},
  Charles W.~Engelbracht\altaffilmark{5},
  Joe~Hora\altaffilmark{8},
  Remy~Indebetouw\altaffilmark{9},
  Marilyn~Meade\altaffilmark{7},
  Karl~Misselt\altaffilmark{5},
  Thomas~Robitaille\altaffilmark{8},
  Marta~Sewi\l o\altaffilmark{10},
  Bernie~Shiao\altaffilmark{1},
  Barbara~Whitney\altaffilmark{11}}

  \altaffiltext{1}{STScI, 3700 San Martin Drive, Baltimore, MD 21218 USA; mboyer@stsci.edu}
  \altaffiltext{2}{Institut d'Astrophysiqe de Paris, CNRS UPR 341, 98bis, Boulevard Arago, Paris, F-75014}
  \altaffiltext{3}{Astrophysics Group, Lennard-Jones Laboratories,
  Keele University, Staffordshire ST5 5BG, UK}
  \altaffiltext{4}{Jodrell Bank Centre for Astrophysics, Alan Turing Building, University of Manchester, M13 9PL, UK}
  \altaffiltext{5}{Steward Observatory, University of Arizona, 933
  North Cherry Avenue, Tucson, AZ 85721 USA}
  \altaffiltext{6}{Academia Sinica Institute of Astronomy and Astrophysics, PO Box 23-141, Taipei 10617, Taiwan}
  \altaffiltext{7}{Department of Astronomy, University of Wisconsin, Madison, 475 North Charter Street, Madison, WI 53706-1582 USA}
  \altaffiltext{8}{Harvard-Smithsonian Center for Astrophysics, 60 Garden Street, MS 65, Cambridge, MA 02138-1516 USA}
  \altaffiltext{9}{Department of Astronomy, University of Virginia, P.O. Box 3818, Charlottesville, VA 22903-0818 USA}
  \altaffiltext{10}{Department of Physics and Astronomy, The Johns Hopkins University, Homewood Campus,
  Baltimore, MD 21218 USA}
  \altaffiltext{11}{Space Science Institute, 4750 Walnut Street, Suite 205, Boulder, CO 80301 USA}

\begin{abstract}
We investigate the infrared (IR) properties of cool, evolved stars in
the Small Magellanic Cloud (SMC), including the red giant branch (RGB)
stars and the dust-producing red supergiant (RSG) and asymptotic giant
branch (AGB) stars using observations from the {\it Spitzer Space
Telescope} Legacy program entitled: ``Surveying the Agents of Galaxy
Evolution in the Tidally-stripped, Low Metallicity SMC", or
SAGE-SMC. The survey includes, for the first time, full spatial
coverage of the SMC bar, wing, and tail regions at infrared (IR)
wavelengths (3.6 -- 160~\micron). We identify evolved stars using a
combination of near-IR and mid-IR photometry and point out a new
feature in the mid-IR color--magnitude diagram that may be due to
particularly dusty O-rich AGB stars. We find that the RSG and AGB
stars each contribute $\approx 20\%$ of the global SMC flux (extended
$+$ point-source) at 3.6~\micron{}, which emphasizes the importance of
both stellar types to the integrated flux of distant metal-poor
galaxies. The equivalent SAGE survey of the higher-metallicity Large
Magellanic Cloud (SAGE-LMC) allows us to explore the influence of
metallicity on dust production. We find that the SMC RSG stars are
less likely to produce a large amount of dust (as indicated by the
$[3.6]-[8]$ color). There is a higher fraction of carbon-rich stars in
the SMC, and these stars appear to able to reach colors as red as
their LMC counterparts, indicating that C-rich dust forms efficiently
in both galaxies. A preliminary estimate of the dust production in AGB
and RSG stars reveals that the extreme C-rich AGB stars dominate the
dust input in both galaxies, and that the O-rich stars may play a
larger role in the LMC than in the SMC.

\end{abstract}


\vfill\eject
\section{INTRODUCTION}
\label{sec:intro}

The recent {\it Spitzer Space Telescope} \citep{werner04,gehrz07}
Legacy program entitled ``Surveying the Agents of Galaxy Evolution in
the Tidally-stripped, Low Metallicity Small Magellanic Cloud''
\citep[SAGE-SMC;][]{gordon11} has provided a spatially and
photometrically complete infrared (IR) survey of the evolved star
population in the SMC. The resulting database allows us to study
thermal emission from circumstellar dust created around stars in the
late stages of evolution and places constraints on the total dust
budget of the SMC. In this work, we present an overview of the cool,
evolved stars in the SMC, specifically Red Giant Branch (RGB) stars,
Asymptotic Giant Branch (AGB) stars, and Red Supergiants (RSGs). We
compare our findings to those of the SAGE survey of the Large
Magellanic Cloud
\citep[SAGE-LMC;][]{blum06,meixner06,bonanos09,bonanos10,srinivasan09,vanloon10}.

The RGB is among the most prominent features of the near-IR
  color--magnitude diagram (CMD). All stars with mass $0.5 \lesssim M
  \lesssim 8~M_\odot$ spend time on the RGB after exhausting core
  hydrogen and before igniting core helium \citep{becker81}. Cool
  effective temperatures (3000 -- 5000~K) cause their bolometric
  luminosities to peak near 1~\micron{}, requiring mid-IR photometry
  to constrain their basic stellar parameters. Little to no dust
  ($\lesssim 10^{-3}~M_\odot$) is expected to form around RGB stars,
  and mass-loss rates are typically lower than
  $10^{-8}~M_\odot~{\rm yr}^{-1}$
  \citep[e.g.,][]{boyer09ngc362,boyer10,mcdonald09,mcdonald11dust,mcdonald11param,momany11,mcdonald11rgb}

\begin{figure*}
\vbox{ \includegraphics[width=0.5\textwidth]{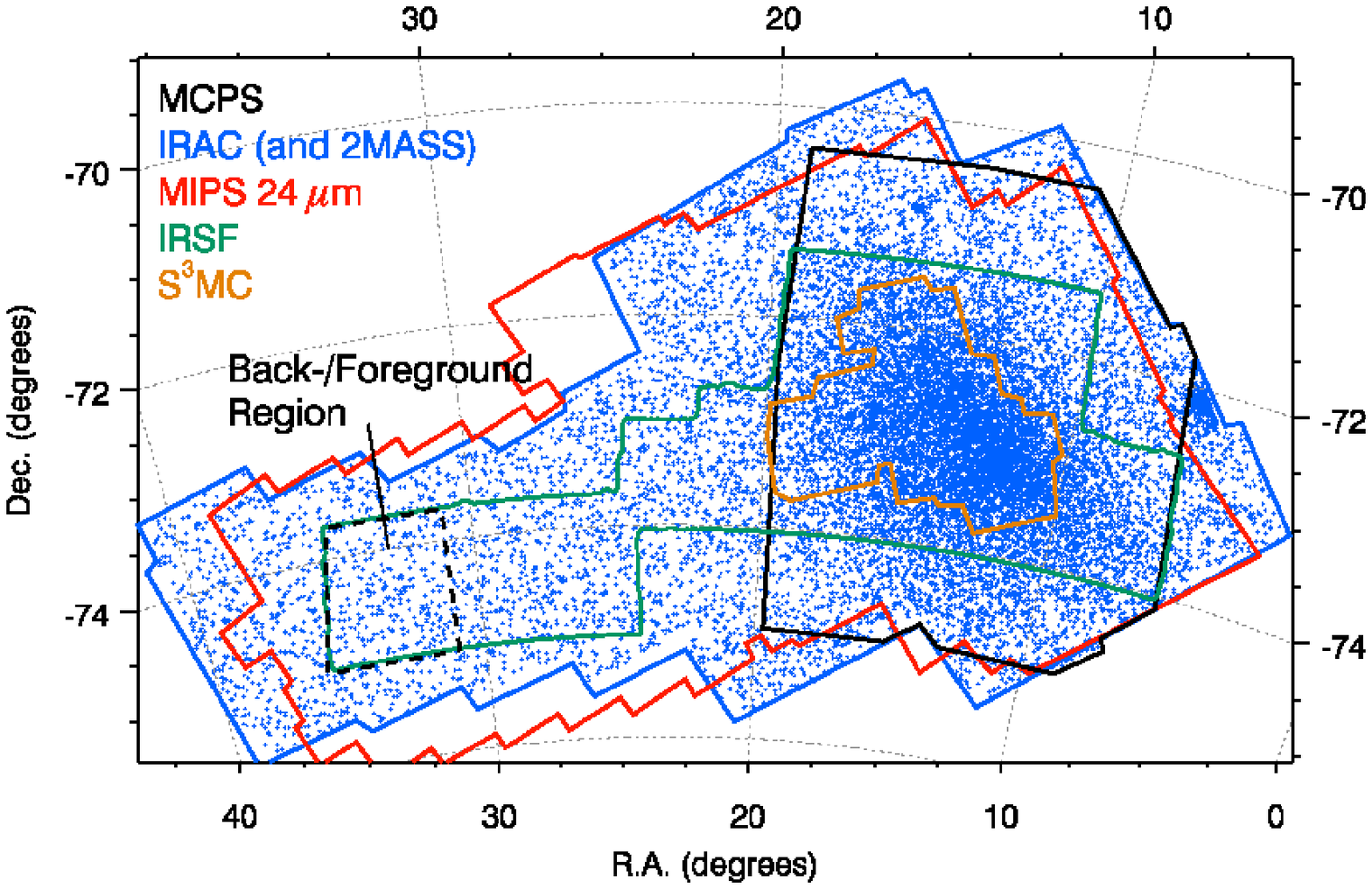}
\includegraphics[width=0.5\textwidth]{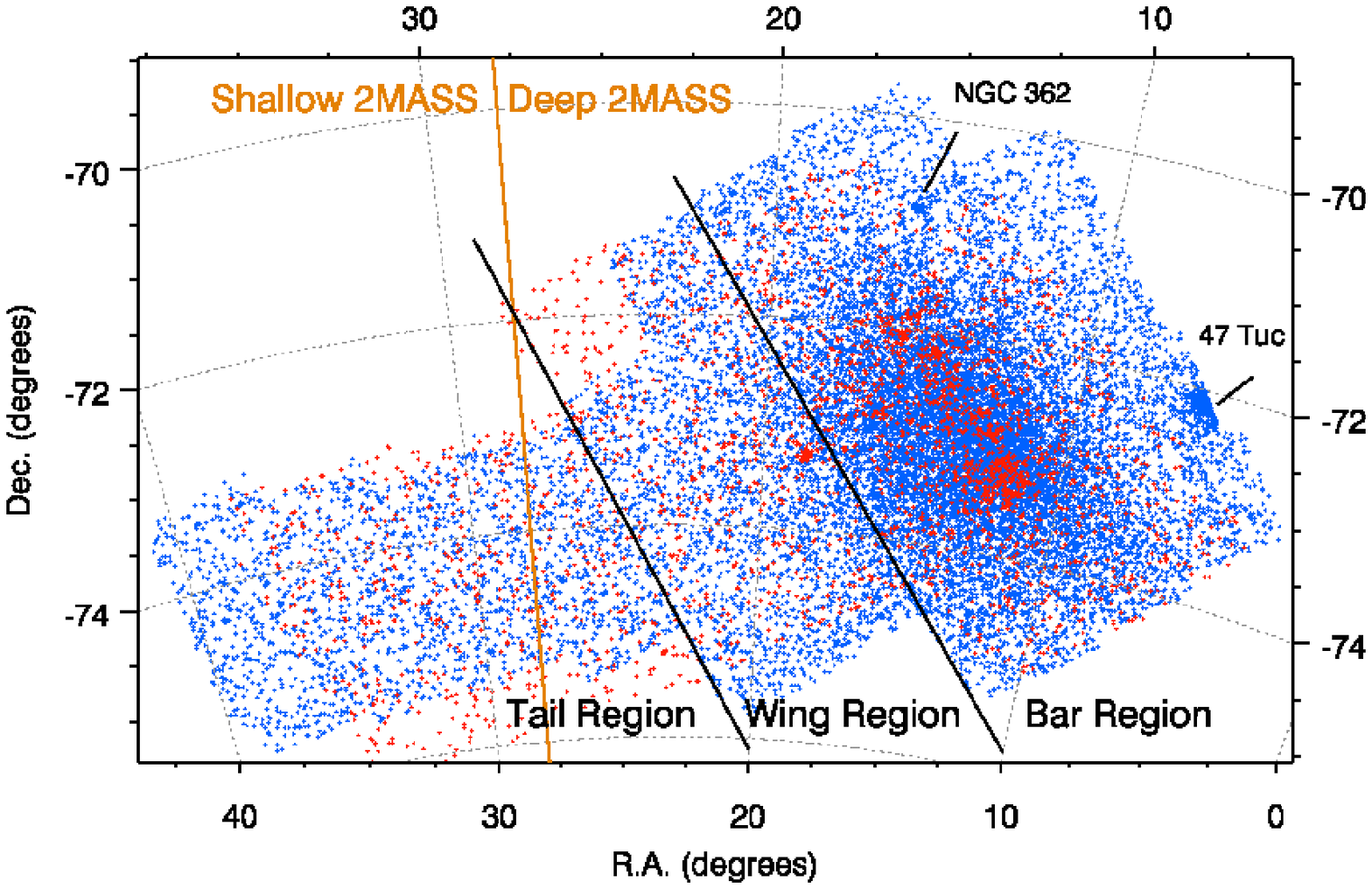} }
\figcaption{Map of the SAGE-SMC catalog coverage. {\it Left:} IRAC
(and 2MASS) coverage is shown in blue, MIPS 24~\micron{} in red,
S$^3$MC in orange, MCPS in black, and IRSF in green. The brightest
3.6-\micron\ point-sources are plotted in blue for reference.  The
region used to estimate the background and foreground point-source
contamination (Section~\ref{sec:contamination}) is shown as a dashed
black line.  {\it Right:} The brightest 24-\micron\ point-sources are
plotted in red. The approximate boundaries between the tail, wing, and
bar regions are shown. Note that 2MASS photometry is shallower in the
tail region than in the wing/bar. \label{fig:coverage}}
\end{figure*}

Following the RGB phase and the subsequent core He-burning phase, the
  low- to intermediate-mass stars ($0.8 \lesssim M \lesssim
  8~M_\odot$) will begin to ascend the AGB. Both RGB and AGB stars
  initially generate stellar winds via acoustic and/or electromagnetic
  chromospheric processes \citep{hartmann80, hartmann84}. In more
  evolved AGB stars, the primary drivers of mass loss are pulsation
  and dust-driving. Pulsations can levitate material from the stellar
  surface and provide density enhancements and shocks, which can
  encourage dust formation and re-processing
  \citep[e.g.,][]{bowen88,winters00,winters03,schirrmacher03,woitke06,woitke07,mattsson08,vanloon08b}. The
  dust composition depends on the atmospheric chemistry (abundance of
  carbon relative to oxygen), which is altered by dredging up newly
  formed carbon to the surface of the star \citep[the third
  dredge-up;][]{iben83}.  Radiation pressure on dust grains and
  grain--gas momentum coupling accelerates the stellar wind from the
  star \citep[e.g.,][ and references therein]{sedlmayr95,
  elitzur01}. This mass loss continues until the stellar envelope has
  been ejected.

Individually, AGB stars produce comparatively small amounts of
  circumstellar dust \citep[$\lesssim$10$^{-9}~M_\odot$~yr$^{-1}$;
  e.g.,][]{vanloon00,groenewegen09}. Nevertheless, they exist in large numbers,
  collectively placing them among the most important known dust
  factories in the Universe \citep[e.g.,][]{gehrz89}.

AGB star luminosities peak in the near-IR, and circumstellar dust
emits in the mid- to far-IR, making IR photometry and spectra
essential for characterizing AGB stellar and dust properties. IR
studies of low mass ($M \lesssim 1~M_\odot$), low metallicity
($0.005\,Z_\odot \lesssim Z \lesssim 0.15\,Z_\odot$) AGB stars in
globular clusters
\citep[e.g.,][]{boyer06m15,boyer08ocen,boyer09ngc362,boyer10,lebzelter06,vanloon06,vanloon08a,ita07,mcdonald09,mcdonald11dust,
mcdonald11param,mcdonald11c} indicate that AGB stars produce dust even
at extremely low metallicities. Similar studies of AGB stars in Local
Group dwarf galaxies
\citep[e.g.,][]{jackson06,jackson07,groenewegen07,matsuura07,boyer09lg,sloan09}
reveal diverse AGB populations depending on star formation history and
show significant dust production at very low metallicities.

Super-AGB stars are the most massive AGB stars (5 --
10~$M_\odot$). These stars undergo efficient hot bottom burning
\citep[HBB;][]{smith85,boothroyd92} and are expected to suffer from weak thermal
pulses, reach lower temperatures, and achieve higher mass-loss rates
than their lower-mass cousins \citep{siess10}. There is evidence that
these stars may be the progenitors of dust-enshrouded supernovae \citep{javadi11}.

Stars with masses 8~$M_\odot$ -- 25~$M_\odot$ become RSGs (or red
helium-burning stars), which generally have warmer effective
temperatures than AGB stars. RSG stars do not undergo a third
dredge-up, so they are exclusively O-rich. Like AGB stars, RSGs show
strong mass loss, enriching the surrounding Interstellar Medium (ISM)
with silicatious \citep[e.g.,][]{verhoelst09} and sometimes also
carbonaceous \citep{sylvester94,sylvester98} material, though it is
unclear whether the carbonaceous material is instead of interstellar
origin. The dust-production rates of RSGs are similar to those of AGB
stars, but RSGs are less numerous. \citet{bonanos09, bonanos10}
examined the SAGE data of optically-selected RSGs in the Magellanic
Clouds. Here, we take the opposite approach and select RSGs by their IR
colors, accounting for very dust-enshrouded examples
\citep[cf.,][]{elias85,wood92,roche93,groenewegen98,vanloon05a,vanloon05b}.

\begin{deluxetable}{lcccc}
\tablewidth{0pc}
\tabletypesize{\normalsize}
\tablecolumns{5}
\tablecaption{SMC and LMC Parameters Adopted in this Work\label{tab:params}}

\tablehead{\colhead{Parameter} & 
\colhead{SMC}&
\colhead{Ref.}&
\colhead{LMC}&
\colhead{Ref.}}
\startdata
Distance,$d$ (kpc)\dotfill & 61$\pm$1 & 1,5 & 51$\pm$1& 1,5\\
Metallicity\tablenotemark{a}, $Z$ ($Z_\odot$)\dotfill & $0.2$$\pm$$0.06$ & 6,7 & $0.5$$\pm$$0.17$& 6,7\\
$A_{\rm V}$ (mag)\dotfill & 0.12 & 4,8\tablenotemark{b} & 0.46 & 2,3\\
$E_{\rm B-V}$ (mag)\dotfill & 0.04 & 4,8 & 0.15 & 9\\
3.6-\micron\ TRGB (mag)\dotfill &12.6&10&11.9&10
\enddata

\tablerefs{\ (1) \citet{cioni00}; (2) \citet{cioni06} (3)
  \citet{glass99}; (4) \citet{harris04}; (5) \citet{keller06};
  (6)\citet{luck98}; (7) \citet{meixner10}; (8) \citet{schlegel98};
  (9) \citet{westerlund97}; (10) This work. }

\tablenotetext{a}{\ The metallicities are likely lower for low-mass ($\approx$$1~M_\odot$) stars.}
\tablenotetext{b}{\ The \citet{schlegel98} extinction map cites the total dust column, so yields an overestimate of the reddening affecting stars that lie somewhere along that column. Because the extinction towards the SMC is low, we do not expect this to have a significant impact on the results.}
\end{deluxetable}

\begin{figure*}
\epsscale{1.1} \plotone{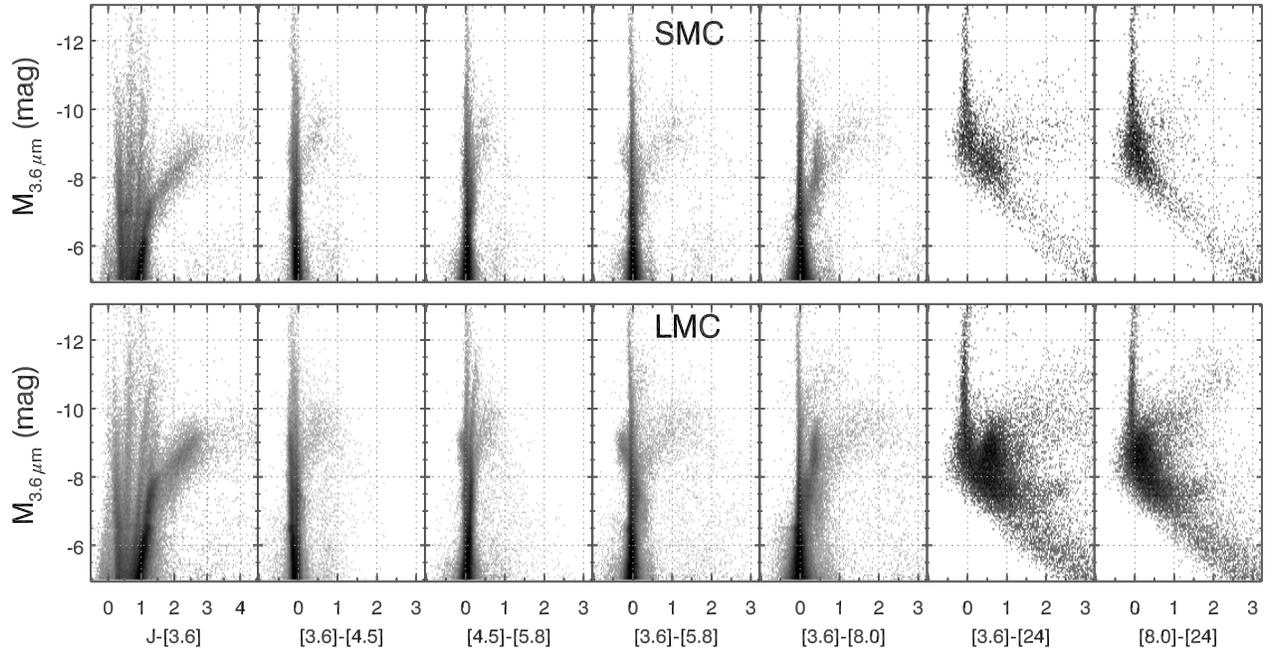}
\figcaption{Color--magnitude diagrams (CMDs) for the SMC and the LMC,
ranging from $J$ to 24~\micron{}.  The y-axis is the absolute
3.6~\micron\ magnitude ($M_{3.6}$) for all panels. The CMDs are
represented as Hess diagrams, with 150 bins on each axis,
corresponding to 0.05~mag $M_{3.6}$ bins and 0.03~mag color
bins. \label{fig:cmds}}
\end{figure*}

Globular cluster studies are limited by their single, low-mass ($0.8 -
0.9\,M_\odot$) stellar populations, and in most dwarf galaxies,
evolved stars are easily confused with unresolved background galaxies
due to the limited resolution of mid-IR imaging ($\gtrsim$$1.7\arcsec{}$).  The
Magellanic Clouds suffer neither of these limitations by containing
multiple populations and being close enough that evolved stars
generally outshine unresolved galaxies. Studies of the
Magellanic Clouds provide valuable insight into the environments of more
distant star-forming galaxies.

Several near-IR and mid-IR surveys of the SMC have been conducted in
recent years, including an $AKARI$ survey \citep[3.2 --
24~\micron{};][]{ita10} of small selected regions within the SMC bar
and the {\it Spitzer} Survey of the Small Magellanic Cloud
\citep[S$^3$MC;][]{bolatto07}, which imaged the SMC bar at 3.6 --
160~\micron{}. The SAGE-SMC survey is unique in its full spatial
coverage of not only the SMC bar, but the wing and tail regions as
well (Section~\ref{sec:obs}), allowing us to examine the SMC structure
using the distribution of the cool evolved stars. The SMC is known to
have an extended halo of old stars \citep{nidever11}.  Stars in the
tail may have been stripped from the SMC bar, but the stars beyond the
tail and into the Magellanic bridge likely formed in situ, from a
tidally-stripped filament of gas \citep{harris07}.

This paper is organized as follows.  In Sections~\ref{sec:obs} and
\ref{sec:selection}, we describe the data and evolved star photometric
classification.  In Section~\ref{sec:agbs}, we present the
observational properties of RGB, RSG, and AGB stars in the LMC and
SMC, and in Section~\ref{sec:probes}, we use the evolved stars as
probes of the SMC environment. Finally, we summarize our findings in
Section~\ref{sec:conclusions}. Throughout this work, we adopt the
parameters listed in Table~\ref{tab:params}. The extinction in the
{\it Spitzer} bands is from \citet{indebetouw05}.

\section{SAGE-SMC Data}
\label{sec:obs}

The photometry presented here is from the SAGE-SMC archive catalog,
available from the {\it Spitzer} Science Center.  Photometric
uncertainties are typically $<$0.1~mag for all wavelengths, but
increase to $<$0.2~mag for the faintest $\approx$2~magnitudes. Mid-IR
sources were matched to 24~\micron{} sources with correlation
thresholds $>$$2\,\sigma$. See \citet{gordon11} for a description of
the observations, data reduction, and point-source extraction.  Two
epochs of SAGE-SMC data were obtained, separated by three months. The
photometry presented here was extracted from a co-addition of both
epochs, limiting spurious detections from transients and artifacts.  A
third epoch of observations from S$^3$MC is also included in the
co-addition where the coverage overlaps
(Fig.~\ref{fig:coverage}). Variable AGB stars show 3.6~\micron{} (or
$L'$, at 3.78~\micron) absolute magnitude amplitudes typically in the
range of $0.1 \lesssim \Delta M_{3.6} \lesssim 2$~mag
\citep[e.g.,][]{lebertre92,mcquinn07,vijh09}. Having two epochs of
data helps to minimize variability effects, and any remaining
systematic effects are minimal since we are looking at a large
population of AGB stars.

The SAGE-SMC catalog includes optical $UBVI$ photometry from the
Magellanic Clouds Photometric Survey \citep[MCPS;][]{zaritsky02},
$JHK_{\rm s}$ photometry from the 2-Micron All Sky Survey
\citep[2MASS;][]{skrutskie06} and the InfraRed Survey Facility survey
\citep[IRSF;][]{kato07}, mid-IR photometry (3.6, 4.5, 5.8, and
8~\micron{}) from {\it Spitzer}'s InfraRed Array Camera
\citep[IRAC;][]{fazio04}, and far-IR photometry (24, 70,
and 160~\micron{}, epoch~1) from the Multiband Imaging Photometer for {\it
Spitzer} \citep[MIPS;][]{reike04}.  Figure~\ref{fig:coverage} shows
the spatial coverage of each survey.

\subsection{Color--magnitude Diagrams}
\label{sec:cmds}

Near-IR to mid-IR stellar density CMDs for the SMC and the LMC, also
known as Hess diagrams, are shown in Figure~\ref{fig:cmds}. We see the
effects of stellar temperature in the near-IR, with several distinct
features apparent in the CMDs. These include branches that trace the
foreground stars, RSGs, hot OB stars, background galaxies, and AGB
stars. These features are labeled in a $J-[8]$ color vs. 8~\micron{}
absolute magnitude ($M_8$) CMD in Figure~\ref{fig:cmd_labeled}, which
is discussed more in Section~\ref{sec:agbid}. Moving into the mid-IR,
the stellar temperature no longer affects the CMD since we are
sampling only the Rayleigh--Jeans tail of the Planck
function. Instead, molecular and dust spectral features cause distinct
photometric features (see Section~\ref{sec:agbid}).

The morphologies of SMC and LMC CMDs look remarkably similar. The
metallicity difference between the galaxies ($Z_{\rm LMC}/Z_{\rm SMC}
= 2-3$) causes only a small difference in the IR colors, with the
higher metallicity LMC appearing slightly redder ($\Delta(J-[8])
\approx 0.1$~mag, $\Delta(J-K_{\rm s}) \approx 0.08$~mag). The
distances to the MCs adopted here (Table~\ref{tab:params}) are
uncertain, so differences in the absolute 3.6-\micron{} magnitudes
(y-axis in Figure~\ref{fig:cmds}; the SMC stars appear slightly
fainter) may not be intrinsic. A downwards 5~kpc shift in the relative
distance of the MCs eliminates the magnitude differences, but this is
well beyond the uncertainty of recent distance measurements
\citep[e.g.,][]{szewczyk09}. Uncertainties in the distance due to
depth along the line-of-sight in both galaxies is small (a few
hundredths of a magnitude), and may be largest in the SMC wing
\citep{subramanian09}. The extinction towards both the Magellanic Clouds is low ($A_{3.6} \lesssim 0.03$~mag), so this cannot explain the difference 3.6~\micron{} magnitude.

\subsection{Foreground and Background Contamination}
\label{sec:fgnd}

\begin{figure}
\vbox{
\includegraphics[width=0.3\textwidth,angle=90]{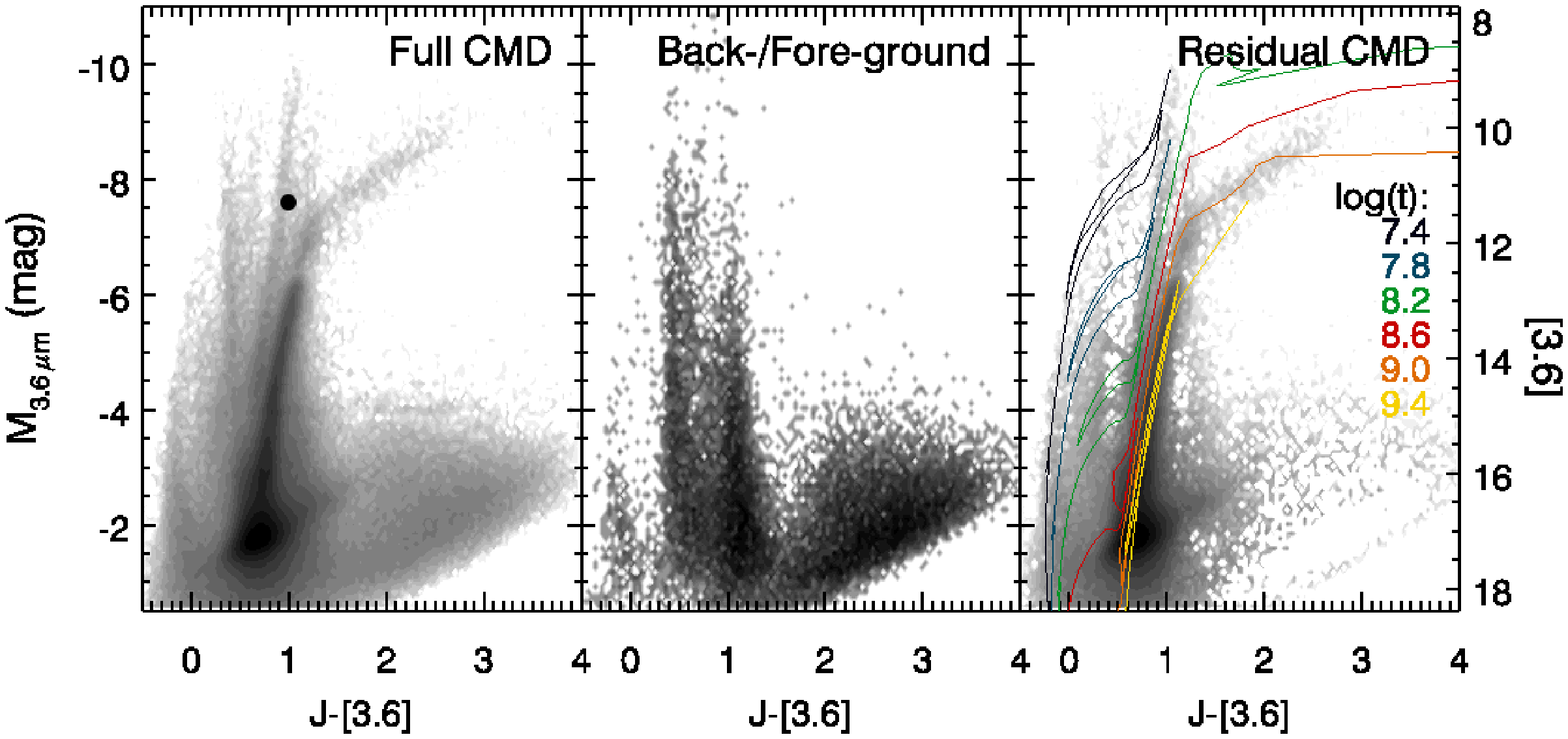}
\includegraphics[width=0.3\textwidth,angle=90]{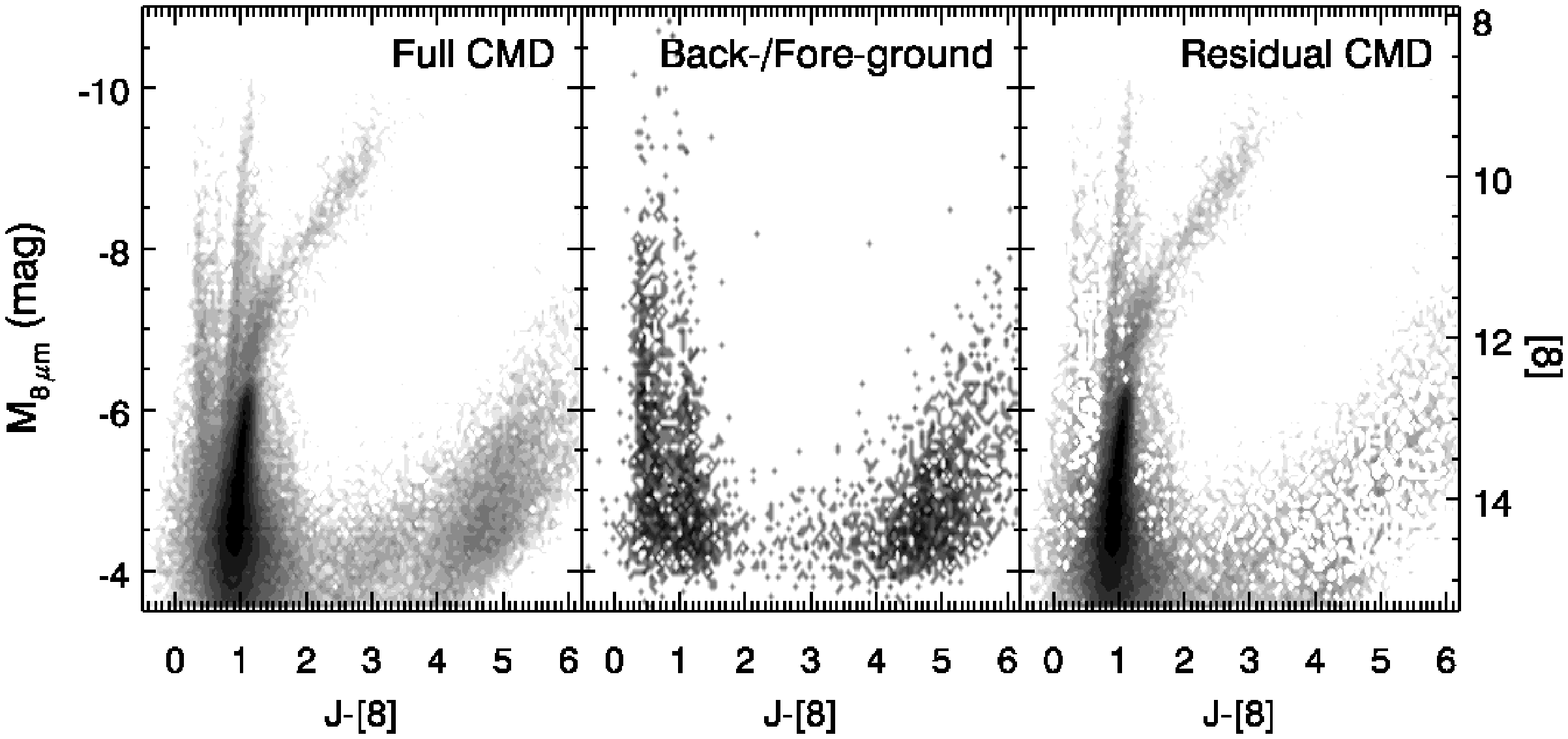}
}
\figcaption{Back-/Foreground-subtracted CMDs. The
background/foreground CMD was created from a 1.6 deg$^2$ field in the
bridge/tail region, centered at R.A. $= 2^{\rm h}13^{\rm m}00\fs0$,
Decl. $= -74\degr19\arcmin30\arcsec$ (Fig.~\ref{fig:coverage}).  The
residual CMD is dominated mainly by RGB, RSG, and AGB stars, though
features from OB stars and YSOs also stand out. The AGB stars are not
strongly affected by foreground/background sources. Padova isochrones
with log($t$) = 7.4 -- 9.4 and 60\% Silicate $+$ 40\% AlOx dust for M
stars and 85\% AMC $+$ 15\% SiC dust for C stars are also shown in the
upper right panel for reference \citep{marigo08}. \label{fig:cmd_sub}}
\end{figure}

Because we have imaged such a large area around the SMC, we can use
the data on the outskirts of the coverage to estimate the approximate
level of foreground and background contamination.
Figure~\ref{fig:cmd_sub} shows the CMD of the full SMC coverage compared
to the CMD of a 1.6~deg$^2$ region on the eastern edge of the SMC
tail/bridge (Fig.~\ref{fig:coverage}). The most prominent feature in
the full CMD (where the source density is highest) is the RGB. Since
no RGB is visible in the back-/fore-ground CMD and since all other CMD
features are vertical, we are confident that the region we have chosen
does indeed contain very few SMC-member sources. 

\begin{figure}
\epsscale{1.2} \plotone{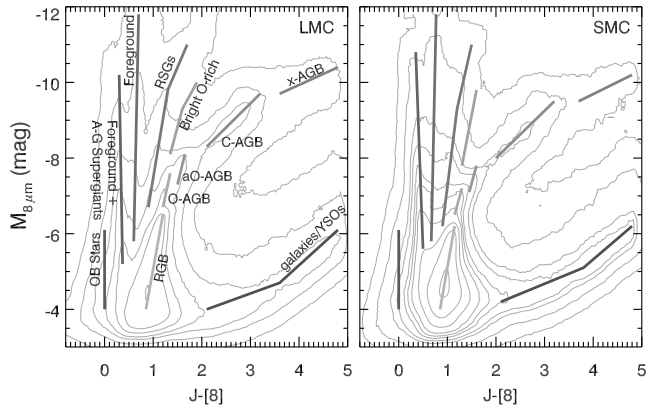} \figcaption{$J-[8]$
vs. $M_8$ CMDs. Contours represent the source density in
color--magnitude space. Each of these branches is also visible in the
$J-[3.6]$ vs. M$_{\rm3.6}$ CMD, except the aO-AGB branch.  A--G
supergiants and OB-stars are identified in \citet{bonanos09,bonanos10}. See the text for identification of RSG, RGB, and
AGB stars. \label{fig:cmd_labeled}}
\end{figure}

Subtracting the back-/fore-ground CMD from the full SMC CMD
(Fig.~\ref{fig:cmd_sub}, right) reveals well-defined branches of RSG and hot
OB-stars in addition to the RGB (see Fig.~\ref{fig:cmd_labeled}). A
set of Padova isochrones \citep{marigo08} are shown for reference. A
population of A--G supergiants also appears
\citep[cf.][]{bonanos09,bonanos10}, along with a population of faint
sources ($M_{3.6} > -5$~mag) redder than $J-[3.6] \approx 1$.  The
latter sources may be dominated by young stellar objects (YSOs;
M.Sewi{\l}o et al.\ 2011, in preparation).

While it is difficult to eliminate individual foreground and
background sources with photometry alone, we can estimate the level of
contamination statistically. In the box used to represent the
back-/fore-ground contamination described above, there is a
point-source density of $1.5 \times 10^4$ sources deg$^{-2}$. However,
this is an overestimate of the contamination in our evolved star
samples since many of these contaminating sources are bluer and/or
fainter than typical cool evolved stars and are not included in our
selection criteria (Section~\ref{sec:selection}). When we apply our
selection criteria to the back-/fore-ground region, we find that our
RGB sample suffers the worst contamination, with 708 sources
deg$^{-2}$, followed by the RSG sample with 37 sources deg$^{-2}$ and
oxygen-rich AGB (O-AGB) sources with 1.9 sources deg$^{-2}$. None of
the other types of AGB stars are detected in this region, so
contamination of those samples is very low. After considering the size
of the IRAC spatial coverage \citep[$\approx$30~deg$^2$;][]{gordon11}, we expect contamination to account for
35\%, 18\%, and 2.5\% of the RSG, RGB, and O-AGB samples,
respectively.

Aside from the spatially uniform contamination from foreground and
background sources, stars belonging to the foreground globular
clusters 47\,Tuc and NGC\,362 (Fig.~\ref{fig:coverage}) have similar
near-IR colors to SMC evolved stars.  To minimize this contamination,
we exclude all stars within 8\arcmin\ and 5\arcmin\ of the centers of
47\,Tuc and NGC\,362, respectively. This corresponds to the
elimination of 120 stars from the RGB sample and 108 stars from the
RSG sample.  Only 5 stars are eliminated from the AGB sample: 4
O-rich, and one C-rich.

\section{Stellar Classification}
\label{sec:selection}

\begin{figure}

\vbox{
\includegraphics[width=0.45\textwidth]{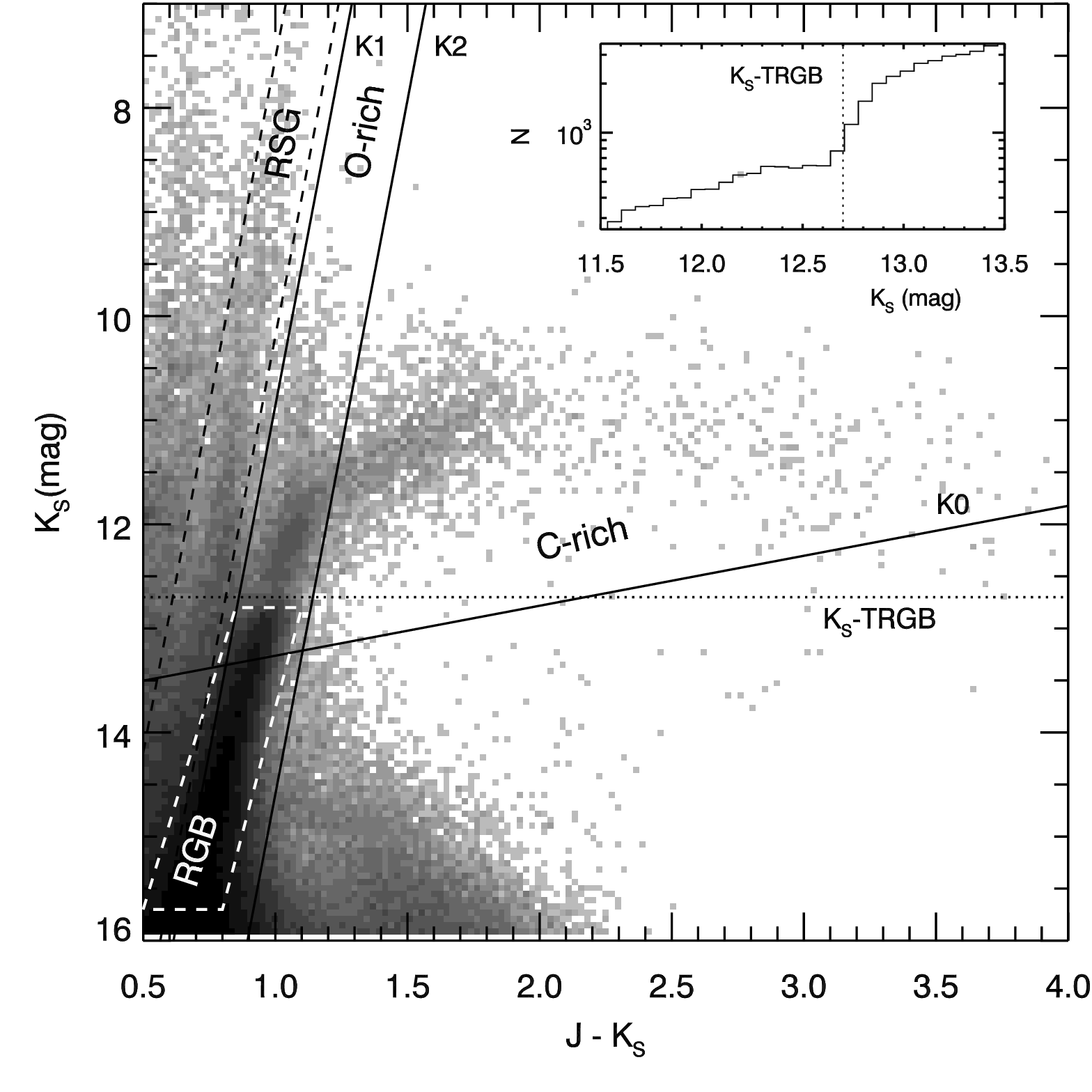}
\includegraphics[width=0.45\textwidth]{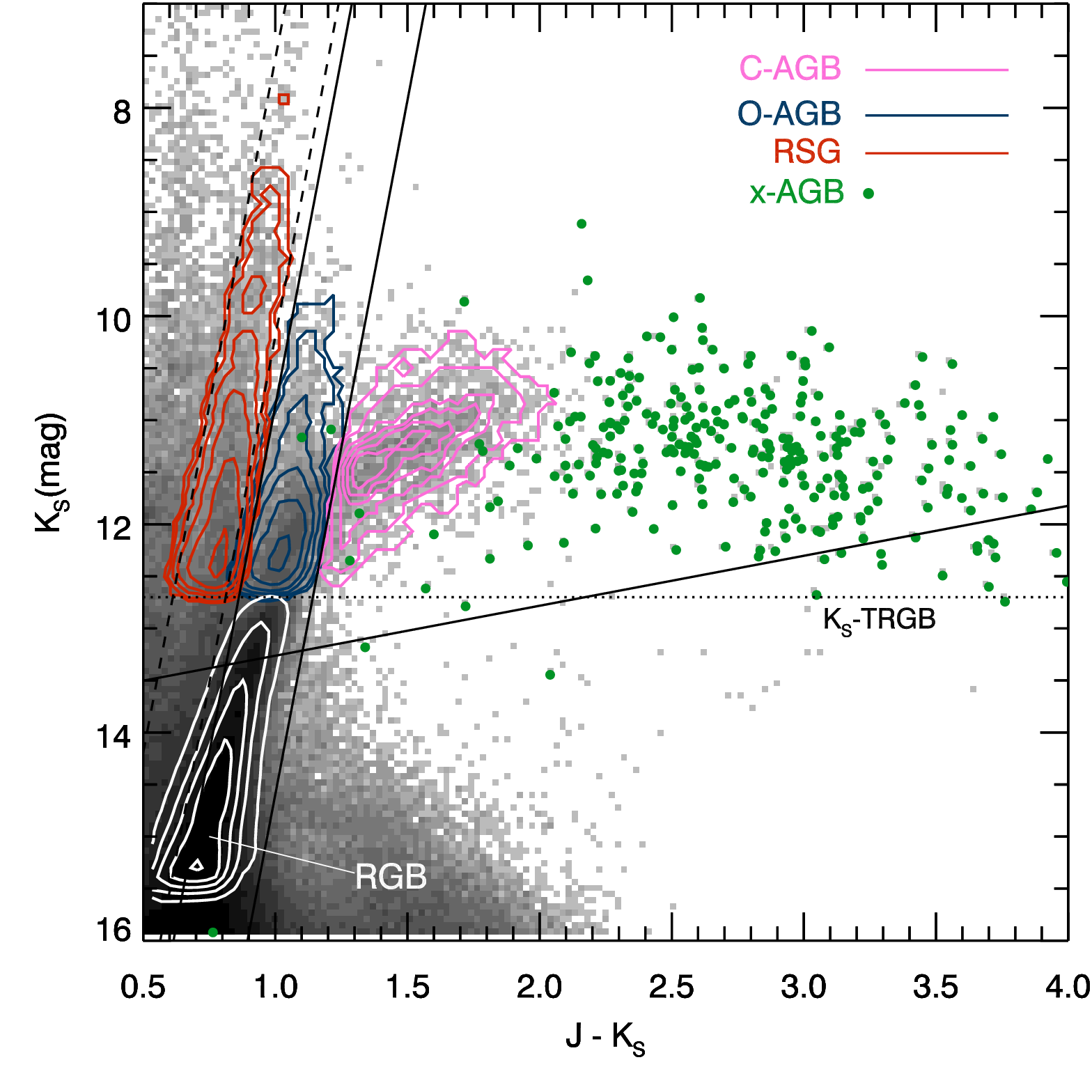}
}

\figcaption{{\it Top:} $J-K_{\rm S}$ CMD showing the separation of
carbon-rich and oxygen-rich AGB stars in the SMC, following
\citet{cioni06,cioni06smc}. The $J-K_{\rm s}$ color is also used to
select RSG and RGB stars. See text. {\it Bottom:} Same as upper
panel, with the selection of C-AGB (pink), O-AGB (blue), RSG (red),
and RGB (white) stars shown in contours. The x-AGB stars are also
plotted in green.  \label{fig:cmd_cioni}}
\end{figure}

\subsection{Color--magnitude selection of cool evolved stars}
\label{sec:agbid}

Figure~\ref{fig:cmd_labeled} labels each branch in the $J-[8]$ vs.\
$M_8$ CMD. AGB stars occupy the reddest and brightest branches,
separating into O-AGB, carbon-rich stars (C-AGB), heavily-extinguished
``extreme'' stars (x-AGB), and a new feature that we are calling the
anomalous O-rich branch (aO-AGB; Section\,\ref{sec:clump}). RGB stars
are the most populous source type, occupying the branch just below the
AGB stars, at $M_8 \gtrsim -6.5$~mag. RSG stars are just to the blue
edge of the O-AGB stars, though 10-\micron{} silicate emission
captured toward the red cutoff of the 8-\micron{} IRAC band results in
redder $J-[8]$ colors, causing overlap between the two branches in
this diagram.

The $J-[8]$ vs $M_8$ CMD is a good diagnostic for classifying the
different cool evolved stars. However, the classification accuracy is
limited in stars that have strong 10-\micron{} silicate features,
which affect the 8-\micron\ flux.  Instead, we use the $J-K_{\rm s}$
vs. $K_{\rm s}$ CMD to separate RGB, RSG, C-AGB, and O-AGB stars and
turn to the mid-IR colors to select x-AGB stars (which are often
undetected in the near-IR) and aO-AGB stars (which are only
discernible by their $J-[8]$ colors). The classification schemes are
outlined below, and the results are listed in Table~\ref{tab:stats1}.

\begin{deluxetable*}{lrrrr}
\tablewidth{0pc}
\tabletypesize{\normalsize}
\tablecolumns{5}
\tablecaption{{\it Spitzer} SAGE SMC and LMC Source Statistics\label{tab:stats1}}

\tablehead{&\multicolumn{2}{c}{SMC}&\multicolumn{2}{c}{LMC}\\ 
\colhead{Population} & 
\colhead{N}&
\colhead{${\rm N}_{24}$\tablenotemark{a}}& 
\colhead{N}&
\colhead{${\rm N}_{24}$}}

\startdata

Point-sources in both $J$ and [3.6]\dotfill                      & 458\,558        & 9\,793   & 2\,301\,842     & 39\,740\\
Point-sources with $[3.6]<$ TRGB$_{[3.6]}$\tablenotemark{b}\dotfill& 19\,290       & 3\,878   & 45\,780         & 18\,337\\
C-AGB stars\tablenotemark{c}\dotfill                             & 1\,729 (54)     & 964 (4)  & 6\,212 (156)    & 5\,246 (32)\\
Faint O-AGB stars ($M_{8}\geq -8.3$~mag)\tablenotemark{c}\dotfill& 2\,251 (1\,190) & 76 (113) & 9\,441 (6\,223) & 1\,288 (2\,880)\\
Bright O-AGB stars ($M_{8}<-8.3$~mag)\dotfill                    & 227             & 173      & 1\,422          & 1\,300 \\
Extreme AGB stars \dotfill                                       & 349             & 323      & 1\,105          & 1\,018\\
aO-AGB stars \dotfill                                            & 1\,244          & 117      & 6\,379          & 2\,912\\
RSG stars \dotfill                                               & 3\,325          & 538      & 4\,604          & 1\,560\\
RGB stars \dotfill                                               & 135\,437        & 41       & 407\,342        & 580\\
FIR objects in AGB and RSG samples\dotfill                       & 57              & 57       & 224             & 224\\
FIR objects in RGB sample\dotfill                                & 303             & 303      & 1\,262          & 1\,262

\enddata

\tablecomments{ \ See Section~\ref{sec:agbid} for a description of the
  stellar classifications. FIR objects were originally classified as
  either AGB, RSG, or RGB stars, but have been split into their own
  category.}

\tablenotetext{a}{ \ ${\rm N}_{24}$ is the number of sources
  with 24-\micron{} counterparts.}
\tablenotetext{b}{ \ The 3.6-\micron{} TRGB is $\approx$12.6~mag and $\approx$11.9~mag in the SMC and LMC, respectively.}

\tablenotetext{c}{ \ The number in parentheses is the number of stars
  in the original C-AGB or O-AGB sample, based on the classification
  using $J-K_{\rm s}$ color, that are re-classified as aO-AGB stars (Section~\ref{sec:clump}).}

\end{deluxetable*}

\subsubsection{C-AGB and O-AGB stars}
\label{sec:CO}

C-AGB and O-AGB stars are selected using color--magnitude cuts in the
$J-K_{\rm s}$ vs.\ $K_{\rm s}$ CMD. Any very dusty RSG stars will be
very red in $J-K_{\rm s}$ due to dust extinction, so these sources may
be included in our O-AGB and C-AGB samples. There is no way to
distinguish very dusty RSG stars from dusty AGB stars with IR
photometry alone, but we expect them to be uncommon compared to their
less-dusty counterparts.

Figure~\ref{fig:cmd_cioni} shows the $J-K_{\rm s}$ color cuts from
\citet{cioni06}, adjusted for metallicity and distance, following
\citet{cioni06smc} and using $Z_{\rm SMC} = 0.2~Z_\odot$ and $d_{\rm
SMC} = 61$~kpc. We shift the K0 line slightly fainter to account for
the difference in the $K_{\rm s}$-band tip of the red giant branch
(TRGB) between the SMC and LMC and to assure that we include the full
C-AGB sample. C-AGB stars fall redward of the K2 boundary, O-AGB stars
fall between the K1 and K2 lines, and all AGB stars are brighter than
the K0 line (except x-AGB stars, see Section~\ref{sec:x}).

To eliminate contamination from RGB stars in our AGB sample, we
exclude stars that are fainter than both the TRGB in $K_{\rm s}$ {\it
and} 3.6~\micron{} (Fig.~\ref{fig:cmd_cioni}, inset).  By checking the
TRGB at two wavelengths, we are sure not to exclude the more
heavily-extinguished stars that might be fainter than the $K_{\rm s}$
TRGB. $K_{\rm s}$-band TRGBs are from \citet{cioni00}, and the
3.6-\micron{} TRGBs are estimated here to be $[3.6]_{\rm TRGB} \approx
11.9$~mag ($M_{3.6}^{\rm TRGB} = -6.6$~mag) for the LMC and $12.6$~mag
($M_{3.6}^{\rm TRGB} = -6.3$~mag) for the SMC, similar to other Local
Group dwarf galaxies
\citep{cioni00,jackson07,jackson06,matsuura07,boyer09lg}. We neglect
the population of non-dusty AGB stars that fall below the TRGB, as
most of these have not yet undergone a third dredge-up and behave
quite differently from their more-evolved counterparts.

We note that the O-AGB stars can also be divided into a bright and
faint population, which occupy very different regions of the
$[8]-[24]$ vs. $M_{3.6}$ CMD (See Fig.~\ref{fig:cmd_agb}). Following
\citet{srinivasan09}, the bright and faint O-rich AGB stars are
divided by $M_8 = -8.3$~mag, and we often treat them separately
throughout this paper. A subset of the bright O-AGB stars show strong
8- and 24-\micron\ excess (Section~\ref{sec:ccds}); it is among these
sources that we might find the more massive super-AGB stars
(Section~\ref{sec:intro}).

The lower panel of Figure~\ref{fig:cmd_cioni} shows the selected AGB stars
overlain on the $J-K_{\rm s}$ CMD. Detection statistics are summarized
in Section~\ref{sec:stats} and Table~\ref{tab:stats1}. We also remind
the reader here that while the SMC C-AGB population includes virtually
no foreground and background sources, such contamination accounts for
2.5\% of the SMC O-AGB population (Section~\ref{sec:fgnd}).

\subsubsection{x-AGB stars}
\label{sec:x}

The x-AGB stars are those that are most likely to be experiencing a
``superwind'', where the mass-loss rate can increase by a factor of
10, and a thick dust envelope obscures the star at optical wavelengths
\citep[cf.][]{vanloon06c}. The mass-loss rate eventually exceeds the
nuclear consumption rate, and so determines the subsequent
evolution of the star. The physical mechanism that causes a star to
enter the superwind phase is not well understood, though x-AGB stars
are known to have longer pulsation periods than other AGB stars \citep{riebel10}.

Due to circumstellar dust extinction, many
x-AGB stars fall below the K0 line and the $K_{\rm s}$-TRGB
(Fig.~\ref{fig:cmd_cioni}). We thus turn to the mid-IR photometry to
recover these sources. As in \citet{blum06} and \citet{srinivasan09},
we identify x-AGB stars as those brighter than the 3.6-\micron\ TRGB
and with $J-[3.6] > 3.1$~mag. Some of the most heavily dust-enshrouded
x-AGB stars are totally undetected in the near-IR. Therefore, if there
is no near-IR detection, but $[3.6]-[8] > 0.8$ and the star is
brighter than the 3.6-\micron\ TRGB, then it is also included in the
initial list of x-AGB stars.

To minimize contamination from YSOs and unresolved background
galaxies, we also apply the following restrictions: x-AGB stars must be
brighter than an empirical boundary in the
$J-[8]$ vs $[8]$ CMD and the $[3.6]-[8]$ vs. $[8]$ CMD, defined
as:

\begin{equation} \label{eq:j8}
[8] = 12 - (0.43 \times J - [8]),
\end{equation}
\begin{equation} \label{eq:38}
[8] = 11.5 - (1.33 \times [3.6] - [8]). 
\end{equation}

\begin{figure}
\epsscale{1.2} \plotone{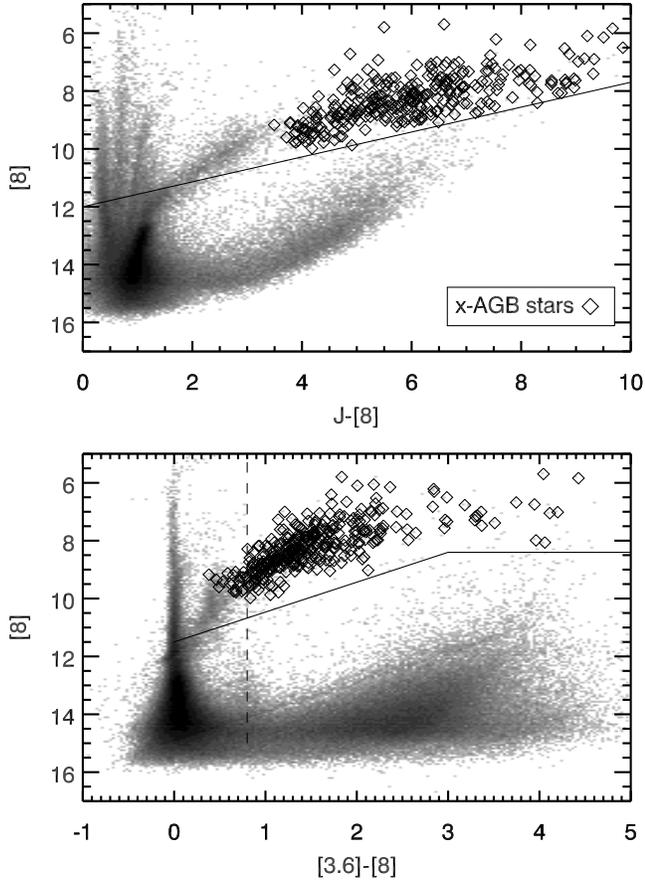}
\figcaption{Contamination from YSOs and unresolved background galaxies
is minimized by requiring that x-AGB stars be brighter than the lines
shown in the two above CMDs (determined empirically). Stars above
these lines that are not x-AGB stars have been classified as FIR
sources (Section~\ref{sec:fir}). Note that some x-AGB stars have colors
bluer than $[3.6]-[8] = 0.8$; these stars are considered x-AGB stars
because their $J-K_{\rm s}$ colors are redder than
3.1~mag. \label{fig:select}}
\end{figure}

\noindent Equation~(\ref{eq:38}) terminates at $[3.6]-[8] = 3$~mag,
and extends horizontally out to redder colors
(Fig.~\ref{fig:select}). The final number of x-AGB stars in both the
LMC and SMC is reported in Table~\ref{tab:stats1}.

There is a small population of very red x-AGB stars with $[3.6]-[8] >
3.5$~mag. We list these sources in Table~\ref{tab:redx}. The reddest
source ($[3.6]-[8] = 4.43$~mag) is not an x-AGB; it is the YSO
S3MC\,01051--7159 \citep{vanloon10b}. The others are not identified
in the literature.

\begin{deluxetable}{llc}
\tablewidth{0.4\textwidth}
\tabletypesize{\normalsize}
\tablecolumns{3}
\tablecaption{x-AGB stars with $[3.6]-[8] > 3.5$~mag\label{tab:redx}}

\tablehead{\colhead{R.A. (J2000)} & \colhead{Decl. (J2000)} &
\colhead{$[3.6]-[8]$}}

\startdata
       00$^{\rm h}$48$^{\rm m}$08.49$^{\rm s}$ & --73$\degr$14$\arcmin$54.7$\arcsec$&4.20\\
       01$^{\rm h}$00$^{\rm m}$41.61$^{\rm s}$ & --72$\degr$38$\arcmin$00.7$\arcsec$&4.06\\
       01$^{\rm h}$04$^{\rm m}$53.13$^{\rm s}$ & --72$\degr$04$\arcmin$03.9$\arcsec$&3.51\\
       01$^{\rm h}$05$^{\rm m}$03.13$^{\rm s}$ & --71$\degr$59$\arcmin$29.7$\arcsec$&3.96\\
       01$^{\rm h}$05$^{\rm m}$03.97$^{\rm s}$ & --71$\degr$59$\arcmin$25.4$\arcsec$&4.11\\
       01$^{\rm h}$05$^{\rm m}$07.26$^{\rm s}$ & --71$\degr$59$\arcmin$42.8$\arcsec$&4.43\tablenotemark{a}\\
       01$^{\rm h}$08$^{\rm m}$17.51$^{\rm s}$ & --72$\degr$53$\arcmin$09.2$\arcsec$&3.75\\
       01$^{\rm h}$24$^{\rm m}$07.95$^{\rm s}$ & --73$\degr$09$\arcmin$04.0$\arcsec$&4.04\\
       02$^{\rm h}$35$^{\rm m}$18.63$^{\rm s}$ & --74$\degr$29$\arcmin$54.0$\arcsec$&3.95

\enddata

\tablenotetext{a}{ \ This source is the YSO S3MC\,01051--7159 \citep[][and references therein]{vanloon10b}.}
\end{deluxetable}

\subsubsection{RSG stars}
\label{sec:rsg}

Stars within the branch just blueward of the K1 line in
Figure~\ref{fig:cmd_cioni} are classified here as RSG stars. We
restrict the branch width to $\Delta (J-K_{\rm s}) = 0.2$~mag and
leave a 0.5~mag gap between the O-AGB stars and the RSG stars to
minimize contamination between the two stellar types. We also restrict
our selection of RSGs to those brighter than the $K_{\rm s}$-band
TRGB, which has the effect of excluding the low-mass RSGs.  This
restriction helps to minimize contamination from foreground sources
and RGB stars (see Section~\ref{sec:fgnd}). However, we estimate that
$\approx$35\% of the final SMC RSG selection is still due to
foreground and background sources (Section~\ref{sec:fgnd}).

\subsubsection{RGB stars}
\label{sec:rgb}

The RGB stars are the most difficult to select, as they are affected
by contamination from foreground sources, and unresolved background
galaxies (Section~\ref{sec:fgnd}), and also from YSOs. We are also
limited by sensitivity, so the RGB sample included in this work is not
complete.  We define RGB stars as those within the box outlined by a
white, dashed line in Figure~\ref{fig:cmd_cioni}, which spans from
(TRGB + 0.1 mag) $< K_{\rm s} <$ (TRGB + 3~mag).  The 0.1~mag buffer
reduces contamination from early-AGB stars.

From this sample of RGB stars, we exclude sources redder than a line in
the $J-[8]$ vs $[8]$ CMD, defined as:

\begin{equation} \label{eq:j8_2}
[8] = A - (11.76 \times J - [8]),
\end{equation}

\noindent where $A=30.29$, to eliminate background sources. In
Section~\ref{sec:fgnd}, we noted that 18\% of the SMC population is
likely contamination from foreground and background sources.

\subsubsection{aO-AGB stars: a new feature in the IR CMD}
\label{sec:clump}

\begin{figure}
\epsscale{1.2} \plotone{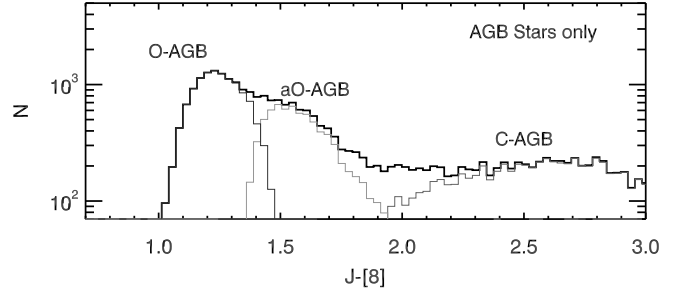} \figcaption{$J-[8]$
histogram showing the O-AGB, C-AGB, and aO-AGB stars.  The aO-AGB
sources can be seen as a bump in the tail of the O-AGB
population. \label{fig:sagb}}
\end{figure}

An unidentified feature is apparent in the $J-[8]$ CMD
(Section~\ref{sec:agbid}, Fig.~\ref{fig:cmd_labeled}). The feature is
stronger in the LMC, but is also present in the SMC, and it suggests
the existence of a population of stars that is distinct from the O-AGB
and C-AGB stars. Figure~\ref{fig:sagb} shows the new population of
stars as a bump in the SMC$+$LMC AGB $J-[8]$ histogram between O-AGB
and C-AGB stars. We note that $\approx$96\% of this new group of stars
(or 98\% in the LMC) are classified as O-rich using the classification
scheme from \citet{cioni06}, and we thus label them anomalous O-rich
AGB stars (aO-AGB) here.

To select stars belonging to the new CMD feature, we have examined the
LMC and SMC $J-[8]$ vs. $M_8$ Hess diagrams
(Fig.~\ref{fig:cmd_labeled}) in color and magnitude space to choose
the boundaries between the C-AGB branch, the O-AGB branch and the
unidentified CMD feature. These boundaries correspond to the regions
of minimum stellar density or prominent density changes between CMD
features. We thus select aO-AGB stars from the original O-AGB sample if
they are redder than the line defined by equation~\ref{eq:j8_2}, with
$A=27.95$, and fainter than $M_8 = -8.3$~mag.  Stars from the original
C-AGB sample that are bluer than the line defined by
equation~\ref{eq:j8_2}, with $A=31.47$, are also classified as aO-AGB
stars.

\begin{figure}[h!]
\epsscale{1.2} \plotone{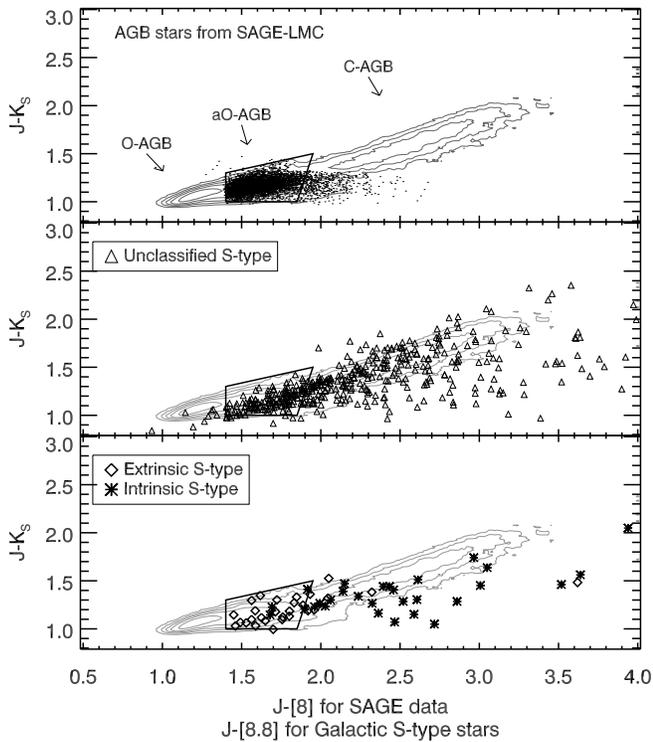}
\figcaption{$J-[8]$ vs. $J-K_{\rm s}$ CCD of LMC AGB stars.  The top
panel shows a contour plot of the LMC AGB stars with the aO-AGB stars
plotted as dots, using 2MASS+IRAC photometry. The middle panel shows
known Galactic S-type stars, which are plotted with 2MASS+MSX
$J-[8.8]$ colors \citep{yang06}. The bottom panel shows known
intrinsic and extrinsic S-type stars, also plotted with $J-[8.8]$
colors \citep{wang02}. The black box in all panels marks the
approximate location of the aO-AGB stars (aO-AGB stars are selected
from the J-[8] vs. [8] CMD). The Galactic S-type stars tend to occupy
the same location on the CCD as our aO-AGB stars. The known extrinsic
S-type stars (blue diamonds) preferentially coincide with our aO-AGB
stars, while the known intrinsic S-type stars (asterisks) have redder
$J-[8.8]$ colors.\label{fig:ccd_j8}}
\end{figure}

\begin{figure*}
\epsscale{1.15} \plotone{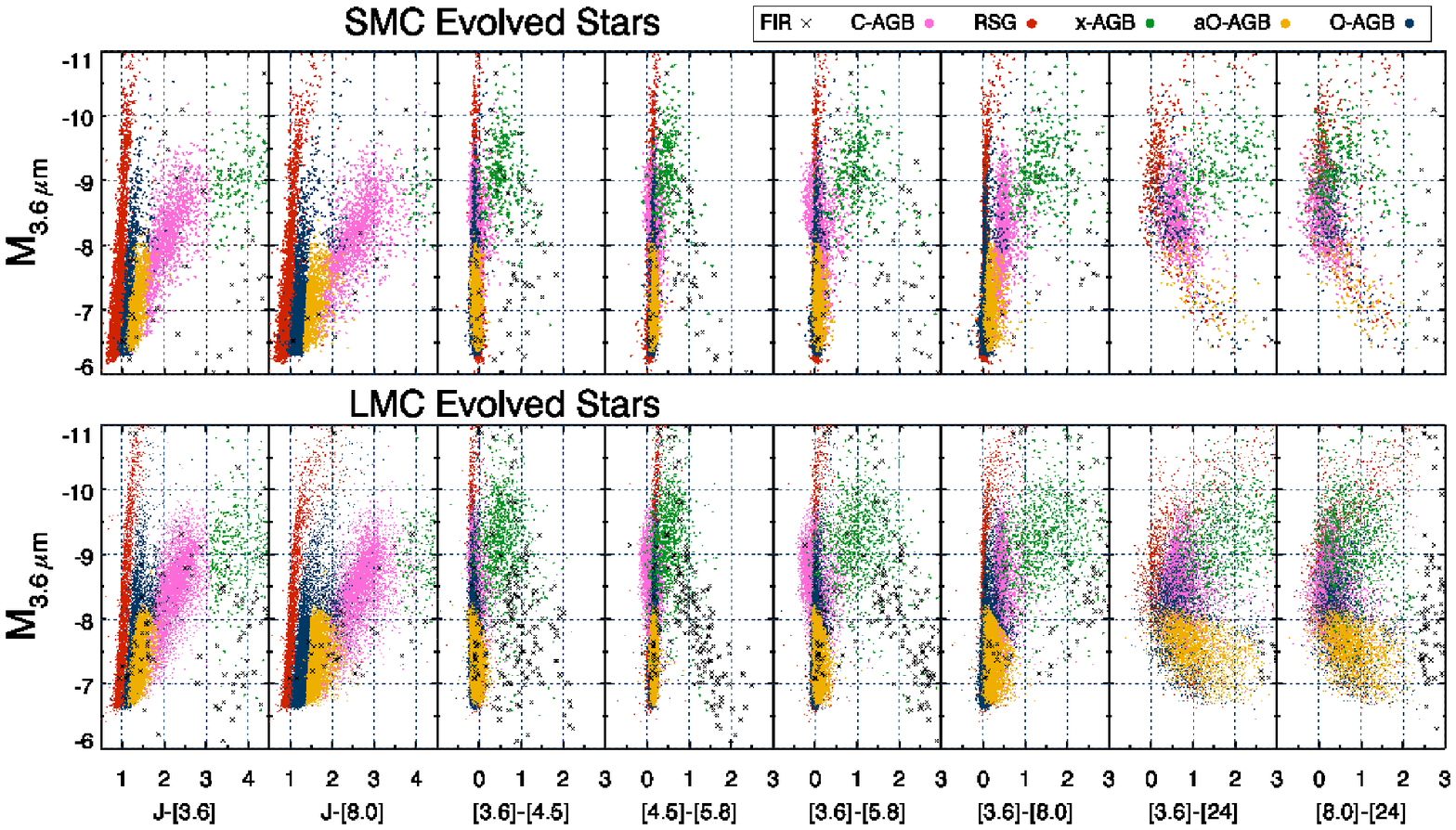} \figcaption{CMDs
showing the FIR objects, AGB and RSG stars: O-AGB (blue), aO-AGB
(orange), C-AGB (pink), x-AGB (green), RSG (red), and FIR objects
(black). The y-axis is the absolute 3.6~\micron\ magnitude ($M_{3.6}$)
for all panels. Several dust and molecular features were captured in
the IRAC and MIPS bands, causing red and blue IR colors.  See text.
\label{fig:cmd_agb}}
\end{figure*}

The aO-AGB stars may simply be a subset of O-rich AGB stars that have
formed (or are just beginning to form) a significant amount of dust.
Low-mass AGB stars or AGB stars that have not yet evolved much along
the AGB may not be very dusty. We see this in globular clusters, which
show that not all stars at the TRGB exhibit excess IR emission
attributed to dust
\citep{boyer09ngc362,mcdonald09,mcdonald11dust}. These aO-AGB stars
may therefore be the dusty siblings of ``naked'' O-AGB stars. Many of
the aO-AGB stars also show excess 24-\micron\ emission (see
Section~\ref{sec:ccds}), which supports this scenario.

It is also possible that the aO-AGB stars are O-rich AGB stars
with particularly strong and/or broad silicate emission, causing a
redder $J-[8]$ color than stars with weaker silicate emission.  It is
unknown what sort of star or what stage of AGB evolution would cause
such a silicate enhancement.

A third possibility is that the aO-AGB stars may be S-type AGB
stars. S-type stars show many spectral features corresponding to
s-process elements, and the C/O ratio in such stars is near unity.  A
collection of near-IR magnitudes from 2MASS and 8.8-\micron{}
magnitudes from the Midcourse Space Experiment ($MSX$) of Galactic
S-type AGB stars was compiled by \citet{wang02} and \citet{yang06},
and has been featured in a number of studies
\citep[e.g.,][]{guandalini08, zhang10}. These studies find that S-type
AGB stars have IR colors very similar to our aO-AGB stars, and
intermediate to O-AGB and C-AGB IR colors. Figure~\ref{fig:ccd_j8}
shows the $J-[8]$ vs. $J-K_{\rm s}$ color--color diagram (CCD) of AGB
stars in the LMC with the S-type $J-A$ and $J-K_{\rm s}$
colors from \citet{yang06} also plotted. We note that
\citet{kastner08} show that $MSX$ $A$-band photometry (centered at
8.8~\micron{}) is systematically $\approx$0.1 -- 0.6~mag brighter than
the {\it Spitzer} 8-\micron{} band for evolved stars over a broad
range of $K-[8]$ color (see their Fig.~6), so some of the Galactic
S-type stars will have slightly redder colors in
Figure~\ref{fig:ccd_j8} than the SAGE stars. Most aO-AGB stars occupy
the region outlined by the black box, and a sizable portion of the
Galactic S-type stars also occupy this space. There is also an
indication that extrinsic S-type stars (those whose s-element
enhancement is due to mass transfer from an AGB binary companion)
preferentially occupy the space belonging to the aO-AGB stars, as
opposed to intrinsic S-type stars (those whose s-element enhancement
is due to their own third dredge-up), which tend to be
redder.\footnotemark \footnotetext{Extrinsic and intrinsic
classifications are from \citet{wang02}, and are based on
Tc-enrichment} However, a lower metallicity may cause SMC and LMC
intrinsic S-type stars to make the transition from O-rich to C-rich
more quickly than in the Galaxy.  This would result in a
less-developed dusty envelope in the Magellanic intrinsic S-type
stars, and thus cause a smaller IR excesses than Galactic intrinsic
S-type stars. It is therefore difficult to estimate whether an
intrinsic or extrinsic classification is more likely if the aO-AGB
stars are indeed S-type stars.  Extrinsic and intrinsic S-type AGB
stars have similar luminosities, so they cannot be distinguished by
their absolute magnitudes.

Spectra showing s-process elements or dust features typical of S-type
stars \citep[e.g.,][]{hony09,smolders10} are required to confidently
identify our aO-AGB stars as S-type.  \citet{hony09} find that IR
features of intrinsic S-type AGB stars (extrinsic S-type stars rarely
show 12-\micron\ excess indicative of dust) are similar to those of
O-rich AGB stars, but the 10-\micron{} silicate features of S-type
stars are smoother, with O-rich AGB stars showing at least three
distinct components due to amorphous silicates and aluminum
oxide. Since the edge of the silicate feature is captured in the IRAC
8-\micron{} band, the differing shape of the feature may be the cause
of the difference in $J-[8]$ color between O-AGB and aO-AGB stars, as
seen here.  However, Figure~\ref{fig:ccd_j8} indicates that the aO-AGB
stars are more likely to be extrinsic than intrinsic, so they should
not exhibit silicate features at all. In any case, if S-type
classification is confirmed, the $J-[8]$ color may prove a useful
photometric diagnostic for identifying S-type AGB star candidates when
spectra are unavailable.  We note, however, that the aO-AGB stars
account for nearly 20\% of the SMC AGB population, which is higher
than what might be expected of S-AGB stars. Inspection of
Figure~\ref{fig:sagb} shows that the aO-AGB stars likely include O-AGB
stars that are in the tail of the $J-[8]$ color distribution. It is
not possible to separate aO-AGB stars and O-AGB stars in this region
with SAGE photometry, so it is possible that we have overestimated the
number of aO-AGB stars and underestimated the number of O-AGB
stars. Therefore, while some of the aO-AGB stars may be S-type, we
expect that many (or indeed most) are instead the dusty cousins of
regular O-AGB stars.

\subsubsection{Far-IR Objects}
\label{sec:fir}

Sources with spectral energy distributions (SEDs) that rise from 8 to
24~\micron\ are typically unresolved background galaxies, compact
H\,{\sc II} regions, planetary nebulae (PNe), or YSOs
\citep{whitney03}. \citet{gruendl09} identify YSOs in the LMC SAGE
data, and 700 of 703 of their YSOs show this type of rising SED. We
call these sources far-IR (FIR) objects.

\begin{figure}[h!]
\epsscale{1.1} \plotone{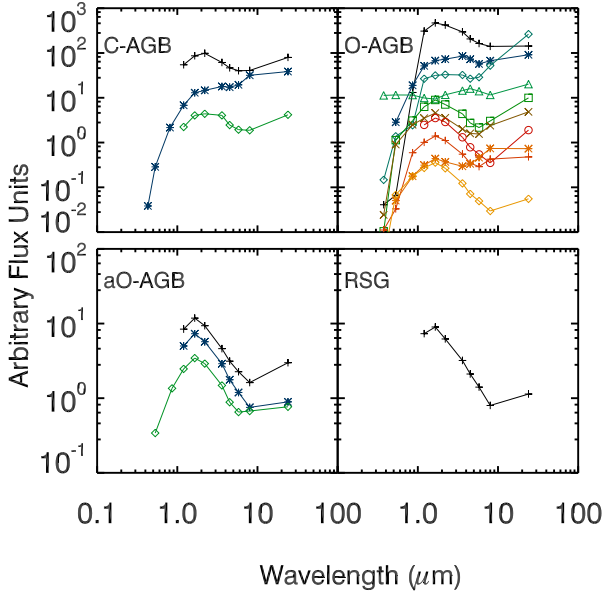} \figcaption{FIR objects
eliminated from the AGB and RSG samples on the basis that the
24-\micron{} flux density exceeds the 8-\micron{} flux density. The
source plotted with blue asterisks in the O-AGB panel is BMB-B\,75, a
confirmed RSG that falls within our O-AGB selection
criteria.\label{fig:ysocands}}
\end{figure}

\begin{figure}[h!]
\epsscale{1.1} \plotone{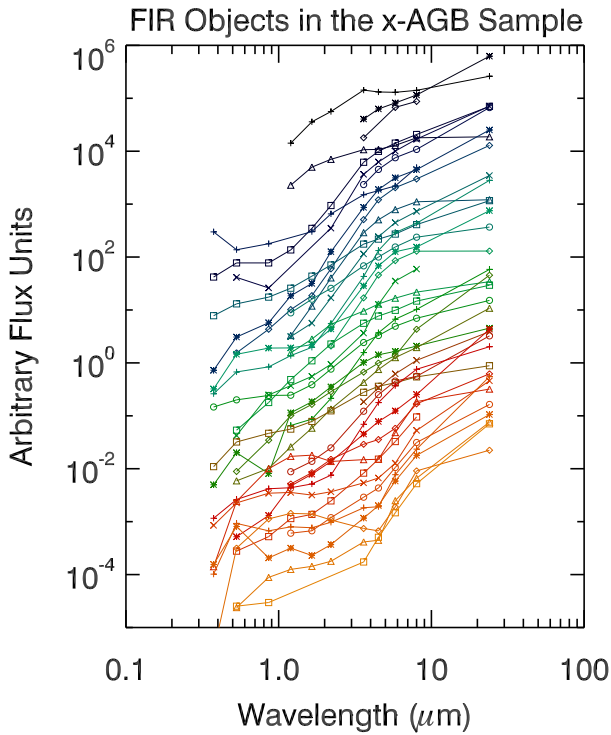} \figcaption{FIR objects
eliminated from the SMC x-AGB sample on the basis that the
24-\micron{} flux density exceeds the 8-\micron{} flux density. Fluxes
have been scaled for clarity. The coordinates and photometry of these
sources are available electronically
(Table~\ref{tab:sample}).\label{fig:xcands}}
\end{figure}

In the SMC, 57 FIR sources fall within the AGB and RSG
photometric selection criteria described in the previous sections,
especially among the O-AGB and x-AGB samples. We show the SEDs of
these FIR sources in Figures~\ref{fig:ysocands} and
\ref{fig:xcands}. The vast majority of the FIR sources are located
within star-forming regions in the bar and wing, with only a handful
located on the outskirts of the bar (see Figs.~\ref{fig:disto} --
\ref{fig:distg}). 

It is possible that there is a small number of extremely enshrouded
evolved stars among these FIR objects; in the LMC, there is an RSG
(IRAS\,05280$-$6910) whose dusty envelope causes its SED to peak near
30~\micron{} and strong silicate self-absorption suppresses the
8-\micron{} flux \citep[e.g.,][]{boyer10sdp}. These objects, which are
on the verge of becoming post-AGB stars, are short-lived
\citep[$\sim$$10^{3-4}$~yr,][]{vanloon10}, so are extremely rare. The
SEDs of known PNe are also similar to those shown in
Figures~\ref{fig:ysocands} and \ref{fig:xcands} \citep{hora08}. It is
unclear how many of the FIR sources are truly evolved stars; since
the exclusion of these sources might mean the exclusion of the
dustiest evolved stars, we retain them in our analysis, albeit in a
separate category of FIR objects.

Our sample of FIR objects contains only one confirmed evolved star;
the RSG BMB-B\,75 was originally classified as an O-AGB by its mid-IR
colors (\ref{sec:CO}). Five FIR sources are classified as YSOs through
modeling of their SEDs \citep{bolatto07,simon07}. In addition, eight
are confirmed as YSOs using far-IR {\it Spitzer} data
\citep{vanloon10b}, and eight more are classified as H\,{\sc II}
regions or emission line stars by \citet[][and references
therein]{wilke03}. Most of the FIR objects (49 of 57) are classified
as YSOs by J.M.Oliveira et al.\ (2011, in preparation), L.Carlson et
al.\ (2011, in preparation), and M.Sewi\l o et al.\ (2011, in
preparation) based on their mid-IR colors.  The coordinates and
photometry of the FIR objects that fall within our evolved star
classifications are listed in Table~\ref{tab:sample}, along with the
other evolved stars.


\subsection{Remaining Contamination}
\label{sec:contamination}

While our selection criteria eliminate most contamination from other
source types, we can estimate the degree to which our evolved star
samples are still contaminated by YSOs and compact \ion{H}{2}
regions. Even after separating the FIR objects from the evolved stars,
there remains some overlap between our AGB sample
(Table~\ref{tab:stats1}) and a catalog of SMC YSO candidates compiled
by Sewi\l o et al.\ (2011, in preparation), who also use IR
color--magnitude cuts to identify sources. According to their catalog,
YSOs make up $<$1\% of our final C- and O-AGB samples and
$\approx$6\% of the x-AGB sample. 

\citet{kastner08} derived $JHK_{\rm s}$- and 8-\micron{} color cuts for
various sources in the LMC (see their Table~3).  Using these cuts, we
find that the x-AGB sample may contain only 3 \ion{H}{2}
regions, even after adjusting the cuts slightly to account for the
metallicity difference between the LMC and SMC. Their RSG cuts are not
usable here as they overlap strongly with our C-AGB and O-AGB sources,
and do not include the bulk of the stars identified here as RSGs
(e.g., Fig.~\ref{fig:cmd_labeled}).

Background galaxies are identified in the SAGE-LMC data by
\citet{gruendl09} through examination of the mid-IR SEDs and
images. Only 5 probable background galaxies from their selection are
included in our LMC x-AGB sample, so we expect the total contamination
from unresolved background galaxies to be low in both the LMC and SMC
samples.

The color-cuts chosen to distinguish O-AGB from C-AGB stars are
approximate, and it is likely that there is cross-contamination
between the two samples.  We discuss this more in
Sections~\ref{sec:ccds} and \ref{sec:seds}.

\subsubsection{Classification Summary}
\label{sec:classsum}

Our catalog of evolved stars and FIR objects for both galaxies is
available electronically (see Table~\ref{tab:sample} for a sample from
the SMC).  Figure~\ref{fig:cmd_agb} presents CMDs spanning optical to
IR wavelengths of the selected AGB and RSG stars. Several dust and
molecular features are visible; the C-AGB stars (pink) show blue
$[4.5]-[5.8]$ and $[3.6]-[5.8]$ colors due to absorption from CO
and/or C$_3$ at $4-6$~\micron{}. The x-AGB stars (green), which tend
to be carbon-rich
\citep[e.g.,][]{vanloon97,vanloon06,vanloon08b,matsuura09}, often show
absorption of HCN $+$ C$_2$H$_2$ near 3~\micron{}, causing very red
$[3.6]-[5.8]$, $[3.6]-[8.0]$, and $[3.6]-[24]$ colors
\citep{matsuura08}. MgS emission in x-AGB stars can inflate the
24-\micron{} flux, causing very red $[3.6]-[24]$ and $[8]-[24]$
colors. Strong silicate features in O-AGB (dark blue) and RSG (red)
stars cause excess 8- and 24-\micron{} emission. There is also a small
subset of x-AGB stars that are O-rich, typically OH/IR stars, whose
silicate features also enhance the 8- and 24-\micron{} flux
\citep{wood92,vanloon01a, vanloon01b,vanloon05b}.  Continuum dust
emission also causes red colors in the IR CMDs. For examples of
typical mid-IR spectra of LMC stars that show the features described
above, see \citet{woods11}.

\begin{deluxetable*}{lrccccc}
\tablewidth{0pc}
\tabletypesize{\normalsize}
\tablecolumns{7}
\tablecaption{Sample of Evolved Stars and FIR objects in the SMC\label{tab:sample}}

\tablehead{
&&
\multicolumn{4}{c}{IRAC}&
\colhead{MIPS}\\
\colhead{IRAC Designation} & 
\colhead{Classification}&
\colhead{$[3.6]$}&
\colhead{$[4.5]$}&
\colhead{$[5.8]$}&
\colhead{$[8.0]$}&
\colhead{$[24]$}
}

\startdata

SSTISAGEMA J001821.19-733601.0& C-AGB&  9.71$\pm$0.04 &  9.71$\pm$0.02 &  9.58$\pm$0.04 &  9.12$\pm$0.02 &  9.24$\pm$0.05\\
SSTISAGEMA J001941.08-732111.0&   RSG&  9.92$\pm$0.02 &  9.92$\pm$0.02 &  9.74$\pm$0.02 &  9.50$\pm$0.03 &  9.51$\pm$0.07\\
SSTISAGEMA J002020.39-735058.7& C-AGB& 10.52$\pm$0.04 & 10.48$\pm$0.04 & 10.45$\pm$0.06 &  9.92$\pm$0.05 &  9.77$\pm$0.06\\
SSTISAGEMA J002124.62-734450.1& x-AGB&  9.85$\pm$0.03 &  9.55$\pm$0.03 &  9.23$\pm$0.02 &  8.82$\pm$0.02 &  8.86$\pm$0.04\\
SSTISAGEMA J002132.72-735222.7& C-AGB& 10.39$\pm$0.03 & 10.09$\pm$0.03 &  9.62$\pm$0.04 &  9.16$\pm$0.04 &  8.88$\pm$0.04\\
SSTISAGEMA J002240.79-733246.9& C-AGB& 10.28$\pm$0.02 & 10.30$\pm$0.02 & 10.06$\pm$0.02 &  9.72$\pm$0.03 &  9.83$\pm$0.08\\
SSTISAGEMA J002249.44-730350.1& C-AGB& 10.86$\pm$0.03 & 10.91$\pm$0.03 & 10.56$\pm$0.04 & 10.40$\pm$0.03 & 10.35$\pm$0.10\\
SSTISAGEMA J002313.05-732835.0&   RGB& 12.66$\pm$0.03 & 12.69$\pm$0.03 & 12.60$\pm$0.06 & 12.52$\pm$0.07 & 10.42$\pm$0.13\\
SSTISAGEMA J002357.70-732542.3&aO-AGB& 11.34$\pm$0.04 & 11.33$\pm$0.04 & 10.96$\pm$0.04 & 10.79$\pm$0.05 & 11.06$\pm$0.23\\
SSTISAGEMA J002358.64-733804.0&   FIR& 12.50$\pm$0.03 & 11.58$\pm$0.03 & 10.12$\pm$0.02 &  8.32$\pm$0.02 &  4.13$\pm$0.01

\enddata

\tablecomments{\ A sample from the SMC evolved star catalog
  is shown here. The full catalog is available electronically, and
  also includes MCPS $UBVI$ and 2MASS $JHK_{\rm s}$
  magnitudes. Magnitudes in this table have not been corrected for
  reddening.}

\end{deluxetable*}

\begin{figure}
\includegraphics[width=0.4\textwidth]{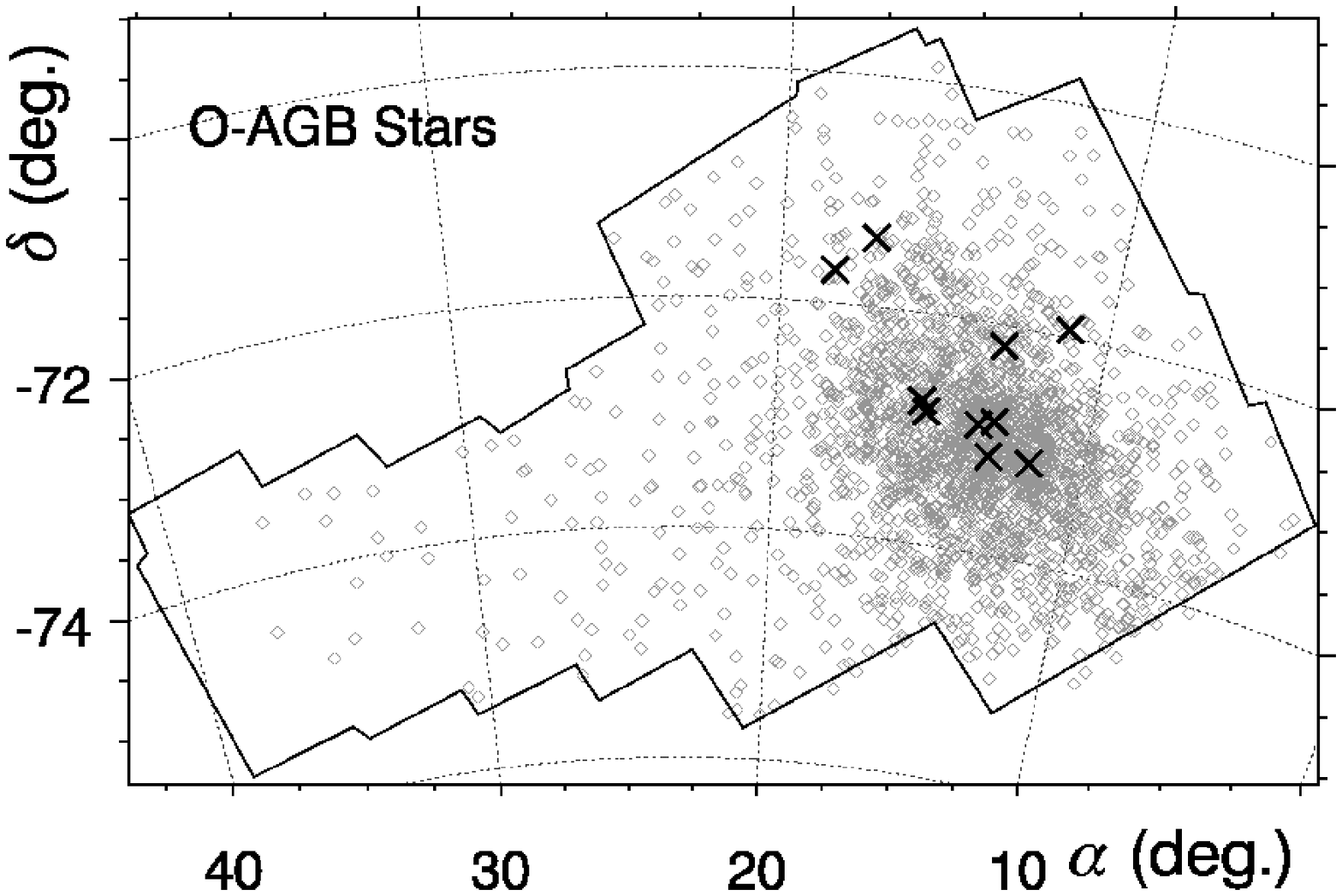}
\includegraphics[width=0.4\textwidth]{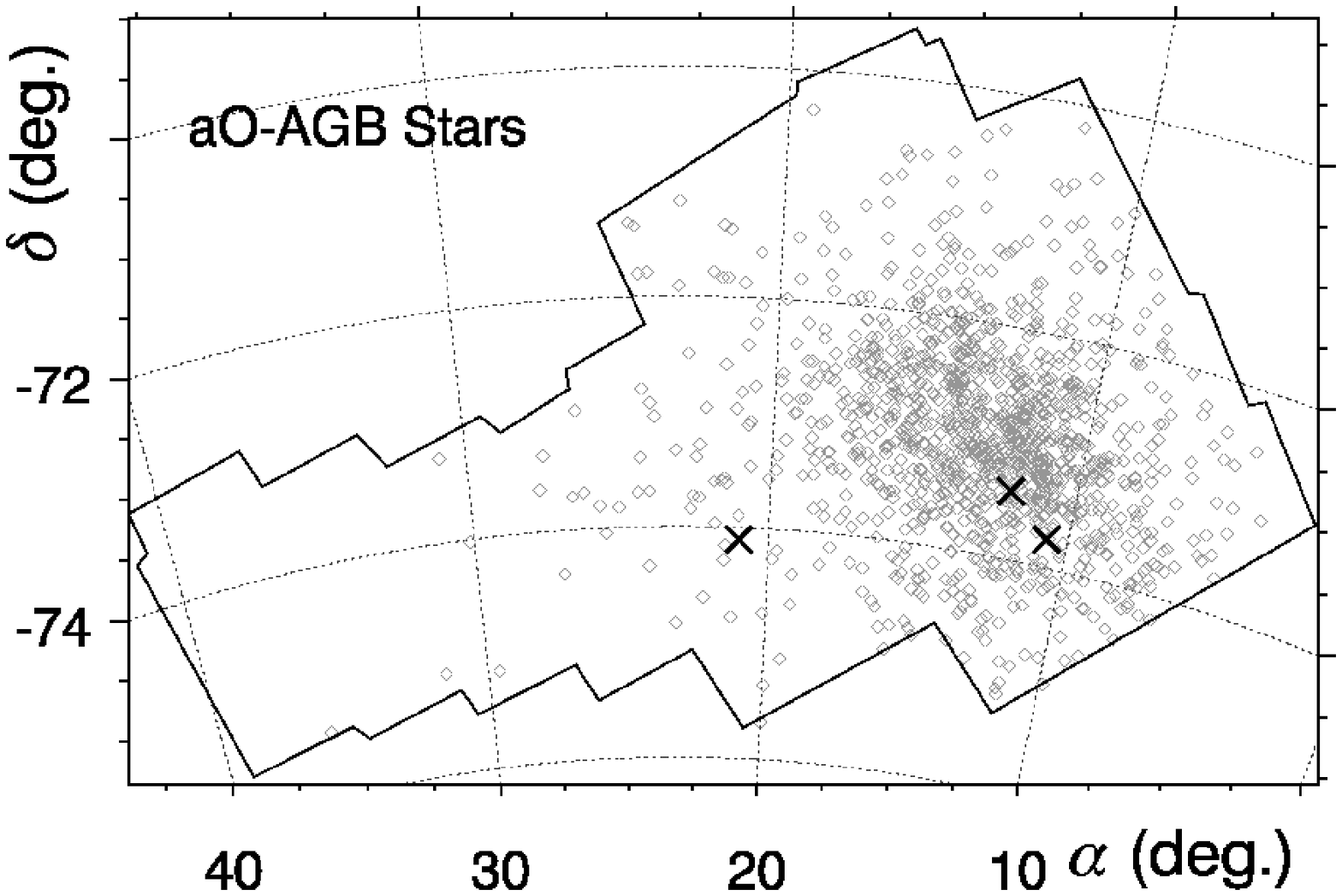}
\figcaption{The spatial distribution of O-AGB (top) and aO-AGB
(bottom) stars in the SMC. The approximate IRAC coverage is outlined
(Fig.~\ref{fig:coverage}). FIR objects selected from each sample are
marked by a black ``x''. Most AGB stars are confined to the
bar. Foreground/background contamination accounts for 2.5\% of the
O-AGB sample (Section~\ref{sec:contamination}).\label{fig:disto}}
\end{figure}
\begin{figure}
\includegraphics[width=0.4\textwidth]{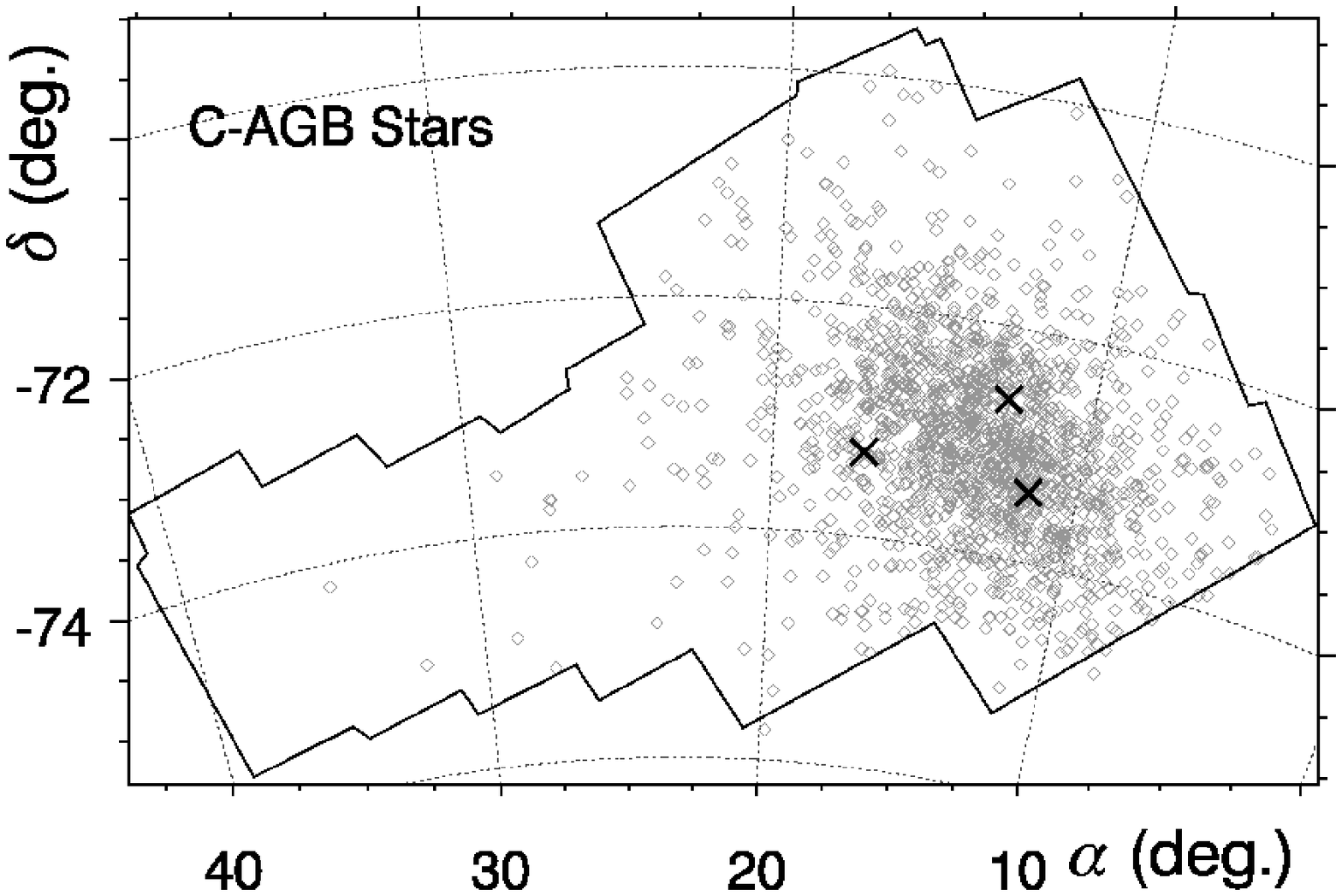}
\includegraphics[width=0.4\textwidth]{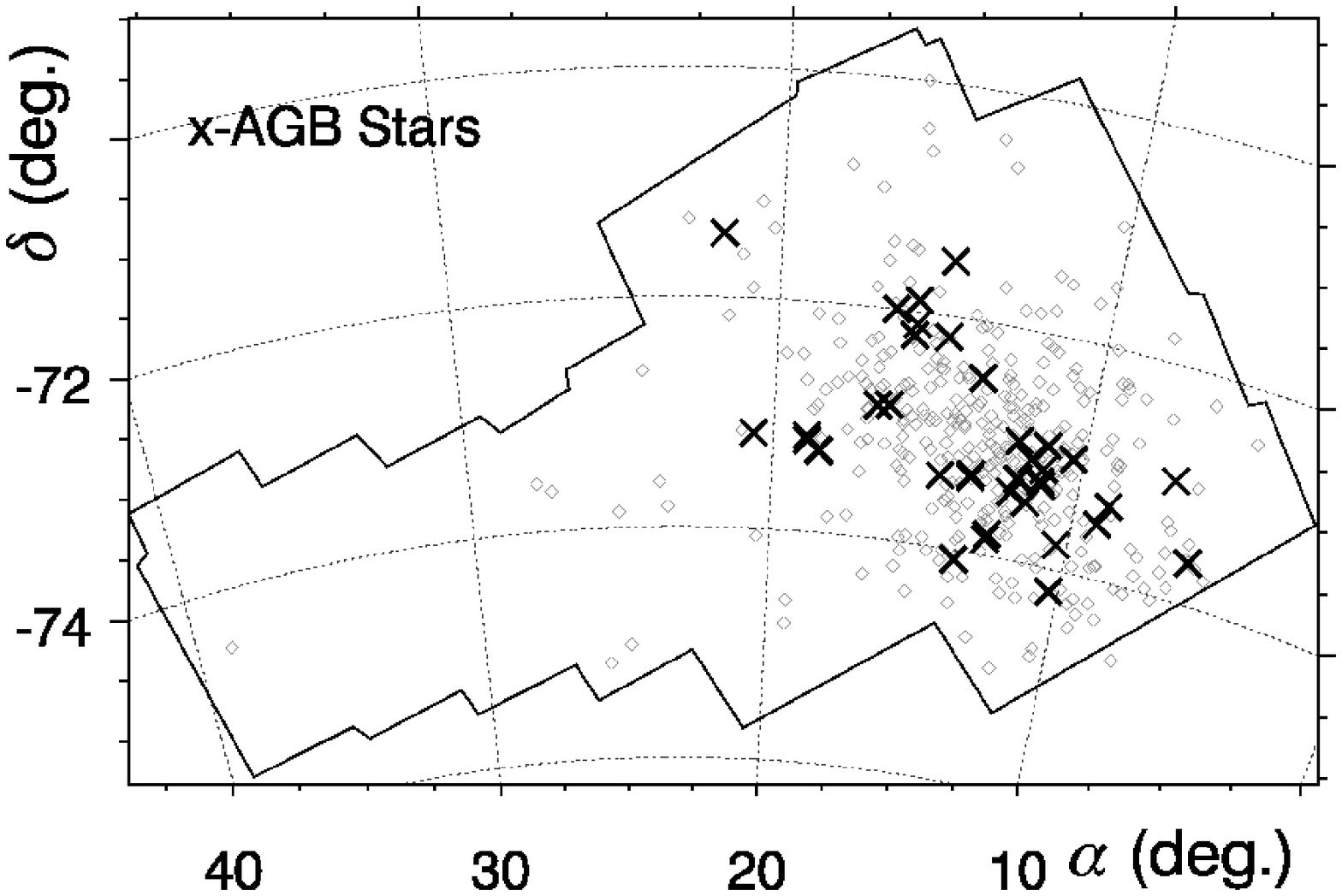}
\figcaption{Same as Fig.~\ref{fig:disto}, for C-AGB (top) and x-AGB
(bottom) stars.\label{fig:distc}}
\end{figure}

\begin{figure}
\epsscale{1.1} 
\plotone{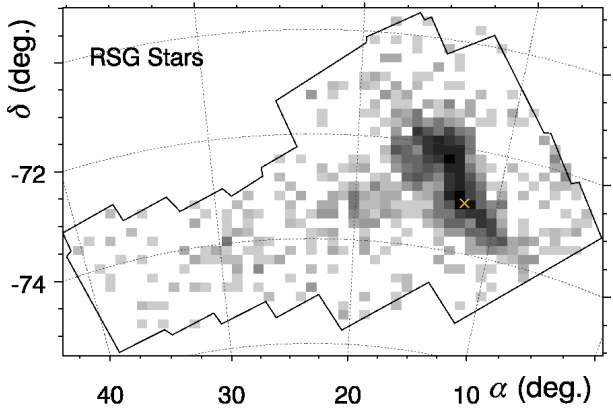} \figcaption{The spatial distributions of
RSG stars, plotted in stellar density, with the level of foreground
contamination estimated in Section~\ref{sec:fgnd} removed. Only one
FIR objects was selected from the RSG sample; it is marked in
yellow. RSG stars show a clumpy distribution that is mostly confined
to the bar and wing. \label{fig:distr}}
\end{figure}
\begin{figure}
\epsscale{1.1} \plotone{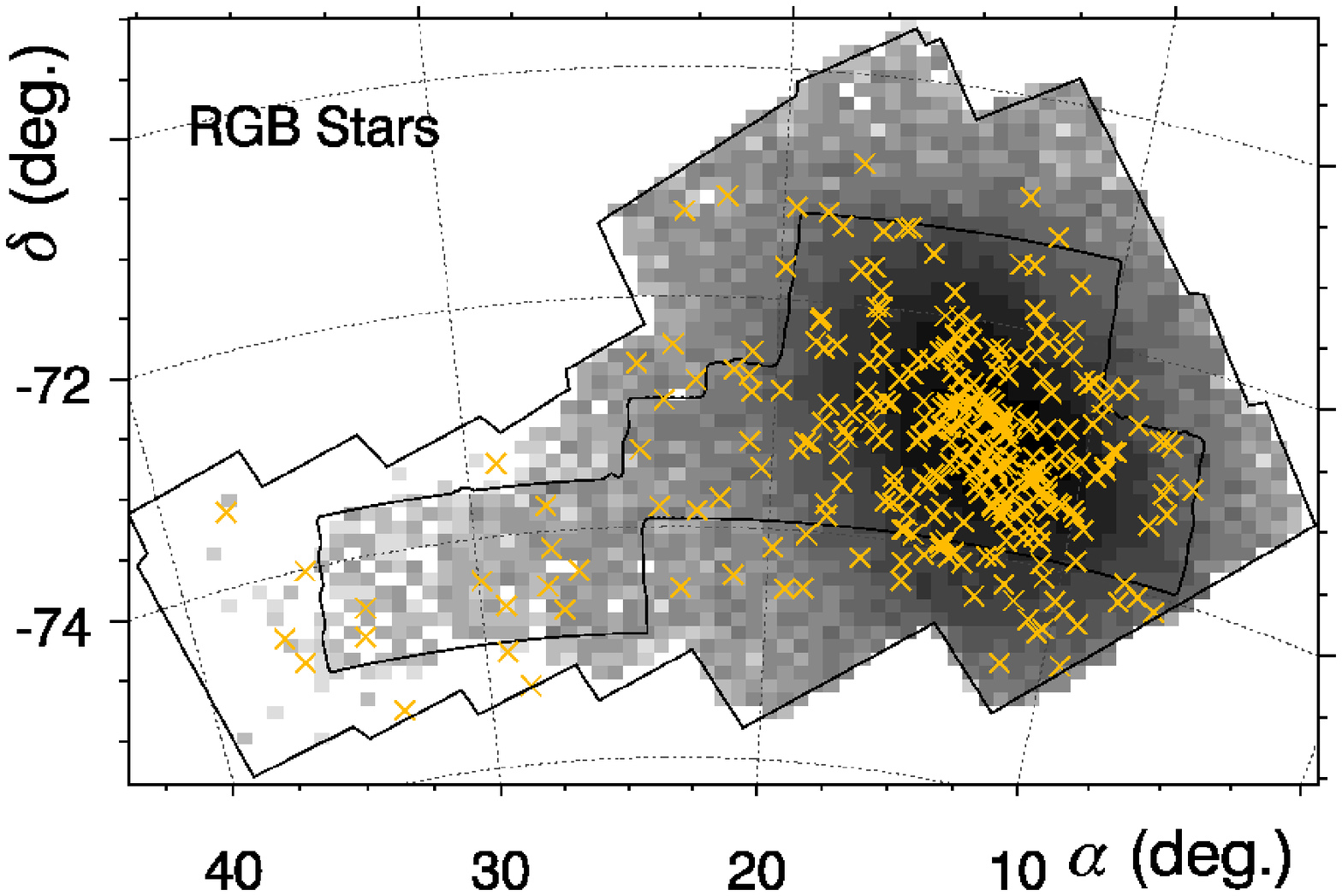} \figcaption{The spatial
distributions of RGB stars, plotted in stellar density, with the level
of foreground contamination estimated in Section~\ref{sec:fgnd}
removed. We outline the IRSF coverage, which is deeper than the 2MASS
coverage in the SMC tail. Since the foreground was estimated using
IRSF data, the regions covered only by the shallow 2MASS survey
(Fig.~\ref{fig:coverage}) have been over-subtracted. This does not
affect the RSG stars, as they are bright enough to be included in the
shallow 2MASS survey.  RGB stars show a smooth distribution from the
bar out to the tail.  FIR objects selected from the RGB sample are
plotted in yellow. \label{fig:distg}}
\end{figure}

The spatial distributions of FIR objects and RGB, RSG, and AGB stars
are presented in Figures~\ref{fig:disto} -- \ref{fig:distg}.  The RGB
stars show a smooth distribution from the bar to the tail, with a drop
in density where the 2MASS data transits from deep to shallow coverage
and there is no IRSF coverage. Most AGB stars are restricted to the
bar, where they constitute 0.4\% of the total 3.6-\micron{}
point-source population (Section~\ref{sec:stats}). In the tail, this
same fraction is 10$\times$ smaller.  RSG stars show a clumpier
distribution than the AGB stars, likely due to recent star formation
\citep[e.g.,][]{harris04,gieles08}. The RSG and RGB branches tend to
be affected by foreground contamination, so we have subtracted the
estimated foreground level from Figures~\ref{fig:distr} and
\ref{fig:distg} (Section~\ref{sec:fgnd}). The resulting distribution
of RSG stars outlines the bar and wing.

FIR sources among each population are also plotted in
Figures~\ref{fig:disto} -- \ref{fig:distg}.  These sources are
preferentially distributed in the bar, suggesting that most of them
are either very dusty evolved stars, \ion{H}{2} regions, or YSOs
belonging to the SMC.  Among the x-AGB sample, the FIR objects are
especially clustered around regions of star formation in the north and
south regions of the bar and in the wing.  FIR objects in the RGB
sample are detected out to edge of the coverage, suggesting that many
of these are unresolved background galaxies.

\subsection{Detection Statistics}
\label{sec:stats}

\begin{deluxetable}{lcc}
\tablewidth{0pc}
\tabletypesize{\normalsize}
\tablecolumns{3}
\tablecaption{Evolved Star Statistics\label{tab:stats3}}

\tablehead{\colhead{AGB Type}&\multicolumn{2}{c}{\% of AGB Stars}\\&\colhead{SMC}&\colhead{LMC}}
\startdata

O-AGB  & 42.7 $\pm$ 1.0\%  &44.2 $\pm$ 0.5\%  \\
C-AGB  & 29.8 $\pm$ 0.8\%  &25.3 $\pm$ 0.4\%\\
x-AGB  & 6.0 $\pm$ 0.3\%  & 4.5 $\pm$ 0.1\%  \\
aO-AGB & 21.5 $\pm$ 0.7\%  &26.0 $\pm$ 0.4\%\\
\hline\\
&\multicolumn{2}{c}{\% of Total Point-Sources\tablenotemark{a}}\\
&\colhead{SMC}&\colhead{LMC}\\
O-AGB  & 0.540 $\pm$ 0.011\% &0.472 $\pm$ 0.003\%\\
C-AGB  & 0.377 $\pm$ 0.009\% &0.280 $\pm$ 0.005\%\\
x-AGB  & 0.076 $\pm$ 0.004\% &0.048 $\pm$ 0.001\%\\
aO-AGB & 0.271 $\pm$ 0.008\% &0.277 $\pm$ 0.003\%\\
RSG    & 0.725 $\pm$ 0.013\% &0.200 $\pm$ 0.001\%\\
RGB    & 29.5  $\pm$ 0.091\% &17.7  $\pm$ 0.030\%\\
FIR    & 0.079 $\pm$ 0.001\% &0.065 $\pm$ 0.001\%
\enddata

\tablenotetext{a}{ \ Percentage of the total number of point-sources detected in both $J$ and [3.6] (Table~\ref{tab:stats1}).}

\end{deluxetable}

The detection statistics for SAGE-SMC are listed in
Table~\ref{tab:stats1} and \ref{tab:stats3}.  The SMC has a higher
fraction of C-rich stars (x-AGB stars tend to have a C-rich
chemistry). This is not unexpected since a generally low oxygen
abundance in low-metallicity stars and in situ carbon enrichment in
thermally-pulsing AGB stars make it easier to achieve C/O $>1$, which
is required to form carbon-rich dust. The same phenomenon can also
explain the higher fraction of aO-AGB stars in the LMC if these stars
are simply dusty O-rich AGB stars.  It should be noted that, though
there is a higher fraction of carbon-rich stars in the SMC, it has
been shown that LMC carbon stars are dustier than their cousins in the
SMC \citep[cf.][]{vanloon00,vanloon06,vanloon08b}. This might be
explained if dust is difficult to form without other metals to act as
nucleation cores.

\begin{figure}[h!]
\epsscale{1.2} \plotone{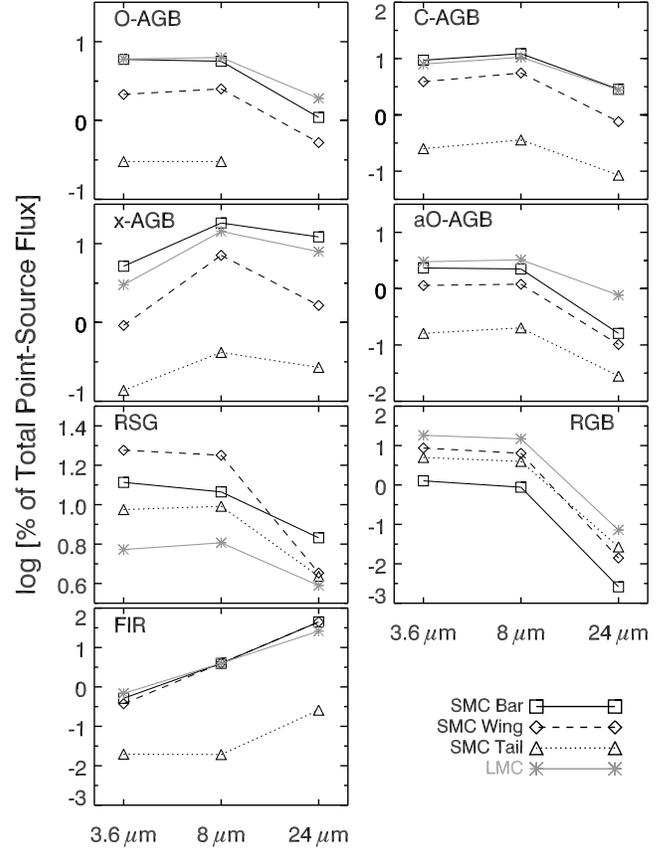} \figcaption{Fraction
of the total point-source flux contributed by AGB stars, RSG stars,
RGB stars, and FIR objects in the SMC and LMC at 3.6, 8, and
24~\micron{}. \label{fig:stats}}
\end{figure}

\begin{figure}[h!]
\epsscale{1.2} \plotone{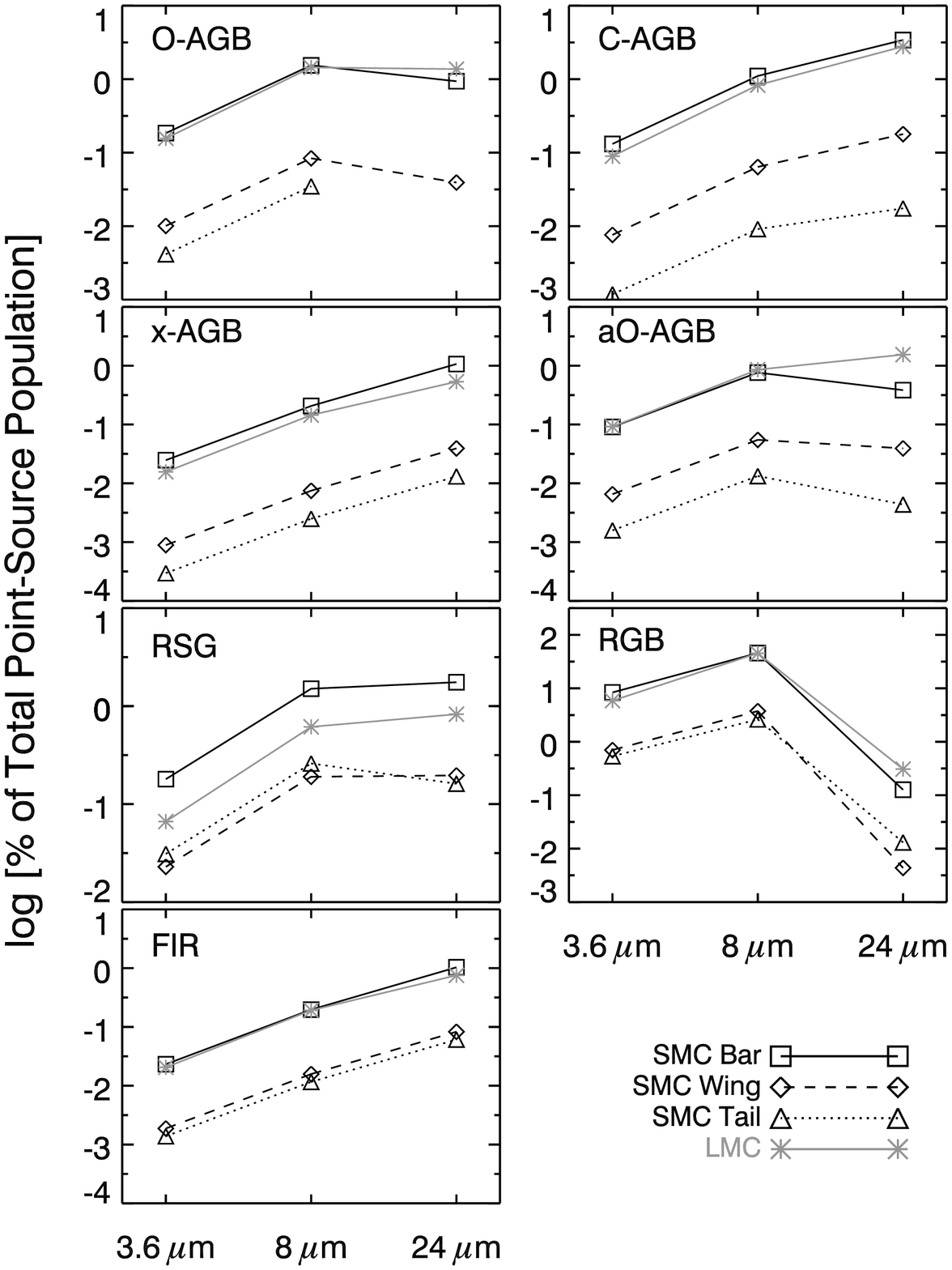} 
\figcaption{ Fraction
of the total point-source population contributed by AGB stars, RSG stars,
RGB stars, and FIR objects in the SMC and LMC at 3.6, 8, and
24~\micron{}.\label{fig:numstats}}
\end{figure}

Figure~\ref{fig:stats} shows the contribution of each stellar type to
the total point-source flux in the SMC and LMC and
Figure~\ref{fig:numstats} shows their contributions to the total
number of point-sources. Below, we point out some interesting
features:

\begin{itemize}

\item In the bar, AGB stars together
contribute 23\%, 38\%, and 16\% of the point-source flux at 3.6, 8, and
24~\micron{}, respectively, despite being only a small fraction of the
population. 

\item At 8 and 24 \micron{}, the x-AGB stars are the most impressive
evolved stars. They contribute $\approx$12\% --18\% of the
point-source flux in the bar while numbering $\lesssim$1\% of the
total stellar population. In the wing, the x-AGB stars appear less
important, though this may be the result of small-number statistics
since only 9 x-AGB stars are detected in the wing.

\item In the LMC and the SMC bar, the AGB stars together out-shine the
RSG stars at 3.6, 8, and 24~\micron{}. The reverse is true in the
wing, where recent star formation may be enhancing the RSG population,
increasing their contribution at 3.6 and 8~\micron. The LMC RSG stars
are less significant, contributing 3$\times$ -- 5$\times$ less to the
total point-source flux than the AGB stars at all wavelengths.

We note that about 16\% of the RSG stars in the SMC have 24-\micron\
counterparts \citep[Table~\ref{tab:stats1}, see also][]{bonanos10};
this fraction rises to 34\% for the LMC. A similar trend is seen in
the O-AGB stars. Some of this discrepancy may be due to the increased
distance to the SMC and limited sensitivity of 24-\micron\ data, but
it may also suggest that O-rich stars form dust more easily in the LMC
\citep{vanloon06,vanloon08b,bonanos10}.

\item Although RGB stars are by far the most numerous stars SAGE-SMC
data, they contribute $<$14\% to the 3.6- and 8-\micron\ point-source
fluxes. This scenario is quite different to that in old stellar
populations such as globular clusters, where RGB stars can contribute
close to 100\% of the mid-IR point-source flux.  The contribution from
RGB stars is higher in the wing and tail than in the bar, perhaps
indicating an aging stellar population in the outskirts of the SMC.

\item The FIR objects are extremely important
contributors to the mid-IR flux, providing up to 45\% of the flux
at 24~\micron{}.  If these sources had not been culled from the
evolved stars list, the result would be a huge overestimate of the
mid-IR flux contribution from evolved stars.

\end{itemize}

\citet{maraston06} and \citet{henriques10} find that thermally-pulsing
AGB stars are important contributors to the SED of galaxies at
redshift $z\sim2-3$, particularly at rest-frame near-IR
wavelengths. The above points and Figure~\ref{fig:stats} reinforce
these results. However, if we are to fully understand the importance
of cool, evolved stars in distant galaxies, we must also consider the
global flux (point-source $+$ extended).  The global flux of the SMC
at IRAC and MIPS wavelengths within 2.5\degr\ of the SMC center (an
area covering most of the bar and wing) is measured by
\citet{gordon11}. At 3.6~\micron{}, we find that the RGB and AGB stars
in this same area each contribute 21\% to the SMC global flux, with
RSG stars trailing close behind at 19.5\%.  The AGB stars do,
therefore, appear to play an important role in the near-IR
SED. However, the contribution from less evolved RGB and more massive
RSG stars is equally important.  This agrees with
recent work by Melbourne et al.\ (2011, in prep), showing the
importance of RSG stars to the near-IR flux of nearby
galaxies. However, we note that RGB and RSG stars are strongly contaminated by
foreground sources (Section~\ref{sec:fgnd}).

The picture changes considerably at longer wavelengths, with the AGB
stars contributing more to the global flux (17\%) than RGB
and RSG stars (each $\approx$7\%) at 8~\micron{}. The ISM emission
dominates at longer wavelengths: $<$3\% of the global
24-\micron{} flux is produced by the cool evolved stars, $2/3$ of which is
due to AGB stars.

\section{Evolved Star Characteristics}
\label{sec:agbs}

\subsection{Infrared colors}
\label{sec:ccds}
\begin{figure}
\epsscale{1} \plotone{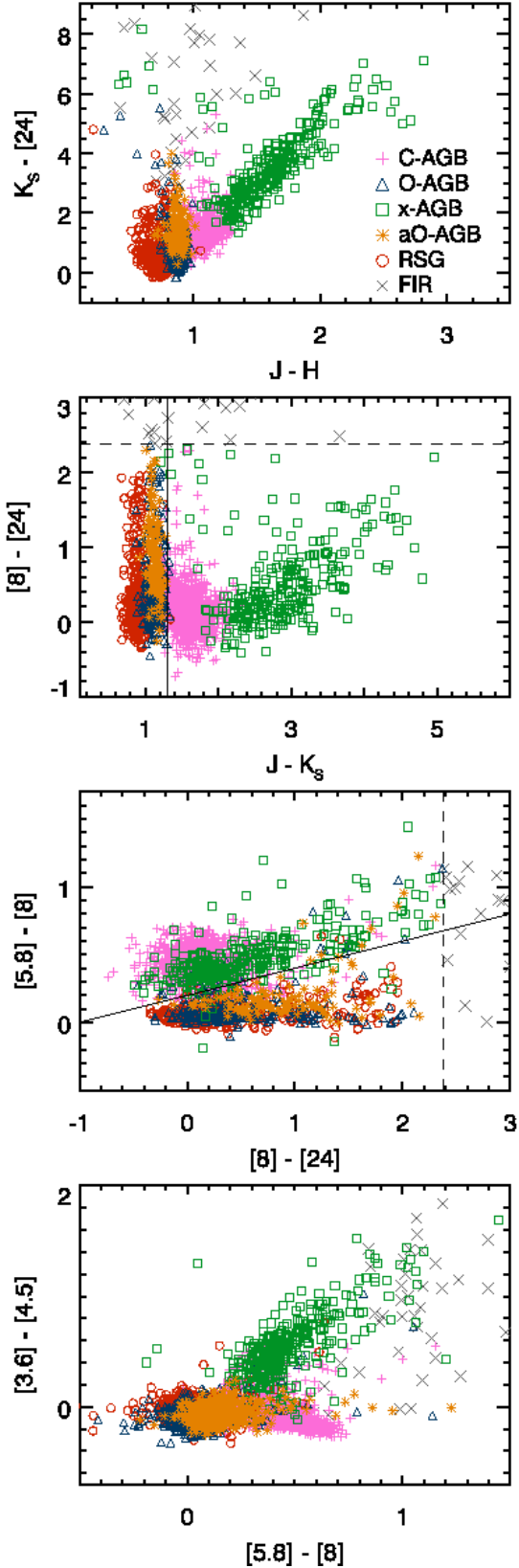} \figcaption{CCDs of the
SMC AGB and RSG stars and FIR objects. The dashed lines indicate the
color where the 24-\micron\ flux equals the 8-\micron\ flux. The solid
lines mark approximate divisions between C-rich and O-rich
stars.\label{fig:ccds}}
\end{figure}

\begin{figure}
\epsscale{1.1} \plotone{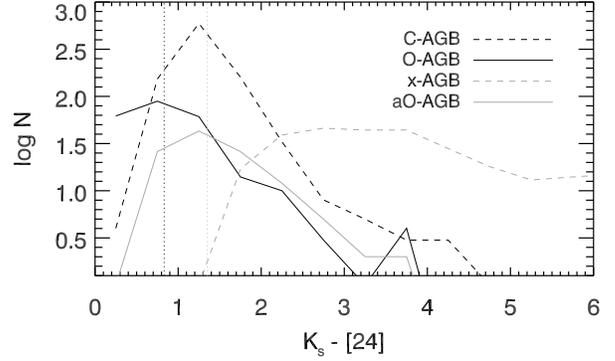} \figcaption{Distribution
of AGB stars among the $K_{\rm s} - [24]$ color. The median color of the aO-AGB stars are is 0.52~mag redder than the O-AGB stars (vertical dotted lines), indicating a
stronger dust excess in the aO-AGB stars.  \label{fig:dust_hist}}
\end{figure}

\begin{figure*}
\epsscale{1.15} \plotone{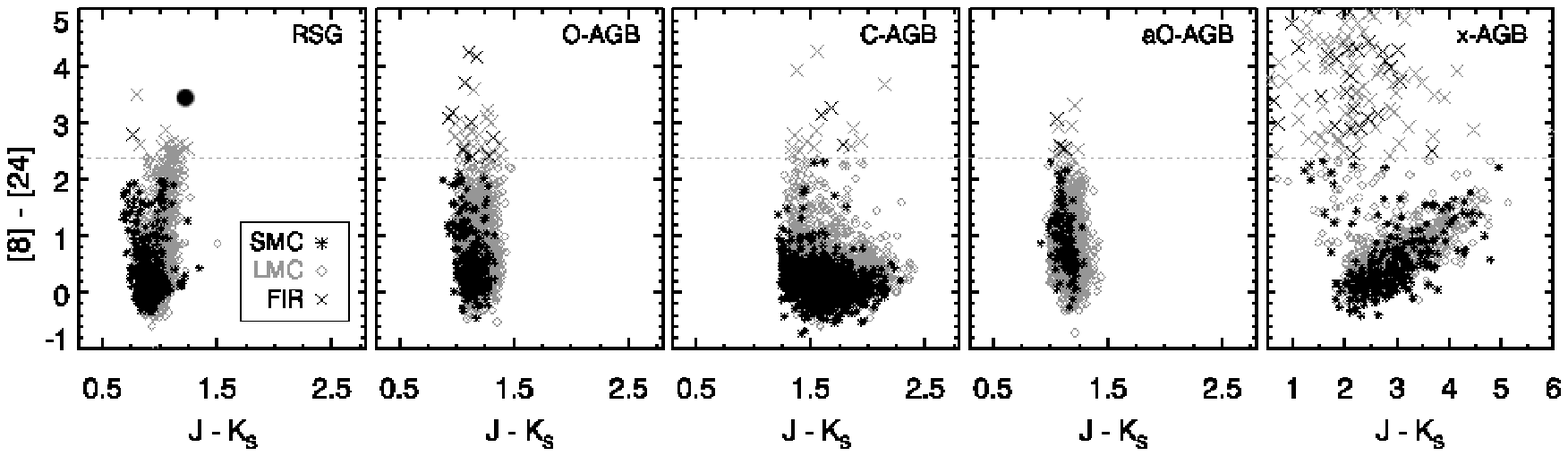} \figcaption{CCDs of the SMC
AGB stars (black) compared to LMC AGB stars (gray). FIR objects are
marked by crosses. The x-AGB samples contain many FIR
objects, illustrating that one must take care when selecting x-AGB
stars by photometric criteria alone.  See \citet{whitney03},
\citet{bolatto07}, and \citet{sloan06} for similar
figures. \label{fig:ccd_compare}}
\end{figure*}

IR colors are good diagnostics of stellar and dust properties when
investigating a large stellar population and/or when lacking IR
spectra. In Figure~\ref{fig:ccds}, we used the AGB and RSG stars
selected in Section~\ref{sec:agbid} to reproduce the CCDs presented in
\citet{sloan06}, \citet{kastner08}, \citet{lagadec07}, and
\citet{groenewegen09} to distinguish different types of AGB stars.

\subsubsection{$J-H$ vs.\ $K_{\rm s}-[24]$}

The top panel of Figure~\ref{fig:ccds} is essentially a comparison of
the photosphere or spectral type ($J-H$) to the dust excess ($K_{\rm
s}-[24]$), except for the most heavily dust-enshrouded stars, where
$J-H$ becomes a measure of the optical depth of the dust envelope
instead. The distribution of AGB stars among $K_{\rm s}-[24]$ color
(Fig.~\ref{fig:dust_hist}) shows that the C-AGB and aO-AGB stars are
dustier than the O-AGB stars. However, we note that since the aO-AGB
stars are selected based on their red $J-[8]$ colors (compared to
O-AGB stars), it is not unexpected to also see red $K_{\rm s}-[24]$
colors. The x-AGB stars show the strongest dust excess and an extremely
extinguished photosphere. We see that there is a large range of
24-\micron{} excess for all evolved stars, even over a small $J-H$
range.

The 24~\micron{} flux is generally dominated by continuum dust
emission, though some O-AGB stars can achieve strong 24-\micron\
excesses since the 24-\micron\ filter clips the red edge of the broad
20-\micron\ silicate feature. While the bulk of O-AGB stars have
$K_{\rm s}-[24] < 3$~mag, a small population of O-rich sources show
colors as red as the x-AGB stars. It is possible that this group of
sources is contaminated by YSOs or that the 24-\micron{} emission is
instead due to illumination of local ISM dust.

\subsubsection{$J-K_{\rm s}$ vs.\ $[8]-[24]$}

In the second panel of Figure~\ref{fig:ccds}, we show a CCD
similar to that presented by \citet{sloan06}, which showed a
separation of stars with spectroscopically-confirmed O-rich and C-rich
dust. The silicate sequence is vertical, with $J-K_{\rm s} \lesssim
1.3$ (solid line), and the carbon-dust sequence is more horizontal,
starting near $[8]-[24] \sim 0$~mag.  Based on this classification,
the majority of the x-AGB stars appear to be dominated by C-rich
chemistry, and the aO-AGB stars do indeed appear to be dominated by
O-rich chemistry. The dashed line shows the limit where $F_\nu({\rm 24
\micron}) = F_\nu({\rm 8 \micron})$.

\subsubsection{$[8]-[24]$ vs.\ $[5.8]-[8]$}

The third panel of Figure~\ref{fig:ccds} is used by
\citet{groenewegen09} and \citet{kastner08} to show the color
differences between O-AGB, C-AGB, and RSG stars. When one takes into
account the difference in nomenclature (in both of those works, the
C-rich AGB and O-rich AGB stars are what we call x-AGB stars and
bright O-rich AGB stars here, respectively), we see that our CCDs are
quite similar to these previous studies.

When restricted to {\it Spitzer} colors, the $[8]-[24]$
vs. $[5.8]-[8]$ CCD is a good choice for distinguishing C-rich and
O-rich AGB stars since 95\% of the stars classified as O-rich in
Section~\ref{sec:agbid} fall below the solid line in the third panel
of Figure~\ref{fig:ccds} and 90\% of the C-rich stars lie above
it. However, the RSG stars remain indistinguishable from the O-rich
AGB stars and x-AGB stars show significant overlap with C-AGB
stars. This color scheme also restricts the selection to those
detected at 24~\micron{}, which is a minority of the O-rich population
in the SMC.  Nonetheless, \citet{lagadec07} show the separation of
spectroscopically confirmed C-rich and O-rich dust-enshrouded sources
in this CCD. Our classification scheme is consistent with what they
find.

\subsubsection{$[5.8]-[8]$ vs.\ $[3.6]-[4.5]$}

In the bottom panel Figure~\ref{fig:ccds}, we can see the effect that
CO and/or C$_3$ absorption has on the $[3.6]-[4.5]$ color. It is clear
that the C-AGB stars show stronger absorption as the 8-\micron{}
excess increases and the x-AGB stars show almost no indication of this
absorption due to dust emission veiling the molecular absorption bands
\citep{vanloon06,vanloon08b}. The IRAC colors are sufficient for
distinguishing the x-AGB stars, but C-AGB and O-rich AGB stars are
difficult to isolate without the addition of near-IR photometry.

\subsubsection{Comparing SMC and LMC Colors}

Figure~\ref{fig:ccd_compare} shows the $J-K_{\rm s}$ vs.\ $[8]-[24]$
CCD from the second panel of Figure~\ref{fig:ccds}, this time
comparing the SMC to the LMC.  The LMC tends to have a larger
population than the SMC at very red $[8]-[24]$ colors, but the {\it
range} in these colors is essentially the same in both galaxies.

The near-IR color tends to be slightly redder for the LMC stars than
the SMC stars, likely due to the difference in metallicity between the
galaxies. The x-AGB stars are the exception to this rule, possibly
because their near-IR stellar flux is extremely extinguished by
circumstellar dust.

The RSG, O-AGB and aO-AGB stars show little, if any, contamination
from C-rich sources, but the scatter in the C-AGB and x-AGB
populations may indicate a population of dusty O-rich sources. This is
especially evident in the LMC. Very obscured O-rich sources (though
rare) can reach colors redder than $J-K_{\rm s} = 2$ \citep[e.g.,
IRAS\,05298$-$6957 or
IRAS\,05280$-$6910;][]{wood92,vanloon01a,vanloon01b,vanloon10,kemper10,boyer10sdp}. These
O-rich interlopers can skew the measurement of the global mass-loss
rate (see Section~\ref{sec:mdot}) for the C-rich stars. However, the
C-rich stars contribute significantly more to the cumulative mass-loss
rate than their O-rich counterparts in the LMC \citep[][also see
Section~\ref{sec:mdot}]{srinivasan09}, so this effect may not be
significant.

The CCDs in Figure~\ref{fig:ccds} indicate that the IR
colors, particularly the combination of $J-K_{\rm s}$ and $[8]-[24]$,
can be used in the absence of IR spectra to identify AGB and RSG stars
with reasonable confidence. The minimal differences between the LMC
and SMC CCDs also suggest that the CCD in Figure~\ref{fig:ccd_compare}
applies to stellar populations over a fairly broad range in
metallicity. However, one must use caution especially when attempting
to distinguish x-AGB stars and YSOs.

\begin{figure}
\epsscale{1.1} \plotone{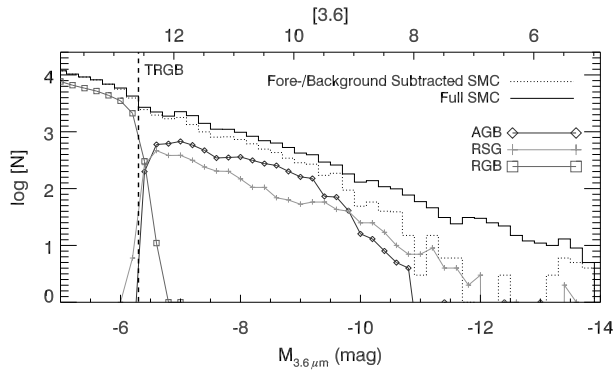} \figcaption{
  3.6-\micron\ luminosity functions for the SMC (dotted line shows the SMC corrected for background and foreground contamination). The luminosity functions of each type of evolved star are also shown. The
  vertical dashed line marks the 3.6~\micron\ TRGB.\label{fig:lumall}}
\end{figure}

\begin{figure}
\vbox{
\includegraphics[width=0.45\textwidth]{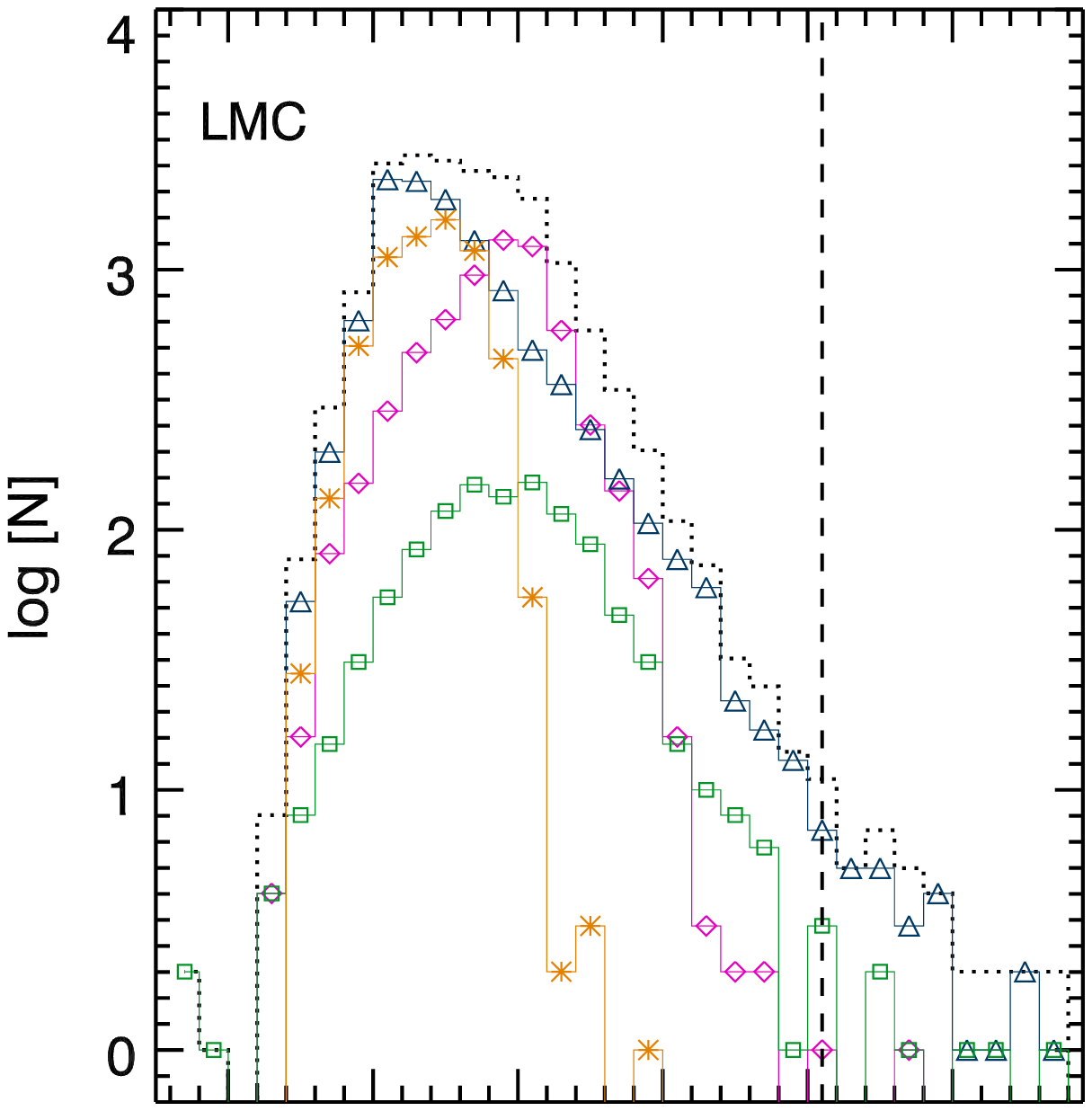}
\includegraphics[width=0.45\textwidth]{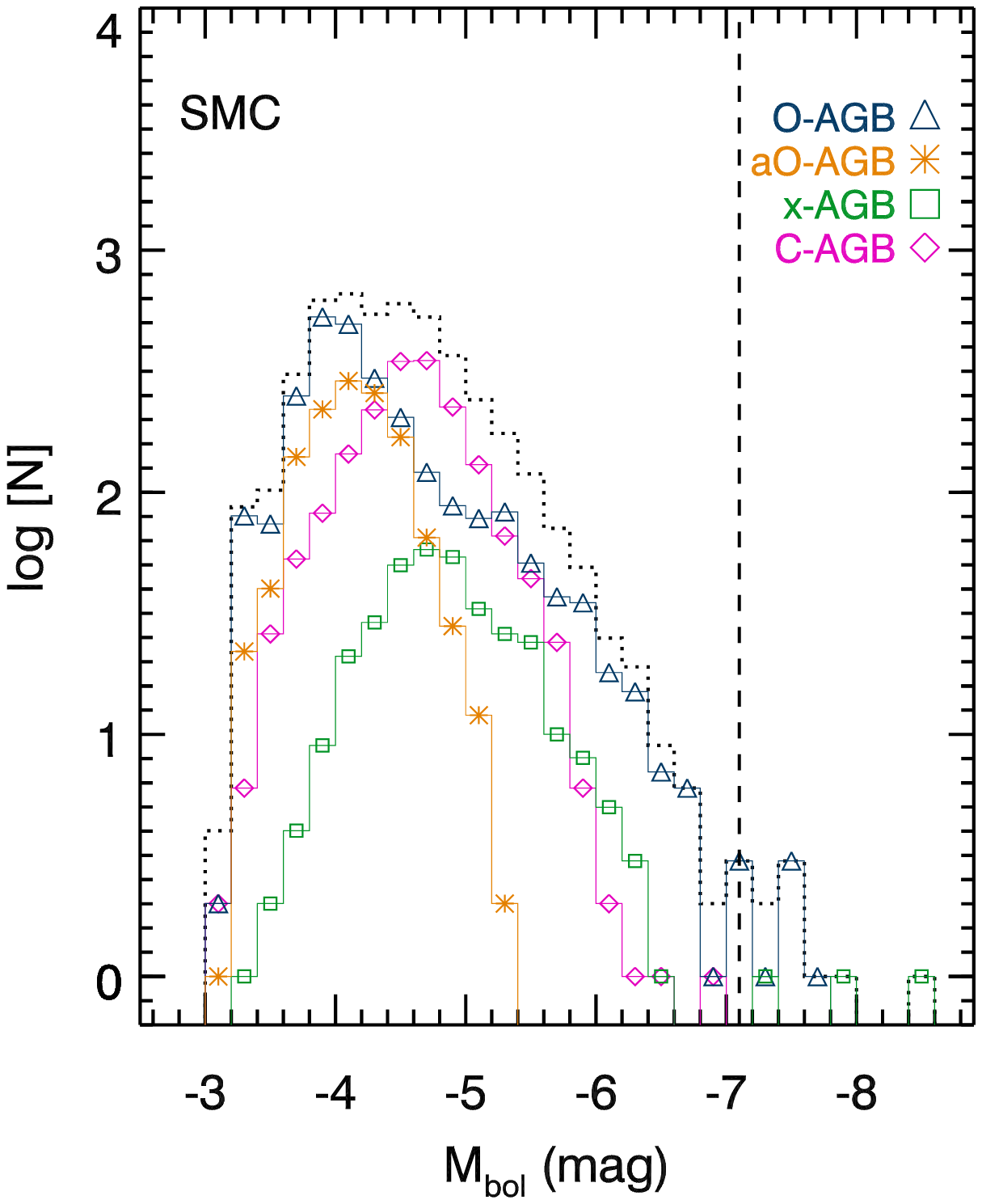}
}

\figcaption{AGB luminosity
  functions for the SMC and LMC. The vertical dashed line marks the classical AGB limit
  ($M_{\rm bol} = -7.1$~mag); only 9 SMC AGB stars ($<0.2\%$) and 34
  LMC AGB stars ($<0.1\%$) exceed this limit. \label{fig:lum}}
\end{figure}

\subsection{Luminosity Functions}
\label{sec:lfunc}

Luminosity functions are useful in constraining the evolutionary
models of evolved stars, providing constraints to the star-formation
history and to several processes and parameters, including
nucleosynthesis, mixing, mass loss, evolutionary rate, stellar
lifetime, and initial stellar mass
\citep{marigo99,javadi11}. Figure~\ref{fig:lumall} shows the
3.6-\micron{} luminosity functions of the SMC and its cool evolved
stars. The RSG stars dominate the luminosity function at the brightest
magnitudes. This points to the importance of RSG
stars to the total integrated near- to mid-IR luminosities of
galaxies, despite their low numbers (Section~\ref{sec:stats}).

The 3.6~\micron\ luminosity function of RSG stars drops smoothly with
magnitude and spans the broadest magnitude range. AGB stars do not
drop smoothly with magnitude, showing an enhancement near $-8 >
M_{3.6} > -10$~mag from C-AGB stars. We show the individual luminosity
functions for AGB stars in Figure~{\ref{fig:lum}. To estimate the
  luminosities of the AGB stars, we performed a simple trapezoidal
  integration from the optical $U$-band through mid-IR 24-\micron{}
  flux. We find 6 O-AGB and 3 x-AGB stars that are brighter than the
  classical AGB limit ($M_{\rm bol} = -7.1$~mag) in the SMC; it is
  possible for AGB stars to exceed the classical limit if they are at
  the peak of their pulsation cycle or are experiencing HBB
  \citep{smith85,boothroyd92,vanloon01a,vanloon05a,vanloon05b}.
  \citet{srinivasan09} found hundreds of AGB stars brighter than the
  classical limit in the LMC, but this is due to an overestimate of
  the luminosities in that work (S. Srinivasan, private
  communication).  We present the revised LMC bolometric magnitudes
  here and find 34 LMC AGB stars brighter than the classical
  limit.

\subsubsection{The Carbon Star Luminosity Function}
\label{sec:cslf}

\begin{figure}
\epsscale{1.2} \plotone{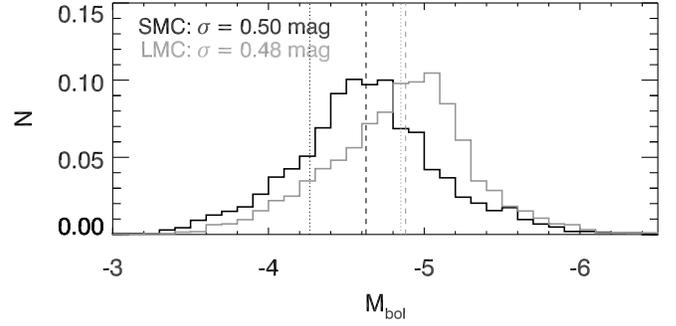} \figcaption{Carbon Star
  Luminosity Functions for the SMC and LMC, normalized to the total
  number of stars in each sample.  Both C-AGB and x-AGB stars are
  included. The medians are marked by dashed lines.  The dotted lines
  mark the medians found by \citet{marigo99}. Excluding the x-AGB
  stars results in a fainter peak ($\Delta M_{\rm bol} = 0.04$~mag for
  the SMC, but no shift for the LMC).  In both galaxies, excluding the
  x-AGB stars results in a dispersion ($\sigma$) that is 0.04~mag
  narrower.\label{fig:cslf}}
\end{figure}

\citet{marigo99} use the carbon star luminosity functions (CSLFs) of
the Magellanic Clouds to constrain the third dredge-up process.  The
CSLF was derived by \citet{groenewegen98}, using $\approx$1\,700 C-rich
AGB stars identified spectroscopically by the Swan C$_2$ bands at
5165\,\AA\ and 4737\,\AA\, and bolometric corrections from
\citet{westerlund86}. These data cover most of the bar and wing of the
SMC, but exclude the heavily enshrouded C-rich x-AGB stars.  They find
that the CSLF peaks at $M_{\rm bol} = -4.265$~mag (adjusting for a
slight difference in adopted distance modulus) and is broad and
roughly symmetric. Here, the median of the combined C-AGB and x-AGB
luminosity function (Fig.~\ref{fig:cslf}) is $M_{\rm bol} =
-4.63$~mag, with a $1\,\sigma$ dispersion of 0.50~mag. The peak is
significantly brighter than the peak observed by \citet{marigo99}.  If
we exclude the x-AGB stars, the result is a slightly fainter and
narrower CSLF ($M_{\rm bol, peak} = -4.59$~mag, $1\,\sigma =
0.46$~mag). \citet{vanloon99a,vanloon99b,vanloon06} also show that the
heavily enshrouded C-rich stars (the x-AGB stars here) tend to be more
bolometrically luminous than the optically-detected carbon stars.

\begin{figure}
\epsscale{1.2} \plotone{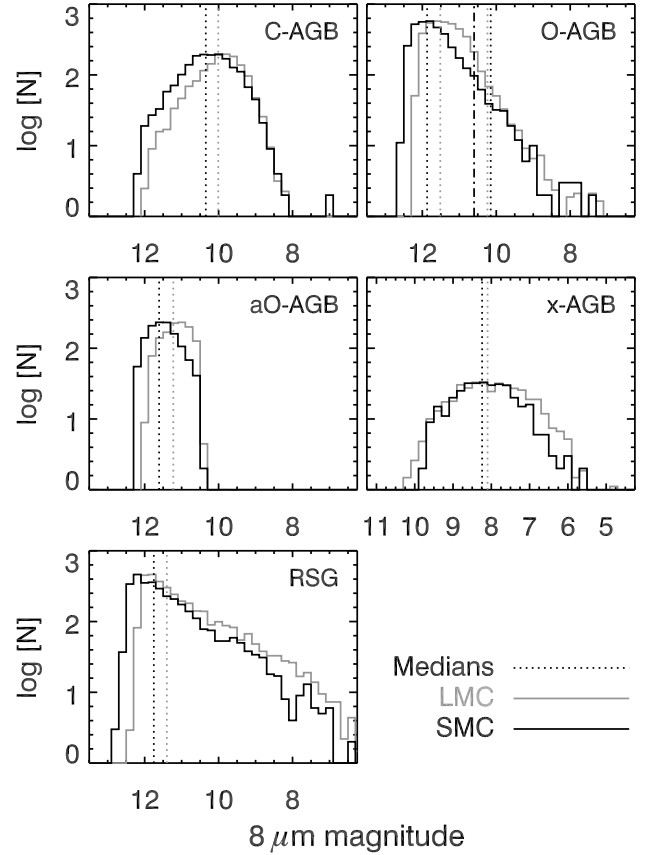} \figcaption{AGB and RSG $[8]$
histograms for the SMC (black lines) and LMC (gray lines). Medians are
marked by dotted lines, and the dot-dash line in the O-AGB panel shows
the adopted division between bright and faint O-AGB stars.  The LMC
has been shifted 0.4 mag fainter to match the SMC distance (assuming
$d_{\rm SMC} = 61$~kpc and $d_{\rm LMC} = 51$~kpc).  The LMC
distributions have also been scaled down so that the peaks of the LMC
and SMC distributions match. Although the aO-AGB stars are shown
separately, they are also included in the C-AGB or O-AGB panels, based
on their original classification using the $J-K_{\rm s}$
color. \label{fig:lfunc_8mag}}
\end{figure}

For the LMC, our CSLF peaks near the \citet{marigo99} CSLF peak, but
is significantly narrower. We find $M_{\rm bol, peak} = -4.88$~mag and
a width $1\,\sigma = 0.48$~mag, whereas \citet{marigo99} find $M_{\rm
bol, peak} = -4.84$~mag and $1\,\sigma = 0.55$~mag.

\citet{marigo99} suggest that a higher efficiency of the third
dredge-up is required in the SMC to explain the fainter peak of the
SMC CSLF compared to the LMC.  Here, we find that the SMC CSLF is
0.37~mag brighter than in that work, though it is still 0.26~mag
fainter than the LMC CSLF.  Therefore, the difference in the
efficiency of the third dredge-up between the Magellanic Clouds may not be as
substantial as they predict.

\begin{figure}
\epsscale{1.2} \plotone{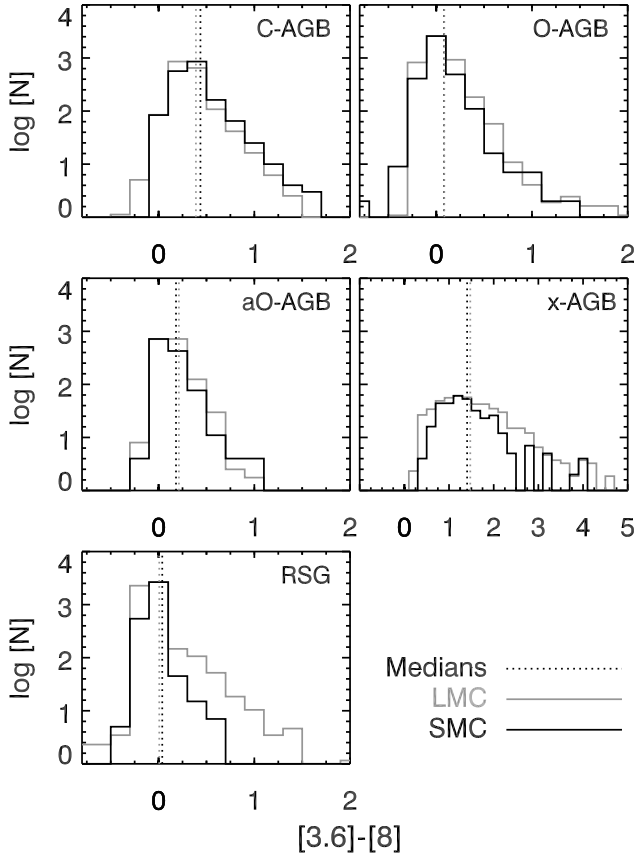} \figcaption{AGB
and RSG $[3.6]-[8]$ histograms for the SMC (black lines) and LMC (gray
lines). Medians are marked by dotted lines.  The LMC distributions
have been scaled down so that the peaks of the LMC and SMC
distributions match. Although the aO-AGB stars are shown separately,
they are also included in the C-AGB or O-AGB panels, based on their
original classification using the $J-K_{\rm s}$
color. \label{fig:lfunc_38}}
\end{figure}

\subsubsection{The 8~\micron{} Luminosity Function}
\label{sec:8lum}

The 8-\micron{} magnitude includes information on the dust emission
rather than on the temperatures of the photospheres. Histograms
showing the distribution of each stellar type over $[3.6]-[8]$ color
and over 8-\micron{} magnitude are shown in
Figures~\ref{fig:lfunc_8mag} and \ref{fig:lfunc_38}.  The LMC
histograms have been shifted down to match the SMC peak in each panel
and shifted 0.4~mag fainter at 8~\micron{} to account for the
difference in distance between the LMC and SMC
(Table~\ref{tab:params}).  The shapes of the resulting LMC histograms
are quite similar to the SMC histograms. The relative difference
between the median 8-\micron\ magnitudes (dotted lines) ranges from
0.32~mag for the bright O-AGB stars to 0.55~mag for the x-AGB stars
and 0.73 -- 0.78~mag for the other stellar types.

\begin{figure*}
\epsscale{1.2} \plotone{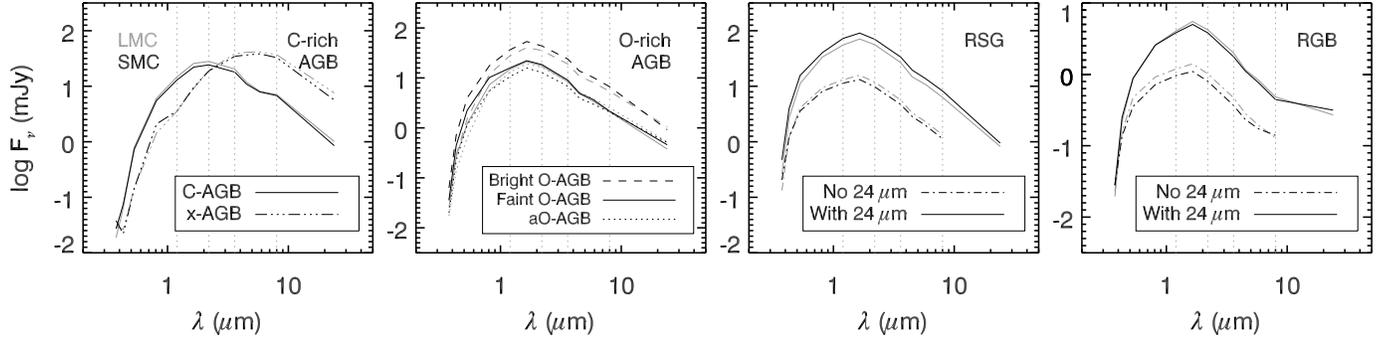} \figcaption{Median
SEDs of AGB stars. LMC SEDs are shifted down to match the
SMC distance. Uncertainties are not shown here for clarity; they
represent the range in flux over which the median was determined.  In
some cases, this range is large. \label{fig:seds}}
\end{figure*}

Figure~\ref{fig:lfunc_38} shows the distribution of the evolved stars
in $[3.6]-[8]$ color, which represents circumstellar dust excess. We
show the IRAC color to ensure that all x-AGB stars are included, since
many x-AGB stars are undetected at near-IR magnitudes. The median
colors of the RSG and x-AGB stars are similar in both galaxies
($\Delta$$([3.6]-[8]) = $ 0.03 and 0.08~mag, respectively), but it is
clear from Figure~\ref{fig:lfunc_38} that a significant number of LMC
RSG and x-AGB stars show strong 8-\micron\ excess \citep[also
see][]{bonanos10}. This suggests that the LMC stars may be more
efficient dust producers, whether the dust is O-rich or C-rich
\citep[cf.][]{vanloon00,vanloon06,vanloon08b}.  It is interesting that
the other evolved stars do not show a strong difference in $[3.6]-[8]$
between the LMC and SMC; it seems as though metallicity does not
strongly affect the dust production except in the more extreme
evolutionary phases.

\subsection{Spectral Energy Distributions}
\label{sec:seds} 

The SAGE-SMC (and SAGE-LMC) catalogs are ideal for investigating the
full evolved star SEDs since they include 12 bands of photometry
and a well-sampled SED near the peak luminosity. The SEDs shown reach
only to 24~\micron{}, but very few stars are expected to show much
emission at longer wavelengths \citep[][also see
Section~\ref{sec:70mic}]{boyer10sdp}. Figure~\ref{fig:seds} shows the
median SEDs for each type of evolved star, with LMC SEDs
scaled down to account for the difference in distance between the
galaxies. We note that these SEDs only include stars in the SMC bar,
as the optical MCPS coverage does not extend to the tail
(Fig.~\ref{fig:coverage}).  The median SEDs are resistant to outliers,
so we show SEDs binned by magnitude and color in
Section~\ref{sec:agb_seds}.

The x-AGB stars are heavily extinguished, showing a peak luminosity
between 4.5 and 8~\micron{}. O-AGB, C-AGB, and aO-AGB stars peak in
the near-IR, with C-AGB stars showing the coolest temperatures of the
three. RSG and RGB stars peak near the $H$-band. A second SED is shown
for RSG and RGB stars undetected at 24~\micron{}. Less than 1\% of the
SMC RGB stars have 24-\micron{} counterparts (Table~\ref{tab:stats1}),
and based on the strong 24-\micron\ excess in these sources, it is
likely that they are either mis-identified or contain a separate
mid-IR source along the line-of-sight.

There are few differences between the LMC median SEDs and the SMC
SEDs. Discrepancies in the peak flux may be due to uncertainty in the
relative distances of the galaxies or to an intrinsic difference
between the luminosity functions of the galaxies (e.g.,
Fig.~\ref{fig:lfunc_8mag}). For AGB stars, the 24-\micron\ point is
slightly brighter in the LMC than in the SMC. This appears to be due
more to the higher fraction of 24-\micron\ counterparts in the LMC,
and less to differences in individual stars. In the previous section,
we demonstrated that the median 8-\micron{} magnitudes of all evolved
stars except the bright O-AGB stars are brighter in the LMC than the
SMC.  Figure~\ref{fig:seds} shows the opposite for the RSG stars,
which is due to separating those that are detected at 24~\micron{}
from those that are not; the large population of faint RSG stars
detected at 24~\micron\ bring down the LMC 8-\micron\ median.

\subsubsection{AGB star SEDs}
\label{sec:agb_seds}
\begin{figure*}

\vbox{
\includegraphics[width=1\textwidth]{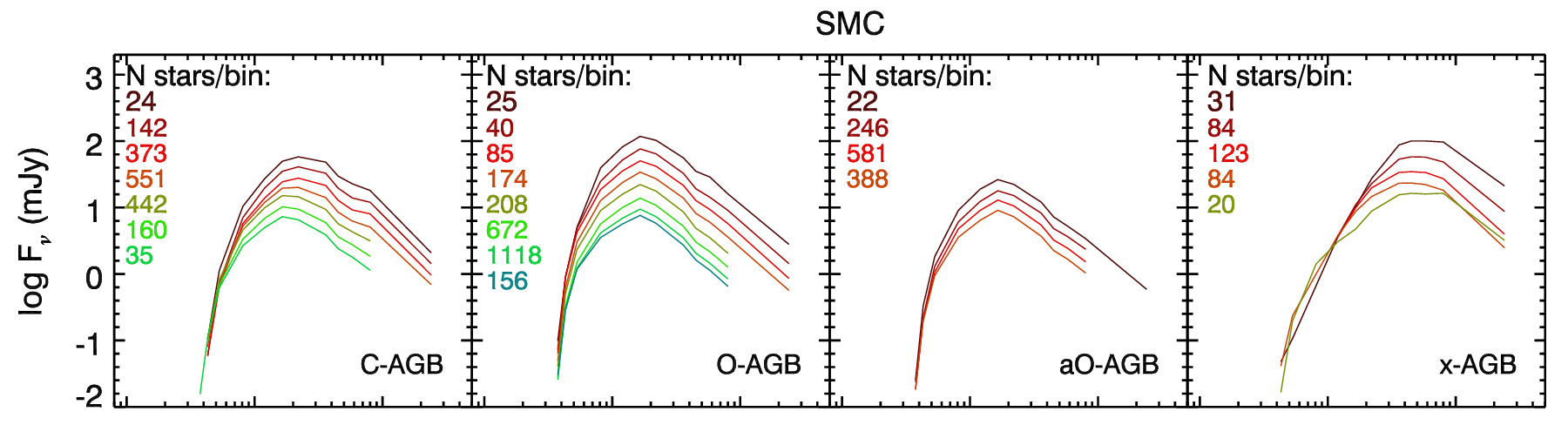}
\includegraphics[width=1\textwidth]{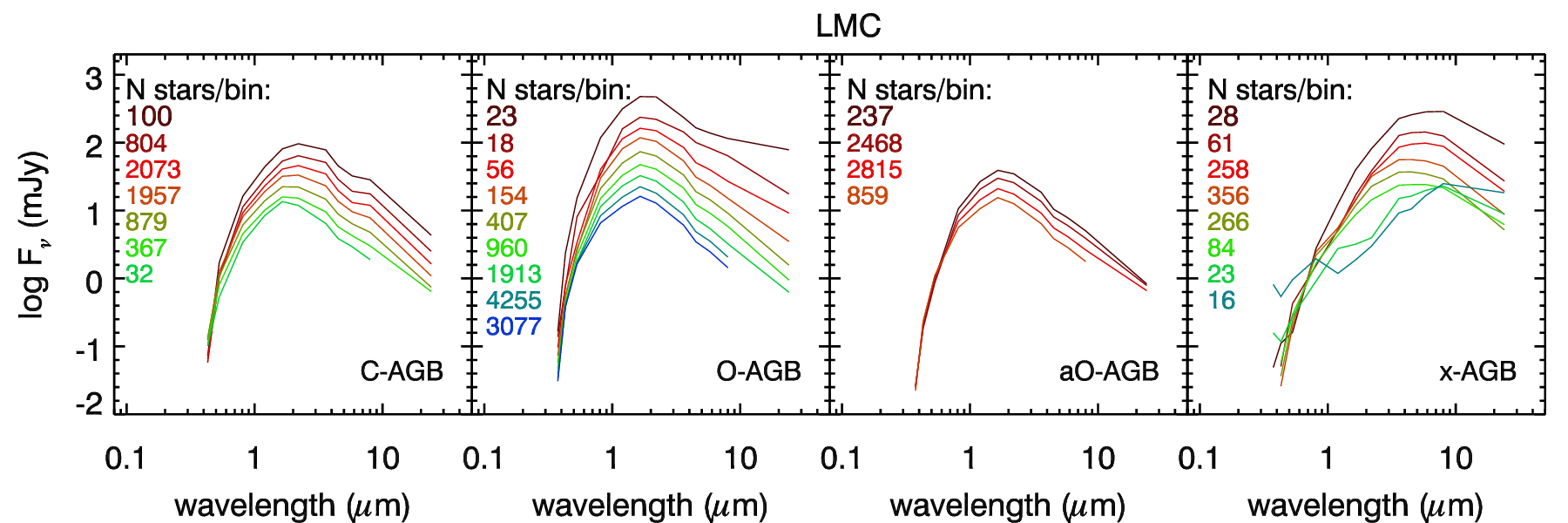}
}

\figcaption{Median SEDs of AGB stars, binned by 3.6-\micron{}
  magnitude.  Bin size is 0.5~mag, and the number of stars within each
  bin is printed in the top left of each panel with the color
  corresponding to the SED color. \label{fig:magbin}}
\end{figure*}

Figure~\ref{fig:magbin} shows the median AGB SEDs, binned in steps of
0.5~mag at 3.6~\micron{}. In the SMC, the faintest and brightest SEDs
of a particular stellar type look quite similar, with no obvious
differences in the SED features (except for non-detections at
24~\micron).  The LMC, on the other hand, shows many differences
between the brightest and faintest sources. LMC O-AGB stars show a
striking increase in 24-\micron\ excess among the brightest
$\approx$100 O-AGB stars. Inspection of the individual SEDs shows that
this excess is real, though it is possible that some of these sources
are RSG or YSO interlopers. In the SMC, we find only one O-AGB
star with 24-\micron\ excess similar to these bright LMC O-AGB stars.

The opposite effect is observed in the LMC aO-AGB stars, with the
brightest stars showing a slightly weaker 24-\micron\ excess.  In this
case, it appears the difference in 24-\micron\ excess is due to the
comparatively small number of stars included in the brightest bin of
aO-AGB stars. The population of aO-AGB stars show a wide range of
24-\micron\ excess, mostly independent of the total luminosity
(Fig.~\ref{fig:cmd_agb}).

In both galaxies, the C-rich stars show an absorption feature from 4
to 8~\micron{} that is likely due to CO $+$
C$_3$. This feature becomes stronger as
C-AGB stars become redder (and brighter), then is veiled by continuum
dust emission in the heavily extinguished x-AGB stars. This
disappearance of the molecular feature in the fainter/bluer C-AGB
stars may indicate some contamination from stars dominated by O-rich
chemistry among the faint/blue edge of the C-AGB branch on the CMD
(Figs.~\ref{fig:cmd_labeled} and \ref{fig:ccd_compare}).  The
disappearance of the feature could also be explained if the faint
C-AGB stars have warm photospheres or if C/O is close to unity.

\subsubsection{RSG and RGB star SEDs}
\label{sec:rsg_seds}

The median SEDs of RSG and RGB stars, binned in steps of 0.5~mag at
3.6~\micron{}, are shown in Figure~\ref{fig:RSGmagbin}. RGB stars show
no distinctive differences between the LMC and SMC or between the
brightest and faintest stars. However, RSG stars show strong
variations as a function of brightness. The brightest $\approx$100
RSGs in the LMC do indeed show enhanced 8- and 24-\micron\ excess, and
the 200 or so RSGs fainter than those also show a broad range of
mid-IR excess.  The same is true for the brightest $\approx$50 RSG
stars in the SMC.  These SEDs indicate that significant amounts of
warm RSG dust form only around the brightest 7\% of LMC RSG stars and
2\% of SMC RSG stars. \citet{bonanos10} find similar trends. Since
this dust emits strongly at 24~\micron{}, its temperature may be
slightly cooler than typical AGB dust due to a larger dust-free inner
envelope \citep[cf.][]{vanloon05a}. Alternatively, these bright RSG
stars may be those with the strongest silicate emission, which would
enhance the flux in both the 8- and 24-\micron{} filters.

\begin{figure}

\vbox{
\includegraphics[width=0.48\textwidth]{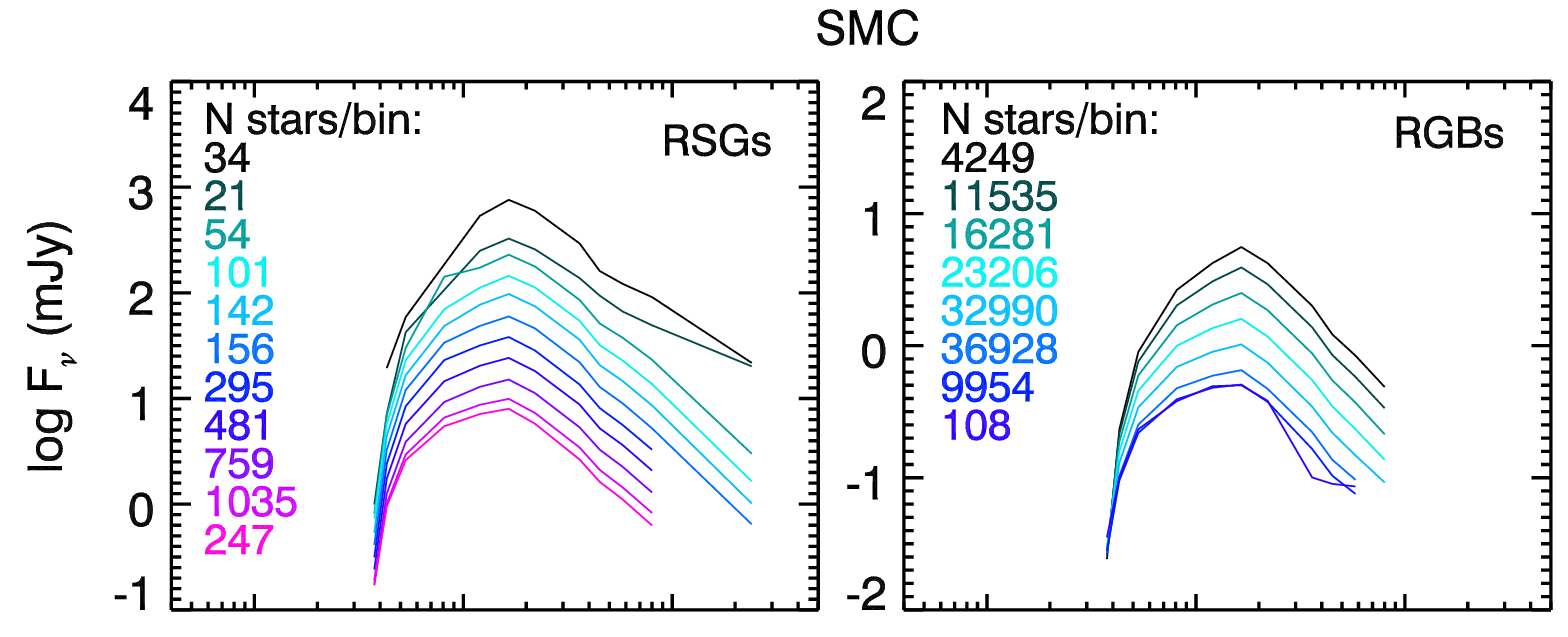}
\includegraphics[width=0.48\textwidth]{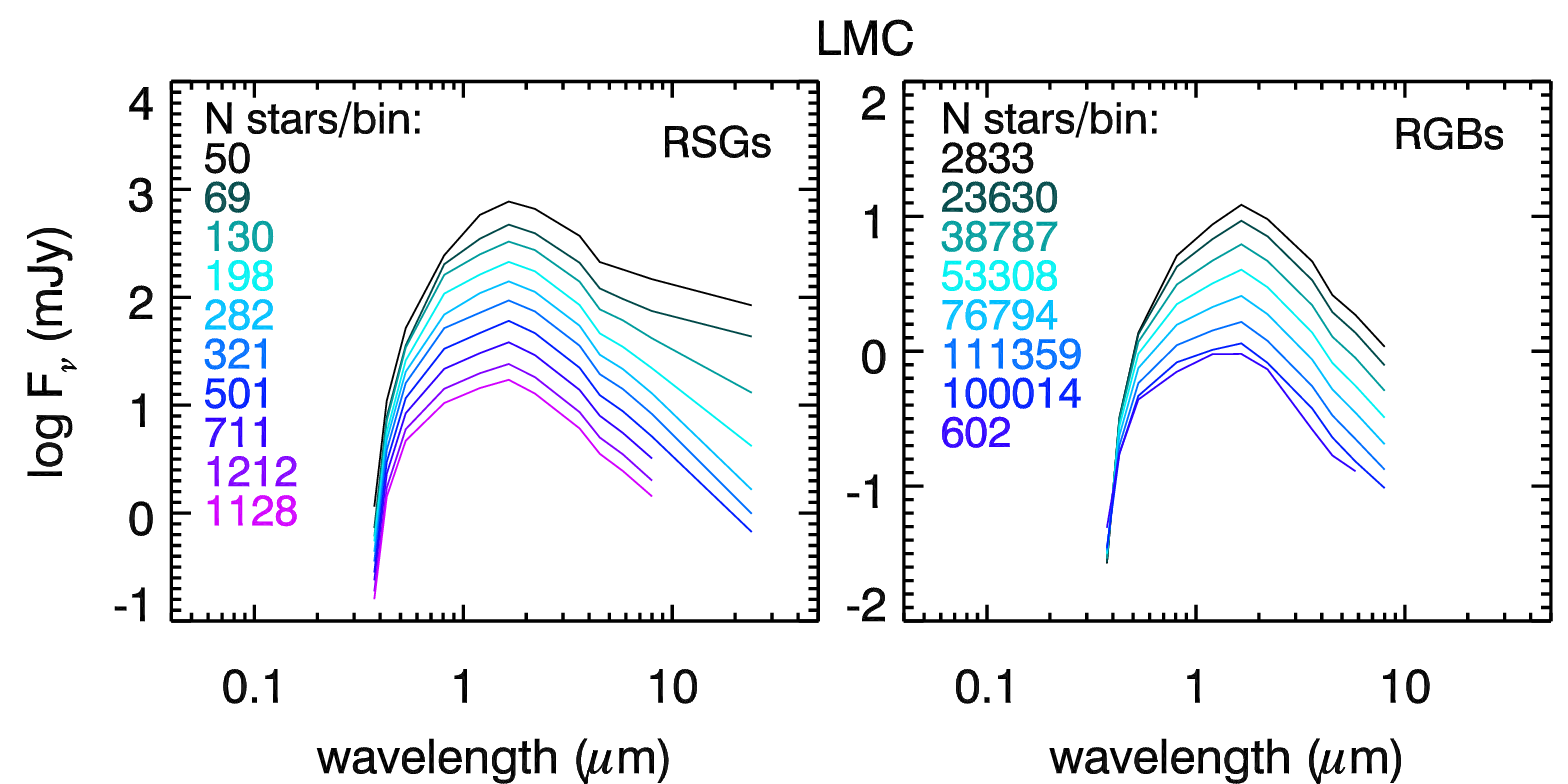}
}

\figcaption{Same as Figure~\ref{fig:magbin}, but showing the RGB and
  RSG stars. Less than 1\% of the RGB stars are detected at
  24~\micron{}, so we do not include the 24-\micron\ point
  here. The range in 8- and 24-\micron\ flux shown in the brightest RSG
  stars is real, i.e., it is not a consequence of having a small
  number of stars in the brightest bins.\label{fig:RSGmagbin}}
\end{figure}

\subsection{70-\micron{} Point-Sources}
\label{sec:70mic}

\begin{deluxetable*}{lrr}
\tablewidth{0pc}
\tabletypesize{\normalsize}
\tablecolumns{3}
\tablecaption{Potential 70-\micron\ Detections\label{tab:70}}

\tablehead{\colhead{SAGE IRAC Designation}&\colhead{Classification}&\colhead{Alternative Name}}
\startdata

\multicolumn{3}{c}{x-AGB Candidates}\\
SSTISAGEMA J003659.53$-$741950.3  & post-AGB [1,2,3] & IRAS 00350$-$7436\\
SSTISAGEMA J004808.49$-$731454.7  & H\,{\sc II} region [4]&N26, IRAS\,00462$-$7331\\
SSTISAGEMA J005640.88$-$725425.2  & \nodata&\nodata\\
SSTISAGEMA J010503.97$-$715925.4  & emission line star [5] & LIN 439\\
SSTISAGEMA J010507.26$-$715942.8  & YSO candidate [6] & S3MC 01051-7159, IRAS 01035$-$7215(?)\\
SSTISAGEMA J012407.95$-$730904.0  & H\,{\sc II} region [7,8]& N88A, IRAS\,01228$-$7324\\
\hline		     
\\		     
\multicolumn{3}{c}{RSG Star Candidates}\\
SSTISAGEMA J004352.42$-$730721.7  & \nodata&\nodata\\
SSTISAGEMA J004717.74$-$730917.5  & \nodata&\nodata\\
SSTISAGEMA J004807.03$-$724612.2  & \nodata&\\
SSTISAGEMA J004955.73$-$730250.8  & KM supergiant [9,10] & SkKM 50, LI-SMC 59\\
SSTISAGEMA J004957.59$-$724815.8  & \nodata&\nodata\\
SSTISAGEMA J005047.65$-$731316.1  & \nodata&\nodata\\
SSTISAGEMA J005553.51$-$731826.8  & Foreground [11] & IRAS F00542$-$7334\\
SSTISAGEMA J005659.94$-$722403.8  & \nodata&\nodata\\
SSTISAGEMA J005749.34$-$723555.6  & \nodata&\nodata\\
SSTISAGEMA J010155.47$-$720029.3  & \nodata&\nodata\\
SSTISAGEMA J010220.80$-$722105.1  & \nodata&\nodata\\
SSTISAGEMA J010905.42$-$723832.4  & Foreground [11] & IRAS 01075$-$7254, HD\,7100\\
SSTISAGEMA J011235.16$-$730935.4  & KM supergiant [9] & SkKM 327

\enddata

\tablecomments{ \ RSG and x-AGB candidates with apparent matches to 70-\micron\ point-sources.  REFERENCES: [1] \citet{whitelock89}; [2] \citet{vanloon08b}; [3] \citet{groenewegen98b}; [4] \citet{indebetouw04}; [5] \citet{lindsay61}; [6] \citet{vanloon10b};  [7] \citet{testor85}; [8] \citet{heydari99}; [9] \citet{sanduleak89}; [10] \citet{loup97}; [11] \citet{massey07}.}

\end{deluxetable*}

The SAGE-SMC and S$^3$MC MIPS 70-\micron{} observations are not
sensitive enough to detect typical AGB and RSG stars in the SMC.
However, a small sample of sources we identify as evolved stars based
on the mid-IR colors (Section~\ref{sec:selection}) are associated with
70-\micron\ point-sources to within one full-width half-maximum (FWHM)
of the 70-\micron\ point-spread function (PSF; 18\arcsec{}). We show
the full SEDs of these sources in Figures~\ref{fig:70xagb} and
\ref{fig:70rsg} \citep[also see][]{vanloon10b}.

Six of the 70-\micron\ sources are in our x-AGB star sample
(Table~\ref{tab:70}). Two sources have the coordinates of known
compact \ion{H}{2} regions \citep[N88A and N26;
e.g.,][]{testor85,heydari99,indebetouw04}. One is 4.6\arcsec\ away
from a known post-AGB star
\citep[IRAS\,00350$-$7436;][]{whitelock89,vanloon08b}, and one is only
0.1\arcsec\ from a YSO candidate
\citep[S3MC\,01051$-$7159;][]{vanloon10b}.  Another source
(SSTISAGEMA\,J010503.97$-$715925.4) is 1.7\arcsec\ from an emission
line star \citep[LIN\,439;][]{lindsay61}.  We expect that the
70~\micron{} emission does in fact originate from these non-AGB
sources or from superimposed background objects rather than from true
x-AGB stars. The remaining 70-\micron\ source in the x-AGB sample
(SSTISAGEMA\,J005640.88$-$725425.2) is not identified in the
literature, and we cannot confidently identify it based on its SED
shape (Fig.~\ref{fig:70xagb}) , especially since it is not matched to an optical or near-IR
source.  This source is centrally located in the bar, and its 24 and
70-\micron\ photometry may be affected by surrounding diffuse
emission.

LIN\,439 and S3MC\,01051$-$7159 are adjacent to one another
(0.38\arcmin) and in a crowded region on the northeast edge of the
bar, surrounded by strong diffuse mid-IR emission.  IRAS\,00350$-$7436
is located on the southern edge of the bar, is isolated, and in a
region of low background.

Several more 70-\micron\ point-sources are associated with FIR objects
that are within the x-AGB locus of the IR CMDs
(Section~\ref{sec:fir}). We excluded these sources from the x-AGB
sample because their 24-\micron\ fluxes exceed their 8-\micron\
fluxes, suggesting that they might be YSOs or background
galaxies. Indeed, if S3MC\,01051-7159, IRAS\,00462--7331,
IRAS\,01228--7324, and LIN\,439 were detected at 24~\micron{}, the
shape of their SEDs (Fig.~\ref{fig:70xagb}) suggest that they should
be classified as FIR objects here instead of x-AGB stars.

\begin{figure}[h!]
\epsscale{1.2} \plotone{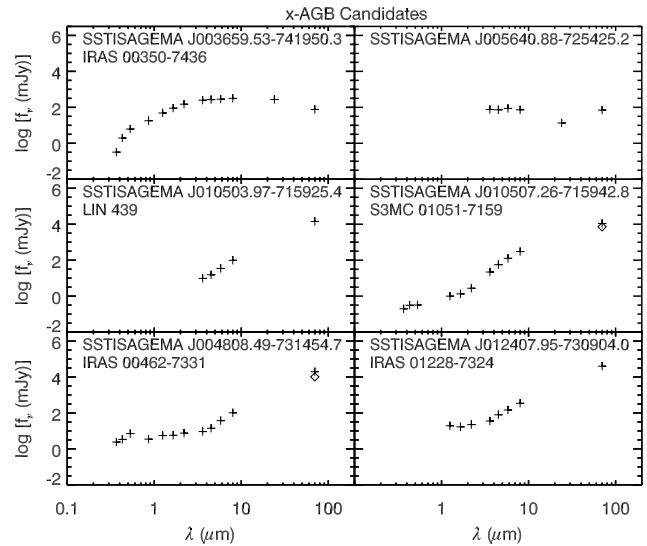} \figcaption{SEDs of
70-\micron{} point-sources in the x-AGB sample.  Diamond points are
S$^3$MC fluxes. \label{fig:70xagb}}
\end{figure}

\begin{figure}[h!]
\epsscale{1.2} \plotone{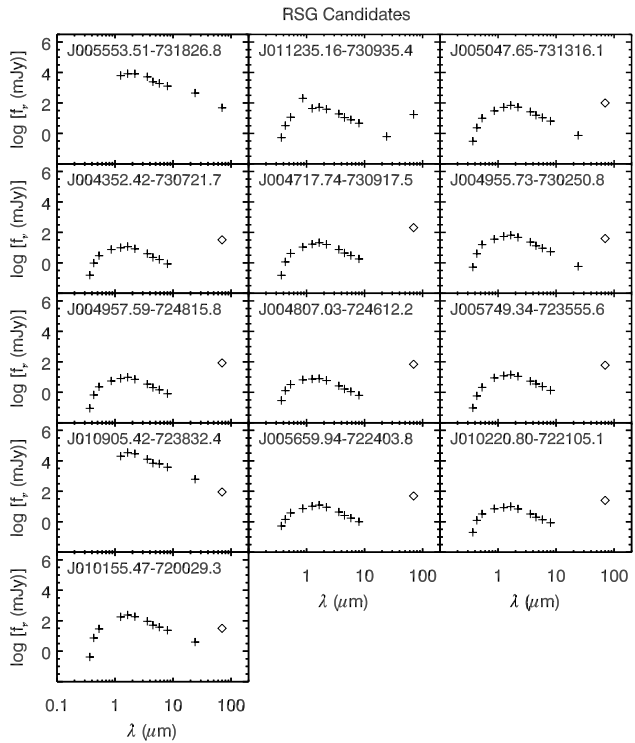} \figcaption{SEDs of RSG
 candidates detected at 70~\micron{}. The fluxes at $\lambda >
 10$~\micron{} may be from background galaxies, YSOs, or compact H\,{\sc ii}
 regions along the line-of-sight. Diamond points are S$^3$MC fluxes.\label{fig:70rsg}}
\end{figure}

Thirteen RSG candidates are also associated with 70-\micron\ sources.
Two of these are likely foreground objects \citep[IRAS\,F00542$-$7334 and
IRAS\,01075$-$7254;][]{loup97,massey07}, and two others are identified as
K or M supergiants \citep[SkKM\,50, SkKM\,327;][]{sanduleak89,loup97}.

The remaining 9 sources (Fig.~\ref{fig:70rsg}) clearly show a
photosphere peaking in the near-IR and what appears to be a secondary
peak at $\lambda > 10$~\micron{}.  SkKM\,50 and SkKM\,327 also fall
within this category. It is possible (and perhaps likely) that the
long-wavelength flux is not from the stars themselves, but from a
superimposed compact \ion{H}{2} region, YSO or background galaxy.  This
possibility is supported by the fact that half of these stars are not
matched to 24-\micron\ sources. If the far-IR flux does indeed originate
from the star, it may be in the form of a detached dusty shell.

No C-AGB, O-AGB, or aO-AGB stars are confidently associated with
70-\micron\ point-sources, though several are just outside the FWHM of
the 70-\micron\ PSF. This includes 5 C-AGB, 2 O-AGB, 3 aO-AGB, and 6
RSG stars. Inspection of the images suggests that the mid-IR sources
are distinct from the far-IR sources.

\section{Evolved stars as probes}
\label{sec:probes}

\subsection{Stellar Spatial Structure}
\label{sec:regional}

Since the SAGE-SMC survey has, for the first time, provided IR imaging
covering the entire extent of the SMC (Fig.~\ref{fig:coverage}), we
can now use the cool evolved stars to investigate the overall
structure of the SMC. Figure~\ref{fig:cmd_bwt} shows the IR CMDs of
the SMC bar, wing, and tail regions (see Fig.~\ref{fig:coverage}),
with foreground/background contamination subtracted
(Section~\ref{sec:fgnd}). It is clear that the bar is home to a
relatively old population of stars, as its CMD is dominated by
cool/red RGB and AGB stars, though a young population also exists.
The wing region resembles the bar, albeit with a smaller population.
The tail region is dominated by foreground and background sources (see
Fig.~\ref{fig:cmd_sub}), with only a faint hint at the presence of the
RGB.  Very few SMC stars reside in the tail, suggesting that the bar
stellar population has not yet been significantly perturbed by
interaction with the LMC and Milky Way.

\begin{figure}[h!]
\epsscale{1.2} \plotone{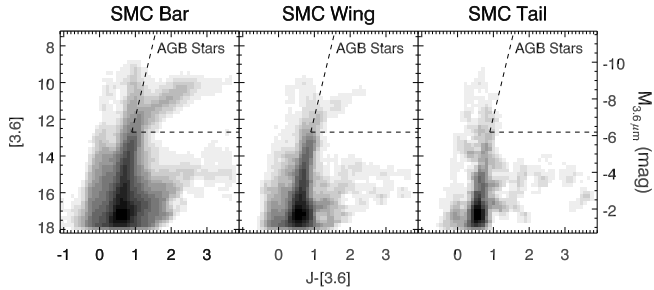} \figcaption{IR
CMD for the SMC bar, wing, and tail regions, with
foreground/background subtracted (see Fig.~\ref{fig:cmd_sub}). Branches
indicative of recent star formation (e.g., the RSG, A--G supergiant,
and OB star branches) are visible in both the wing and bar.  The tail
is dominated by the RGB. \label{fig:cmd_bwt}}
\end{figure}

The population characteristics implied by the CMDs are corroborated by
the radial distributions of different stellar types. We measured
radial profiles of AGB, RSG, OB, and RGB stars by computing the source
density within concentric elliptical annuli centered where the bulk
stellar population peaks, with the semi-minor axes of the annuli in
steps of $\approx$10\arcmin{}, increasing along the direction of the
tail (Fig.~\ref{fig:radtech}). We remove the foreground and background
contamination using the estimates from Section~\ref{sec:fgnd} and
present the results in Figure~\ref{fig:agbprofiles}.

All profiles show a smooth decline in source density into the wing
region.  Subtracting the foreground from the O-AGB, RGB, OB, and RSG
sources causes distributions to vary significantly in the tail, due to low
source counts in that region. However, it is also possible that the
source density of young OB and RSG stars is enhanced in the tail due
to continued star formation in this gaseous filament
\citep{mizuno06,gordon09}. This is indicated by an increase in the OB
and RSG star populations in the tail ($\approx$4$\degr$ from the center),
but this increase is within 3\,$\sigma$ (error bars in
Fig.~\ref{fig:agbprofiles} are 1\,$\sigma$). The AGB and RGB profiles
do not have this enhancement in the tail, instead showing a smooth
decline in source density from the bar, through the wing, and to the
tail \citep[cf.][]{harris07}.

The RGB profile decreases steadily out to at least 5\degr\ from the
center of the bar. This suggests the presence of a very extended halo,
as indicated by \citet{nidever11} and supporting $\Lambda$CDM
simulations of galaxy formation. However, since we sample the extended
population along the wing and tail, the profile at large radius may
not be representative of the entire population.

\begin{figure}
\epsscale{1} \plotone{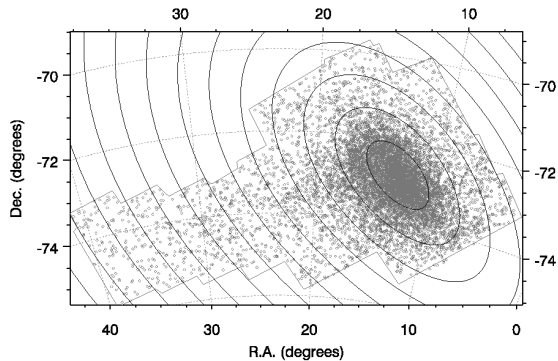} \figcaption{To
  measure the stellar radial density profiles, we compute the source
  density in concentric elliptical annuli in steps of 10\arcmin{}
  (steps of 1\degr\ are shown here for clarity), centered on the bar
  where the stellar population peaks, with semi-minor axes increasing
  along the direction of the tail.\label{fig:radtech}}
\end{figure}

\begin{figure}
\epsscale{1.1} 
\plotone{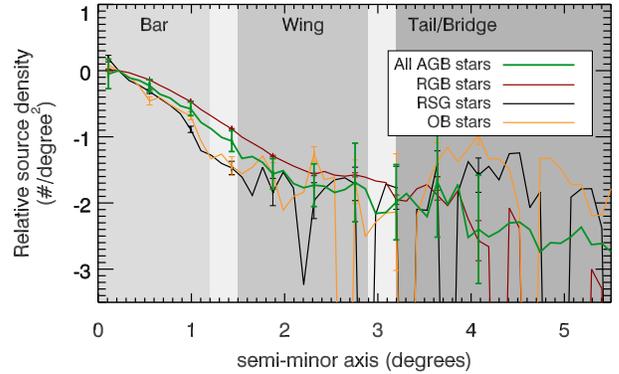} \figcaption{Stellar radial
density profiles, normalized to the second bin. We show 1\,$\sigma$
error bars derived from Poisson statistics, which represent the degree
to which small number statistics affect the trend (some error bars are
excluded for clarity). The foreground and background contamination was
subtracted according to
Section~\ref{sec:fgnd}. \label{fig:agbprofiles}}
\end{figure}

\begin{figure}
\epsscale{1.15} \plotone{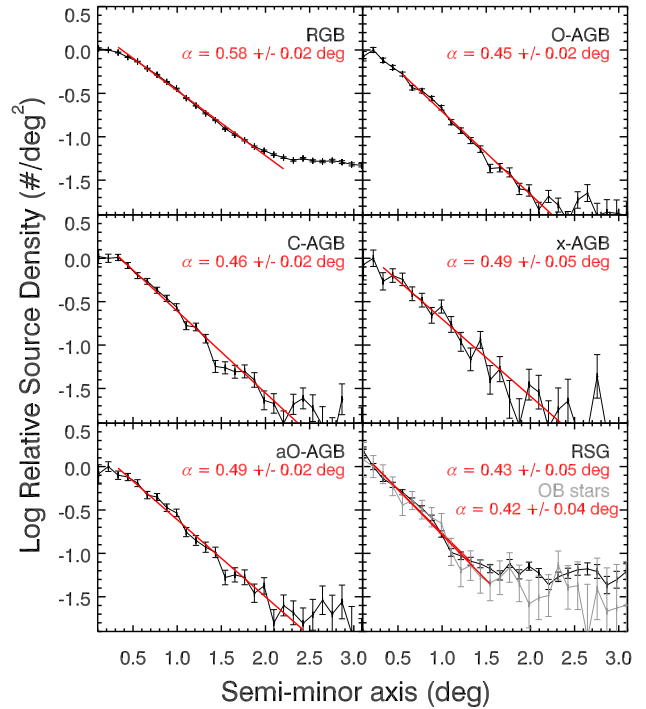} \figcaption{Radial
profiles, fit by the function shown in Equation~(\ref{eq:proffit}),
where $\alpha$ is the scale length. The oldest stellar population
(RGB stars) have the longest scale length, while the youngest
populations (OB and RSG stars) have the shortest. \label{fig:proffit}}
\end{figure}

The spatial distribution of different stellar types provides an
excellent probe of the radial age gradient within a galaxy. In
Figure~\ref{fig:proffit}, we show that the young stellar populations
(OB stars and RSGs) have a smaller scale length ($\alpha$) than the
intermediate-aged AGB stars and old RGB stars.  We fit each profile to
an exponential function of the form:

\begin{equation}
N(a) = N_0 e^{-b/\alpha}, \label{eq:proffit}
\end{equation}

\noindent where $b$ is the semi-minor axis and find that the oldest
stars are the most radially extended population. Similar scenarios
have been observed in several other dwarf galaxies
\citep[e.g.,][]{aparicio97,minniti97,hidalgo03,hidalgo09,vansevicius04,battinelli06,battinelli07,tikhonov06}.

Three scenarios might explain this stellar age gradient: (1) dynamical
relaxation as a result of encounters between stars over time, (2)
outside-in growth, or a shrinking of the star forming region, or (3)
tidal interactions.  In low-mass, isolated galaxies, \citet{stinson09}
use smoothed particle hydrodynamics simulations to show that it is
possible to produce extended old stellar halos without merging by
including a combination of options (1) and (2) above. However, the SMC
is clearly not an isolated galaxy; recent evidence shows that the LMC
in fact contains a population of stars that originated in the SMC
\citep{olsen11}.  Moreover, the star formation histories of other
nearby galaxies points to inside-out growth
\citep[e.g.,][]{williams09,gogarten10}. It therefore seems likely that
the stellar age gradient is due to dynamical relaxation of the SMC
\citep[also see][]{gieles08}.

\subsection{The C/M Ratio}
\label{sec:cmratio}

The ratio of C-rich to O-rich AGB stars, usually called the C/M ratio,
is often used as a tracer of metallicity, where a high C/M ratio
corresponds to a low metallicity, though the C/M ratio also depends on
the star formation history. Using near-IR photometry of AGB stars in
the SMC, \citet{cioni06smc} showed the C/M ratio to be a good tracer
of metallicity if the underlying stellar population is of
intermediate-age.

In Figure~\ref{fig:cm}, we show the C/M ratio for the SMC bar region,
using our selection of AGB stars. C-type stars include the C-AGB and
x-AGB stars. \citet{cioni06smc} exclude very extinguished x-AGB stars
($J-K_{\rm s} > 2.5$~mag) in their analysis, but we include them here
since the IRAC data provides us with a complete x-AGB sample.  M-type
sources include O-AGB and aO-AGB sources. \citet{cioni06smc} have 40\%
more stars in their sample due to their inclusion of sources fainter
than the $K_{\rm S}$-band and 3.6-\micron{} TRGBs (see
Section~\ref{sec:agbid}). Their M-star sample thus includes
significant contamination from RGB stars and our M-star sample
excludes the early AGB stars.

\begin{figure}[h!]
\epsscale{1.2} 
\plotone{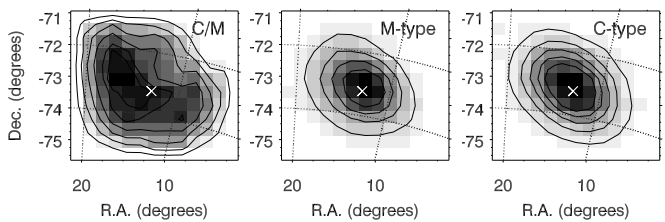}
\figcaption{C/M ratio for the bar and wing regions of the SMC. The
tail is not included because the number of AGB stars in the tail
region is not sufficient to reliably compute a C/M ratio. M-type stars
include aO-AGB and O-AGB stars, and C-type include C-AGB and x-AGB
stars. {\it Left panel:} C/M ratio in gray scale, with contours at
0.1, 0.2, 0.3, 0.4, 0.5, and 0.55. {\it Middle panel:} M-type
sources. Contours range from 25 to 100 stars, in steps of 25. {\it
Right panel:} C-type sources. Contours range from 10 to 50 stars, in
steps of 10. In all panels, the bin size is 0.4 degrees, and the white
cross marks the center of the RGB population.\label{fig:cm}}
\end{figure}

The procedure used here mimics \citet{cioni06smc}. We use a bin size
of 0.4 deg, yielding a 0.16 deg$^2$ area. The results are boxcar
smoothed with width $= 2$. The number count of AGB stars in the tail
region of the SMC ($<$$5$ stars per bin; Figs.~\ref{fig:disto} -- \ref{fig:distc} is
too small to include here without introducing substantial errors. In
principle, such an analysis can be done with careful consideration of
the uncertainties, but this is beyond the scope of this paper. We show
only the bar/wing area, and only include bins with more than 5 C-type
and 5 M-type stars.

The resulting C/M map (Fig.~\ref{fig:cm}, left panel) is similar to
the \citet{cioni06smc} map, but not identical. We confirm a region
with an enhanced C/M ratio near ${\rm R.A.} = 15\degr$, ${\rm Dec.} =
-73\degr$, though our map shows this feature to be more elongated to
that found by \citet{cioni06smc}. The strong peak near ${\rm R.A.} =
10\degr$, ${\rm Dec.} = -73\degr$ seen by \citet{cioni06smc} is much
weaker here, but they show this peak to be at a low confidence
level. Assuming that the C/M ratio corresponds directly to
metallicity, our C/M map suggests that there is a minimum in the
metallicity in the NE corner of the bar/wing, and that this minimum
stretches to the SMC center. The metallicity appears to increase
towards the south edge of the bar. This is consistent with the findings in
\citet{cioni06smc}, despite our exclusion of sources below the TRGB.

\subsection{Mass Loss}
\label{sec:mdot}
\begin{figure}[h!]
\epsscale{1} 
\plotone{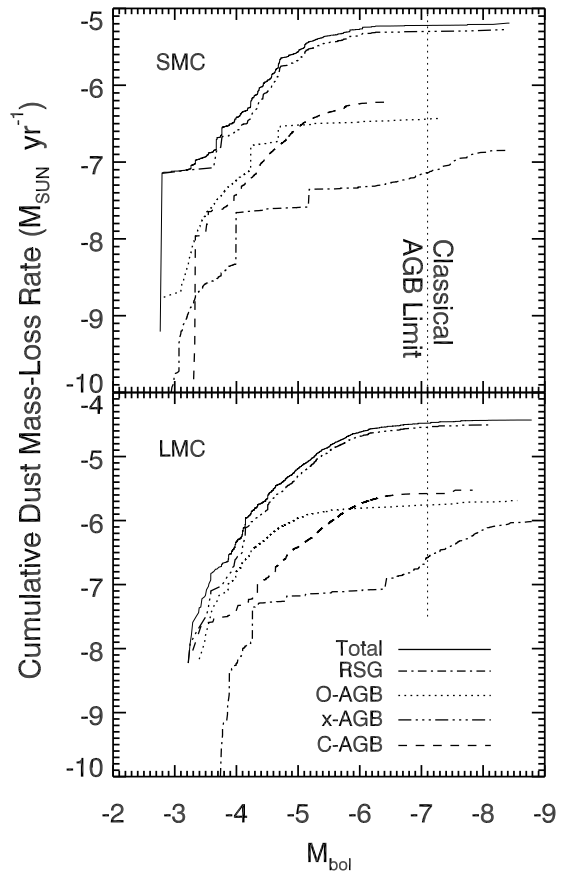}
\figcaption{Cumulative dust mass-loss rates for AGB and RSG stars in
  the Magellanic Clouds.  The mass-loss rates were estimated based on
  the $[3.6]-[8]$ color, using the radiative transfer grids from
  \citet{groenewegen06}. See text.\label{fig:mdot}}
\end{figure}

The total amount of dust input into the ISM from cool evolved stars
must be measured if we are to obtain a complete picture of the
lifecycle of dust within galaxies. The most reliable way to measure
accurate dust mass-loss rates is by radiative transfer modeling of
individual sources, though other methods using the IR excess
\citep{srinivasan09} and the IR color \citep[cf.][]{groenewegen06} provide
a good first-order estimate of the mass-loss rate especially when
considering large populations of stars. 

A detailed analysis of the dust mass loss from AGB and RSG stars is
beyond the scope of this paper.  Here, we provide a simple analysis of
the mass-loss rates using the IR colors and refer the reader to a
forthcoming follow-up paper for a more thorough analysis (M.L.\,Boyer
et al., in preparation).

We estimate dust mass-loss rates here by applying the
\citet{groenewegen06} radiative transfer grids to the measured
$[3.6]-[8]$ color. For C-AGB and x-AGB stars, we assume $T_{\rm eff} =
3600$~K and a dust composition of 85\% amorphous carbon and 15\%
SiC. We lump the aO-AGB stars with the O-AGB stars and assume $T_{\rm
eff} = 3297$~K and a dust composition of 60\% silicates and 40\%
AlOx. The \citet{groenewegen06} models were computed with AGB stars in
mind, but we also apply the O-rich models to the RSG stars. The rates
are scaled according to \cite{vanloon06b}, where $\dot{M}_{\rm dust}
\propto \tau \psi^{0.5} L^{0.75}$, $L$ is the stellar luminosity, and
$\tau$ is the dust optical depth at 11.75~\micron{}. The dust-to-gas
ratio ($\psi$) scales as $\psi = \psi_\odot 10^{\rm [Fe/H]}$ and
$\psi_\odot = 0.005$ \citep{vanloon05b}. However, we note that the
dust-to-gas ratio is uncertain, and may not be the same for O-rich and
C-rich stars. It has been suggested that C-rich stars may have
dust-to-gas ratios similar to Galactic values, even in low metallicity
environments \citep{habing96,groenewegen07}. Using the Galactic
dust-to-gas ratio would result in higher mass-loss rates for x-AGB and
C-AGB stars. The expansion velocity is also uncertain; here, we assume
$v_{\rm esc} = 10~{\rm km~s^{-1}}$, which may be an overestimate for
the carbon stars \citep[e.g.,][]{lagadec10}, thereby overestimating
their mass-loss rates.

Figure~\ref{fig:mdot} shows the resulting cumulative dust mass-loss
rates for the LMC and SMC. The LMC rates are slightly higher than
those presented by \citet{srinivasan09} using IR excesses in that
galaxy, but the overall trends are similar.  In both galaxies, the
x-AGB stars dominate the dust input by an order of magnitude despite
their small numbers (Table~\ref{tab:stats1}). The O-AGB and C-AGB
stars have similar total dust mass-loss rates, but the rates of
individual C-AGB stars are higher than the O-AGB star rates on
average. RSG stars do not contribute as strongly to the total dust
input, but their contribution increases significantly above the
classical AGB limit.

In the LMC, the O-AGB stars input more dust overall than C-AGB stars
at $M_{\rm bol} > -6$~mag.  The picture is different in the SMC, where
the C-AGB and O-AGB stars contribute similar amounts of dust until
$M_{\rm bol} \approx -5$~mag, where the C-AGB stars finally surpass
the O-AGB stars. This suggests that fainter O-rich sources in the SMC
may have difficulty producing enough dust to drive a wind.

\section{Summary and Conclusions}
\label{sec:conclusions}

The SAGE-SMC survey is the first to image the entire spatial extent of
the SMC (including bar, wing, and tail) at mid-IR to far-IR
wavelengths with high sensitivity and spatial resolution, thus
providing the first opportunity to study the full SMC population of
cool evolved stars at wavelengths where circumstellar dust
emits. Using near-IR and mid-IR photometric criteria, we find 2\,478
O-AGB, 1\,729 C-AGB, and 349 extreme (x-AGB) star candidates, along
with 1\,244 stars belonging to a new class of O-rich AGB stars
(aO-AGB). These stars represent the complete census of AGB stars in
the SMC.  We also classify 3\,325 RSG stars and 135\,437 RGB stars,
which represent the brighter, least contaminated portions of the full
populations. To compare the SMC evolved stars to those in the LMC, we
apply the same classification criteria to the SAGE-LMC data. Our
findings are summarized below:

\begin{itemize}

\item We find that O-rich sources have a higher occurrence of strong 8
and 24-\micron\ excess in the LMC, suggesting that O-rich dust is
produced more efficiently or that silicate emission is more prominent
in higher-metallicity environments. In fact, O-rich stars in general
are less numerous in the SMC, with a higher fraction of stars showing
evidence of a C-rich chemistry. 

\item The $[3.6]-[8]$ colors
indicate that SMC C-rich stars are as efficient at producing dust as
their higher-metallicity LMC counterparts.

\item The RSG, AGB, and RGB stars contribute near equal amounts of
flux to the global (extended + point-source) 3.6~\micron{} flux within
the bar and wing area. However, the RSG stars show a stronger
contribution in the wing. At 24~\micron{}, the x-AGB stars dominate
the total point-source flux even though they are $<$3\% of the
population.

\item In general, the characteristics of the AGB stars in the SMC are
similar to those in the LMC, showing only small differences in the
median SEDs and in the distributions among 8-\micron\ flux and
$[3.6]-[8]$ color. However, the RSG stars in the LMC reach much redder
$[3.6]-[8]$ colors than in the SMC, indicating more efficient RSG dust
production at higher metallicity. This is not the case for the O-rich
AGB stars.

\item Among the evolved stars, there is a population of far-IR sources
  in both the LMC and SMC, whose 24-\micron\ flux exceeds the
  8-\micron\ flux.  Most of these are likely to be YSOs and compact
  \ion{H}{2} regions, and it is unclear what portion may be very dusty
  evolved stars.

\item Very few evolved star candidates are detected in the SAGE-SMC
70-\micron{} images, and the SEDs of those detected indicate that they
are likely to be YSOs or other 70~\micron\ sources along the
line-of-sight.

\item The bulk of the evolved star population is restricted to the bar
and wing regions, though the old RGB distribution extends to the
tail. The distribution of young RSG and OB stars may also indicate
continued star formation in the gas-rich tail.

\item In the bar, the ratio of C-type to M-type AGB stars indicates only
small fluctuations in metallicity, with a peak in C/M to the
northeast of the center of the RGB population.

\item A preliminary estimate of the dust mass-loss rates in AGB and
  RSG stars suggests that the very dusty x-AGB stars dominate the dust
  return in both galaxies, despite their very small numbers. The dust
  input in both galaxies is therefore currently dominated by a C-rich
  chemistry. O-AGB and RSG stars appear to play a larger role in the
  LMC than in the SMC, particularly for the fainter O-AGB stars and
  the brighter RSG stars.
\end{itemize}

\acknowledgements

This work is supported by NASA via JPL contracts 1309827 and
1340964. We thank the referee for careful reading of the manuscript.

\bibliographystyle{astroads}
\bibliography{resubmitted}

\end{document}